\documentclass[11pt,a4paper]{article}
\usepackage[margin=2.5cm]{geometry}
\usepackage{iftex}
\usepackage{amsmath}
\ifPDFTeX
  \usepackage[T1]{fontenc}
  \usepackage[utf8]{inputenc}
  \usepackage{newtxtext}                  
\fi
\usepackage{newtxmath}                   
\ifXeTeX
  \usepackage[no-math]{fontspec}
  \DeclareSymbolFont{operators}{OT1}{ntxtlf}{m}{n}   
  \SetSymbolFont{operators}{bold}{OT1}{ntxtlf}{b}{n}
\fi
\usepackage{bm}
\usepackage{graphicx}
\usepackage{booktabs}
\usepackage[section]{placeins}
\usepackage[authoryear,round]{natbib}
\usepackage{xcolor}
\usepackage{xspace}
\usepackage{tikz}
\usetikzlibrary{arrows.meta,positioning,fit,backgrounds,calc,shapes.geometric,shapes.misc,decorations.pathreplacing,patterns,matrix}
\definecolor{cBlue}{HTML}{2A78D6}
\definecolor{cOrange}{HTML}{EB6834}
\definecolor{cGreen}{HTML}{1BAF7A}
\definecolor{cYellow}{HTML}{EDA100}
\definecolor{cPink}{HTML}{E87BA4}
\definecolor{cPurple}{HTML}{6250D6}
\definecolor{cRed}{HTML}{E34948}
\definecolor{cInk}{HTML}{0B0B0B}
\definecolor{cInkTwo}{HTML}{52514E}
\definecolor{cGrid}{HTML}{D9D8D4}
\definecolor{cPhob}{HTML}{86B6EF}
\definecolor{cPhil}{HTML}{EC835A}
\definecolor{cLiq}{HTML}{CFE0F6}
\tikzset{
  >={Stealth[length=2.4mm]},
  blk/.style={draw=#1!75!black, fill=#1!10, rounded corners=3pt, line width=0.8pt, align=center,
              minimum height=0.95cm, inner sep=4pt, font=\small},
  blk/.default=cBlue,
  sblk/.style={blk=#1, minimum height=0.7cm, font=\footnotesize, inner sep=3pt},
  sblk/.default=cBlue,
  op/.style={circle, draw=cInk, fill=white, line width=0.8pt, inner sep=1pt, minimum size=0.55cm, font=\small},
  arr/.style={->, line width=0.9pt, draw=cInk},
  garr/.style={->, line width=1.0pt, dashed, draw=cRed},
  lbl/.style={font=\footnotesize, text=cInkTwo, align=center},
  grp/.style={draw=#1!70!black, dashed, rounded corners=6pt, line width=0.7pt, inner sep=7pt},
  grp/.default=cInkTwo,
  ttl/.style={font=\small\bfseries, text=cInk},
}

\usepackage{algorithm}
\usepackage{algpseudocode}
\definecolor{linkblue}{RGB}{0,84,180}
\usepackage[colorlinks=true,linkcolor=linkblue,citecolor=linkblue,urlcolor=linkblue,filecolor=linkblue]{hyperref}
\newcommand{\xref}[1]{\ifcsname xr@#1\endcsname{\color{linkblue}\csname xr@#1\endcsname}\else\textbf{??}\fi}
\expandafter\def\csname xr@sec:notation\endcsname{S1}
\expandafter\def\csname xr@tab:notation\endcsname{S1}
\expandafter\def\csname xr@sec:archdetails\endcsname{S2}
\expandafter\def\csname xr@tab:hyper\endcsname{S2}
\expandafter\def\csname xr@fig:si_inputs\endcsname{S1}
\expandafter\def\csname xr@fig:si_haar\endcsname{S2}
\expandafter\def\csname xr@alg:sampler\endcsname{S1}
\expandafter\def\csname xr@sec:runs\endcsname{S3}
\expandafter\def\csname xr@tab:si_runs\endcsname{S3}
\expandafter\def\csname xr@fig:si_sched\endcsname{S3}
\expandafter\def\csname xr@fig:si_hist\endcsname{S4}
\expandafter\def\csname xr@sec:acc\endcsname{S4}
\expandafter\def\csname xr@tab:si_acc\endcsname{S4}
\expandafter\def\csname xr@tab:si_fam\endcsname{S5}
\expandafter\def\csname xr@tab:si_uniform\endcsname{S6}
\expandafter\def\csname xr@tab:si_pol\endcsname{S7}
\expandafter\def\csname xr@fig:si_constraints\endcsname{S5}
\expandafter\def\csname xr@fig:si_force\endcsname{S6}
\expandafter\def\csname xr@fig:si_refine\endcsname{S7}
\expandafter\def\csname xr@sec:gal\endcsname{S5}
\expandafter\def\csname xr@fig:si_g2t\endcsname{S8}
\expandafter\def\csname xr@fig:si_g2w\endcsname{S9}
\expandafter\def\csname xr@fig:si_g3t\endcsname{S10}
\expandafter\def\csname xr@fig:si_g3w\endcsname{S11}
\expandafter\def\csname xr@sec:stiff\endcsname{S6}
\expandafter\def\csname xr@alg:stiff\endcsname{S2}

\usepackage{caption}
\graphicspath{{figures/}}
\newcommand{\upi}{\mathrm{\pi}}
\newcommand{\dfeno}{\texorpdfstring{\textbf{DropFENO}}{DropFENO}\xspace}
\newcommand{\dfenotitle}{\texorpdfstring{DropFENO}{DropFENO}}
\newsavebox{\fitb}
\newcommand{\fitbox}[1]{\sbox{\fitb}{#1}\ifdim\wd\fitb>\linewidth\resizebox{\linewidth}{!}{\usebox{\fitb}}\else\usebox{\fitb}\fi}
\algrenewcommand\textproc{\textrm}
\algrenewcommand\algorithmiccomment[1]{\hfill{\color{cInkTwo}\footnotesize$\triangleright$ #1}}

\newcommand{\NAblFnoBase}{6.7\times10^{-4}}

\newcommand{\NAblFnoLong}{1.8\times10^{-4}}

\newcommand{\NAblNoPrior}{3.2\times10^{-3}}

\newcommand{\NAblParRankFour}{13.9}
\newcommand{\NAblParRankSixtyFour}{75.8}
\newcommand{\NAblParRankThirtyTwo}{42.8}

\newcommand{\NAblRankEight}{2.0\times10^{-3}}

\newcommand{\NAblRankFour}{2.3\times10^{-3}}

\newcommand{\NAblRankSixtyFour}{1.7\times10^{-3}}

\newcommand{\NAblRankThirtyTwo}{2.0\times10^{-3}}

\newcommand{\NAblSeedOne}{6.2\times10^{-4}}

\newcommand{\NAblTfnoLong}{9.0\times10^{-4}}

\newcommand{\NAnisoMaxFnoThree}{1}
\newcommand{\NAnisoMaxTfnoThree}{1}
\newcommand{\NAnisoMaxWnoThree}{0}
\newcommand{\NAnisoRefMaxThree}{36}
\newcommand{\NDeltaVapNinety}{0.0107}
\newcommand{\NDeltaVapNinetyThree}{0.0375}
\newcommand{\NEdropA}{0.311}
\newcommand{\NEdropB}{0.124}
\newcommand{\NEdropC}{0.398}
\newcommand{\NFerrStrThreeFno}{0.04}
\newcommand{\NFerrStrThreeTfno}{0.06}
\newcommand{\NFerrStrThreeWno}{0.07}
\newcommand{\NFerrThreeFno}{0.15}
\newcommand{\NFerrThreeTfno}{0.19}
\newcommand{\NFerrThreeWno}{0.21}
\newcommand{\NFmaxThree}{1.80}
\newcommand{\NFpinErrFnoTwo}{0.15}

\newcommand{\NFstrAgreeFnoTwo}{25}
\newcommand{\NFstrAgreeTfnoTwo}{24}
\newcommand{\NFstrAgreeWnoTwo}{22}
\newcommand{\NFstrErrFnoTwo}{3.9\times10^{-6}}

\newcommand{\NGTratioMean}{1.01}
\newcommand{\NGTratioSpread}{0.09}
\newcommand{\NHFstraddleDev}{7\times10^{-4}}

\newcommand{\NHigherStepFnoThree}{1}
\newcommand{\NHigherStepFnoTwo}{1}

\newcommand{\NHigherStepTfnoTwo}{1}

\newcommand{\NHigherStepWnoTwo}{1}
\newcommand{\NHigherStrFnoTwo}{2}
\newcommand{\NHigherStrTfnoTwo}{2}
\newcommand{\NHigherStrWnoTwo}{2}

\newcommand{\NIoUStepFnoThree}{0.874}

\newcommand{\NKinRatioHMax}{1.002}
\newcommand{\NKinRatioHMin}{1.000}
\newcommand{\NKinRatioMax}{1.081}
\newcommand{\NKinRatioMin}{1.078}

\newcommand{\NLtwoRandFnoTwo}{0.057}

\newcommand{\NLtwoRandTfnoTwo}{0.120}

\newcommand{\NLtwoRandWnoTwo}{0.118}

\newcommand{\NNotBestRandFnoTwo}{1}

\newcommand{\NNpinTwo}{13}
\newcommand{\NNstrTwo}{26}
\newcommand{\NNstraddleTwo}{26}

\newcommand{\NParamsFnoTwo}{6.30}

\newcommand{\NParamsTfnoTwo}{42.8}

\newcommand{\NPhiMaxFortyThree}{-0.89}

\newcommand{\NRayOpFno}{0.09}
\newcommand{\NRayOpTfno}{0.23}
\newcommand{\NRayOpWno}{0.26}
\newcommand{\NRaySmoothFno}{0.56}

\newcommand{\NRayTranslFno}{0.031}

\newcommand{\NRayWhiteFno}{3.80}

\newcommand{\NSlopeRatioHMax}{1.001}
\newcommand{\NSlopeRatioHMin}{1.000}
\newcommand{\NSlopeRatioMax}{1.070}
\newcommand{\NSlopeRatioMin}{1.063}
\newcommand{\NSpeedRandFnoThree}{188}
\newcommand{\NSpeedRandFnoTwo}{123}

\newcommand{\NSrStepFnoOne}{1.0\times10^{-3}}
\newcommand{\NSrStepFnoTwo}{9.2\times10^{-3}}

\newcommand{\NSrUniFnoOne}{2.5\times10^{-5}}
\newcommand{\NSrUniFnoTwo}{2.7\times10^{-3}}

\newcommand{\NStrThreeDev}{1.0}
\newcommand{\NStrThreeN}{5}
\newcommand{\NThOpRangeHiFnoTwo}{107}

\newcommand{\NThOpRangeLoFnoTwo}{101}

\newcommand{\NThParRefWFour}{104}
\newcommand{\NThParRefWSix}{103}

\newcommand{\NThParRefWTwelve}{94}

\newcommand{\NThPerpRefWFour}{116}
\newcommand{\NThPerpRefWSix}{93}

\newcommand{\NThPerpRefWTwelve}{130}

\newcommand{\NThRefRangeHiTwo}{124}
\newcommand{\NThRefRangeLoTwo}{74}
\newcommand{\NThRmsStrFnoTwo}{13}
\newcommand{\NThRmsStrTfnoTwo}{16}
\newcommand{\NThRmsStrWnoTwo}{16}
\newcommand{\NTinfRandFnoThree}{628}
\newcommand{\NTinfRandFnoTwo}{47}
\newcommand{\NTinfRandTfnoThree}{575}

\newcommand{\NTinfRandWnoThree}{778}

\newcommand{\NTpolRandFnoTwo}{6.4}

\newcommand{\NTrefRandFnoThree}{117.8}
\newcommand{\NTrefRandFnoTwo}{5.8}

\newcommand{\NVapFracNinetyThree}{0.34}

\newcommand{\NVrelRandFnoTwo}{1.7\times10^{-4}}

\newcommand{\NWsCapThree}{313}
\newcommand{\NWsGainFive}{1.5}
\newcommand{\NWsGainFour}{1.8}
\newcommand{\NWsGainThree}{6.0}
\newcommand{\NWsN}{30}
\newcommand{\NWsOpFive}{598}
\newcommand{\NWsOpFour}{363}
\newcommand{\NWsOpThree}{25}
\newcommand{\NXpinA}{0.6}

\newcommand{\NXrelA}{1.6}

\newcommand{\NXsnapA}{-1.0}

\newcommand{\NYoungDevMaxThree}{3.9}
\newcommand{\NYoungDevMaxTwo}{2.3}
\newcommand{\NYoungDevRmsTwo}{1.3}

\newcommand{\NdEmaxRandFnoTwo}{4.8\times10^{-2}}

\newcommand{\NdEmedRandFnoThree}{1.0\times10^{-3}}
\newcommand{\NdEmedRandFnoTwo}{6.7\times10^{-4}}
\newcommand{\NdEmedRandTfnoThree}{2.0\times10^{-3}}
\newcommand{\NdEmedRandTfnoTwo}{2.0\times10^{-3}}
\newcommand{\NdEmedRandWnoThree}{1.8\times10^{-3}}
\newcommand{\NdEmedRandWnoTwo}{1.9\times10^{-3}}
\newcommand{\NdEmedStepFnoThree}{8.5\times10^{-3}}
\newcommand{\NdEmedStepFnoTwo}{1.0\times10^{-3}}
\newcommand{\NdEmedStepTfnoThree}{1.2\times10^{-2}}

\newcommand{\NdEmedStepWnoThree}{1.4\times10^{-2}}
\newcommand{\NdEmedStepWnoTwo}{8.0\times10^{-3}}
\newcommand{\NdEmedStrFnoThree}{2.7\times10^{-3}}
\newcommand{\NdEmedStrFnoTwo}{4.2\times10^{-3}}

\newcommand{\NdEmedUniFnoThree}{6.9\times10^{-5}}
\newcommand{\NdEmedUniFnoTwo}{6.0\times10^{-5}}

\newcommand{\NdEmedUniTfnoTwo}{1.9\times10^{-4}}

\newcommand{\NdEmedUniWnoTwo}{3.1\times10^{-4}}

\newcommand{\NGAlphaCHalf}{30.8}
\newcommand{\NGAlphaCOne}{14.9}
\newcommand{\NGBoMin}{0.256}
\newcommand{\NGCritN}{4}
\newcommand{\NGCritOpDev}{0.01}
\newcommand{\NGCritRefDev}{1.3}
\newcommand{\NGDeltaCos}{0.805}
\newcommand{\NGFlatBOneForty}{68}
\newcommand{\NGFlatBSixty}{26}
\newcommand{\NGFlatHOneForty}{24}
\newcommand{\NGFlatHSixty}{27}
\newcommand{\NGForceBal}{0.07}
\newcommand{\NGHeldMaxHalf}{30}
\newcommand{\NGHeldMaxOne}{15}
\newcommand{\NGHsixtyOne}{13}

\newcommand{\NGNheld}{16}
\newcommand{\NGOpBerr}{0.64}
\newcommand{\NGOpHerr}{0.29}
\newcommand{\NGPinErrFno}{0.17}
\newcommand{\NGPinOne}{1.4}
\newcommand{\NGPinZero}{1.0}
\newcommand{\NGPlatOne}{1.9\times10^{-3}}

\newcommand{\NGSlidMinHalf}{32}
\newcommand{\NGSlidMinOne}{16}
\newcommand{\NGSnapOne}{-1.4}
\newcommand{\NGSnapZero}{-1.0}
\newcommand{\NGTinfFno}{14}

\newcommand{\NGTrainFno}{0.9}
\newcommand{\NGTrainTfno}{0.7}
\newcommand{\NGTref}{5.8}
\newcommand{\NGXeqErr}{0.09}
\newcommand{\NGXeqErrMax}{0.19}
\newcommand{\NGYLdevB}{4.8}
\newcommand{\NGYLdevBzero}{5.5}
\newcommand{\NGYLdevH}{4.4}
\newcommand{\NGYLdevHzero}{1.6}
\newcommand{\NGdERandomFno}{9.0\times10^{-4}}

\newcommand{\NGdEStepFno}{2.1\times10^{-3}}

\newcommand{\NGdEUniformFno}{2.7\times10^{-4}}

\title{\Large \dfenotitle: A Data-Free, Physics-Based Operator-Learning Framework for
Droplet Equilibria on Mixed-Wettability Patterned Surfaces}

\author{Ganesh Sahadeo Meshram$^{1,*}$, Partha Pratim Chakrabarti$^{2}$ and Suman Chakraborty$^{1}$\\[4pt]
\small $^{1}$Department of Mechanical Engineering, Indian Institute of Technology Kharagpur,\\ \small Kharagpur 721302, India\\[2pt]
\small $^{2}$Department of Computer Science and Engineering, Indian Institute of Technology Kharagpur,\\ \small Kharagpur 721302, India\\[2pt]
\small $^{*}$Corresponding author: ganeshmeshram.iitkgp@gmail.com}

\date{}

\begin{document}
\maketitle

\begin{abstract}
We study whether the equilibrium of a drop on a chemically patterned wall, and the capillary forces that the pattern
exerts on it, can be learned from the free energy alone, without solution data. We introduce \dfeno (Droplet
Free-Energy Neural Operator), in which Fourier (FNO), tensor-factorised Fourier (TFNO) and wavelet (WNO) neural
operators are trained in two and three dimensions on the diffuse-interface free energy of the drop, with its volume and
centroid constraints imposed as penalties. The predictions are tested with the virtual-work (Hellmann--Feynman)
theorem: the force exerted by the pattern on a held drop is a wall integral of the equilibrium field which, for a drop
straddling a wettability step, reduces identically to the contact-line law $F=\sigma\Delta\cos\theta\,L$. \dfeno
reproduces this law and the energy landscape of a drop crossing a step, and reproduces the pinning of the receding
contact line less accurately. The residual error lies in the soft modes of the capillary energy (small displacements of
the interface and the contact line). This explains why the energy is predicted accurately while contact-line positions
are not, and why on striped walls the operators return Cassie-like states instead of the edge-pinned states whose
anisotropic apparent angles the reference minimiser shares with lattice Boltzmann simulations. On held-out patterned
walls the median energy of the predictions lies $\NdEmedRandFnoTwo$ of the free-drop energy above the best known
minimum (FNO, two dimensions), at about one hundredth of the cost of a per-instance minimisation, and a refinement by a
classical solver started from the prediction returns the reference state in almost all cases. With gravity included,
the virtual-work force also predicts where a drop comes to rest on an inclined wall with a wettability step and the
inclination at which it slides off, $\sin\alpha_c=\Delta\cos\theta/\upi Bo$ in two dimensions; \dfeno predicts this
threshold accurately and the rest position less accurately, because the rest position is set by the pinned contact
line. We also show that
imposing the volume through the linear liquid fraction supersaturates the vapour by the Gibbs--Thomson effect, so that
about one third of the liquid is held in the vapour in three dimensions; an indicator that is flat in the bulk phases
removes this artefact exactly.
\end{abstract}


\section{Introduction}\label{sec:intro}

A drop on a chemically patterned wall settles into a shape that minimises its free energy under the constraint of
fixed volume. The minimiser encodes most of what one wants to know about the wetting of the pattern: the apparent
contact angles and their anisotropy on stripes \citep{Brandon2003,Jansen2012}, the pinned or depinned state of the
contact line at a wettability edge \citep{Gibbs1906,Oliver1977,Kusumaatmaja2007}, and, through the dependence of
the minimum energy on the position of the drop, the capillary force that drives a drop across a wettability
step or gradient \citep{Brochard1989,Subramanian2005}. Computing it is a constrained, nonconvex variational problem
that must be solved anew for every pattern and every position. Surface Evolver \citep{Brakke1992}, diffuse-interface
energy minimisation and the long-time limit of lattice Boltzmann or phase-field dynamics are the standard tools
\citep{Meshram2023FMFP,Meshram2024AIP,Meshram2025JAST};
each solve costs from seconds to hours, and parametric studies over pattern families multiply that cost.

Neural operators learn maps between function spaces from samples of their action and evaluate them at a cost
independent of the solver that produced the samples \citep{Lu2021,Li2021,Kovachki2023}. The Fourier neural
operator (FNO) of \citet{Li2021}, its tensor-factorised variant (TFNO) \citep{Kossaifi2023} and the wavelet neural
operator (WNO) \citep{Tripura2023} are three representative architectures, which parametrise the global kernel
respectively by dense spectral weights, by a low-rank canonical-polyadic factorisation of those weights, and by a
mixing of wavelet coefficients. Trained on data, they are surrogates of a numerical solver and inherit both its
cost (the training set) and its errors. Physics-informed training replaces or supplements the data by the
residual of the governing equations \citep{Raissi2019,Wang2021,Li2024PINO}, and data-driven and physics-informed
networks have been applied to the classification of wetting states and to droplet wetting and spreading on rough
surfaces \citep{Meshram2025Procedia,Meshram2026Langmuir,Meshram2026AAAI}; for problems that derive from a
variational principle the residual can be replaced by the energy itself, as in the deep Ritz method for single
instances \citep{E2018} and in variational operator learning \citep{Goswami2022}. For a free-boundary problem with a
contact line, the energy formulation is convenient: the Euler--Lagrange equation is a nonlinear elliptic equation with
a nonlinear boundary condition at the wall and Lagrange multipliers for the volume and centroid constraints, whereas the
energy is a sum of local terms that involve at most first derivatives of the field, and its constrained minimiser
satisfies all of these conditions.

Two questions remain open, and they are the subject of this paper; we address them with a framework that we call
\dfeno (Droplet Free-Energy Neural Operator). The first is whether an operator trained
\emph{only} on the free energy (with no solution field ever computed, stored or shown to it) reaches the
accuracy of a converged per-instance minimiser over a broad family of chemically heterogeneous walls, in two and in
three dimensions, and how the three architectures compare when the solution is a sharp interface meeting a
patterned wall. The second, which is more important for the physics, is how to verify such a prediction
when no data exist. Pointwise agreement with a reference field is necessary but says little about the physics,
because the quantities of interest in wetting (forces, angles, pinning) are functionals of the field that
can be sensitive to small errors near the contact line. We show that the constrained-minimisation structure itself
provides a precise test. By the Hellmann--Feynman (virtual-work) theorem the force exerted by the pattern on a
drop held at equilibrium is an explicit wall integral of the equilibrium field, and for a drop straddling a
wettability step this integral reduces identically to the contact-line force law $F=\sigma\Delta\cos\theta\,L$, with
$L$ the wetted chord on the step. The force law is thus an exact property of the true minimiser which an approximate field does not satisfy in
general; testing an operator against it, and against the slope of its own energy landscape, measures whether it has
learnt the physics, not only how close it is to a reference solution.

The problem we consider is the static counterpart of the dynamics in our two- and three-dimensional lattice
Boltzmann simulations of drops that migrate across hybrid and striped wettability patterns. There, the force that holds a drop on a step was found to follow
the contact-line law with the chord of the footprint, and stripes produced a pronounced anisotropy of the apparent
angles. Here the same questions are asked of the equilibrium problem, without any dynamics and without any data.
We show that (i) \dfeno reproduces the contact-line force law, the energy landscape of a drop crossing a step and,
less accurately, the pinning of the rear contact line at the step, as emergent properties of predictions made
independently at each position; (ii) all three \dfeno operators, trained only on the energy, predict equilibrium fields
whose median energy lies within a fraction of a per cent of the converged minimum, almost always in the basin of the correct
state, at a cost about two orders of magnitude lower than the per-instance solve; (iii) their residual error is
concentrated in the soft modes of the energy (displacements of the interface and of the contact line), which
explains why the energy is accurate while contact-line positions are not, why on striped walls the operators
return Cassie-like averaged states instead of the edge-pinned states that set the apparent angles, and why a
classical solver started from a prediction converges faster only to a coarse tolerance; (iv) the architectures differ:
the dense FNO is the most accurate, the factorised TFNO comes within a factor of three of it with $150$ times fewer
parameters, and the local wavelet kernel of the WNO brings no advantage for a problem whose response to the pattern is
non-local; and (v) with gravity, the same force balance gives the rest position of a drop on an inclined wall with a
wettability step and the inclination at which it slides off. A by-product of the
analysis is a constraint artefact of general relevance to diffuse-interface and physics-informed calculations: a
volume constraint imposed through the linear liquid fraction supersaturates the vapour by the Gibbs--Thomson
(Kelvin) effect, which in three dimensions can hold a third of the liquid or evaporate the drop altogether and which
biases the force inferred from the energy landscape; an indicator that is flat in the bulk phases removes it
exactly.

The paper is organised as follows. \hyperref[sec:formulation]{Section~\ref*{sec:formulation}} states the variational problem, derives the
virtual-work force and the contact-line law, and describes the discretisation. \hyperref[sec:methods]{Section~\ref*{sec:methods}} describes
the data-free training, the three operators and the reference solver. \hyperref[sec:verification]{Section~\ref*{sec:verification}} verifies the
reference minimiser against Young's law, the vapour state and the force identity, with both choices of the liquid
indicator. \hyperref[sec:accuracy]{Section~\ref*{sec:accuracy}} assesses the accuracy, basin selection and cost of the operators over the
pattern family. \hyperref[sec:physics]{Section~\ref*{sec:physics}} turns to the physics: the force on a drop crossing a step in two and three
dimensions, and the apparent angles on stripes. \hyperref[sec:gravity]{Section~\ref*{sec:gravity}} adds gravity and treats a
drop on an inclined wall. \hyperref[sec:ablation]{Section~\ref*{sec:ablation}} examines the factorisation rank, the prior,
the training budget and the transfer to a finer grid; \hyperref[sec:discussion]{Section~\ref*{sec:discussion}} interprets the results through the
stiffness of the operator error, and \hyperref[sec:conclusions]{Section~\ref*{sec:conclusions}} concludes.

\section{Problem formulation}\label{sec:formulation}

\subsection{Free energy of a drop on a chemically patterned wall}
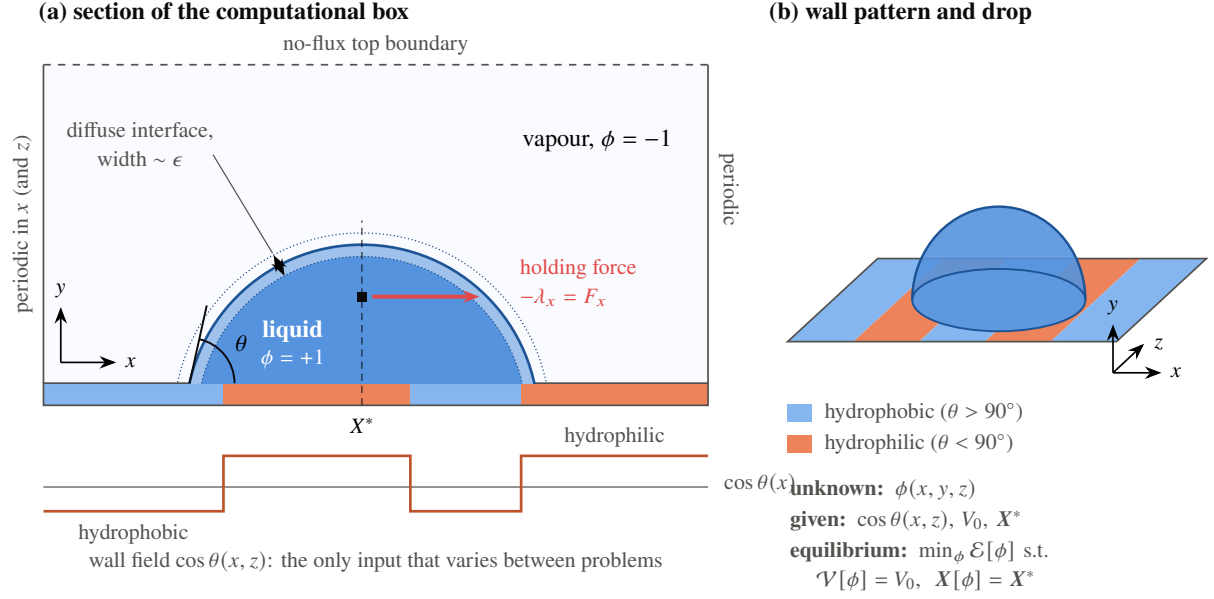
\begin{figure}
\centering
\fitbox{
\begin{tikzpicture}[font=\small]
\begin{scope}
  \node[ttl] at (-0.2,5.35) [anchor=west] {(a) section of the computational box};
  \fill[cBlue!3] (0,0) rectangle (9.6,4.6);
  \draw[cInkTwo, line width=0.7pt, dashed] (0,4.6) -- (9.6,4.6) node[midway, above, lbl] {no-flux top boundary};
  \draw[cInkTwo, line width=0.7pt] (0,0) -- (0,4.6) (9.6,0) -- (9.6,4.6);
  \node[lbl, rotate=90] at (-0.3,2.3) {periodic in $x$ (and $z$)};
  \node[lbl, rotate=-90] at (9.9,2.8) {periodic};
  \fill[cPhob] (0,-0.32) rectangle (2.6,0);
  \fill[cPhil] (2.6,-0.32) rectangle (5.3,0);
  \fill[cPhob] (5.3,-0.32) rectangle (6.9,0);
  \fill[cPhil] (6.9,-0.32) rectangle (9.6,0);
  \draw[cInkTwo, line width=0.7pt] (0,-0.32) rectangle (9.6,0);
  \begin{scope}
    \clip (0,0) rectangle (9.6,4.6);
    \fill[cBlue!80] (4.6,-0.55) circle (2.38);
    \fill[cBlue!45] (4.6,-0.55) circle (2.55) ;
    \fill[cBlue!80] (4.6,-0.55) circle (2.38);
    \draw[cBlue!60!black, line width=0.5pt, densely dotted] (4.6,-0.55) circle (2.72);
    \draw[cBlue!60!black, line width=0.5pt, densely dotted] (4.6,-0.55) circle (2.38);
    \draw[cBlue!70!black, line width=1.1pt] (4.6,-0.55) circle (2.55);
  \end{scope}
  \node[text=white, font=\small\bfseries] at (3.6,0.75) {liquid};
  \node[text=white, font=\footnotesize] at (3.6,0.35) {$\phi=+1$};
  \node[font=\small] at (8.0,3.5) {vapour, $\phi=-1$};
  \draw[<->, line width=0.7pt] ($(4.6,-0.55)+(118:2.38)$) -- ($(4.6,-0.55)+(118:2.72)$);
  \draw[cInkTwo, line width=0.5pt] ($(4.6,-0.55)+(118:2.75)$) -- (2.45,3.3);
  \node[lbl, anchor=west] at (0.2,3.45) {diffuse interface,\\ width $\sim\epsilon$};
  \coordinate (cl) at (2.11,0);
  \draw[cInk, line width=0.8pt] (cl) -- ++(77.5:1.15);
  \draw[cInk, line width=0.8pt] (cl) ++(0:0.65) arc (0:77.5:0.65);
  \node at ($(cl)+(38:0.95)$) {$\theta$};
  \draw[cInk, densely dashed] (4.6,-0.32) -- (4.6,2.35);
  \fill[cInk] (4.6,1.25) ++(-0.07,-0.07) rectangle ++(0.14,0.14);
  \node[anchor=north, font=\footnotesize] at (4.6,-0.36) {$X^*$};
  \draw[->, cRed, line width=1.6pt] (4.75,1.25) -- (6.3,1.25);
  \node[font=\footnotesize, text=cRed, align=left, anchor=west] at (6.75,1.45) {holding force\\$-\lambda_x=F_x$};
  \draw[->, line width=0.7pt] (0.25,0.3) -- (1.05,0.3) node[right, font=\footnotesize] {$x$};
  \draw[->, line width=0.7pt] (0.25,0.3) -- (0.25,1.1) node[above, font=\footnotesize] {$y$};
  \begin{scope}[yshift=-1.5cm]
    \draw[cInkTwo, line width=0.5pt] (0,0) -- (9.6,0);
    \draw[cOrange!80!black, line width=1.1pt] (0,-0.35) -- (2.6,-0.35) -- (2.6,0.45) -- (5.3,0.45) -- (5.3,-0.35) -- (6.9,-0.35) -- (6.9,0.45) -- (9.6,0.45);
    \node[lbl, anchor=west] at (9.7,0.05) {$\cos\theta(x)$};
    \node[lbl] at (1.3,-0.68) {hydrophobic};
    \node[lbl] at (8.25,0.73) {hydrophilic};
  \end{scope}
  \node[lbl, anchor=north] at (4.8,-2.3) {wall field $\cos\theta(x,z)$: the only input that varies between problems};
\end{scope}
\begin{scope}[xshift=10.75cm, yshift=0.6cm]
  \node[ttl] at (-0.4,4.75) [anchor=west] {(b) wall pattern and drop};
  \coordinate (O) at (0,0);
  \def\dx{0.95}
  \foreach \i in {0,...,4} {
    \pgfmathsetmacro{\c}{mod(\i,2)==0 ? 1 : 0}
    \ifnum\i=1 \def\col{cPhil}\else\ifnum\i=3 \def\col{cPhil}\else\def\col{cPhob}\fi\fi
    \fill[\col] ($(O)+(\i*\dx,0)$) -- ++(\dx,0) -- ++(1.3,1.2) -- ++(-\dx,0) -- cycle;
  }
  \draw[cInkTwo, line width=0.7pt] (O) -- ++(5*\dx,0) -- ++(1.3,1.2) -- ++(-5*\dx,0) -- cycle;
  \fill[cBlue!35] (3.05,0.6) ellipse (1.25 and 0.45);
  \fill[cBlue!80, opacity=0.9] (1.8,0.6) arc (180:0:1.25 and 1.35) arc (0:-180:1.25 and 0.45);
  \draw[cBlue!70!black, line width=1.0pt] (1.8,0.6) arc (180:0:1.25 and 1.35);
  \draw[cBlue!70!black, line width=0.8pt] (3.05,0.6) ellipse (1.25 and 0.45);
  \draw[->, line width=0.7pt] (4.7,-0.45) -- (5.4,-0.45) node[right, font=\footnotesize] {$x$};
  \draw[->, line width=0.7pt] (4.7,-0.45) -- (4.7,0.25) node[above, font=\footnotesize] {$y$};
  \draw[->, line width=0.7pt] (4.7,-0.45) -- (5.15,-0.03) node[right, font=\footnotesize] {$z$};
  \fill[cPhob] (0.0,-1.15) rectangle ++(0.35,0.25); \node[lbl, anchor=west] at (0.4,-1.03) {hydrophobic ($\theta>90^\circ$)};
  \fill[cPhil] (0.0,-1.6) rectangle ++(0.35,0.25); \node[lbl, anchor=west] at (0.4,-1.48) {hydrophilic ($\theta<90^\circ$)};
  \node[lbl, anchor=north west, text width=5.0cm, align=left] at (-0.1,-1.85)
     {\textbf{unknown:} $\phi(x,y,z)$\\[1pt] \textbf{given:} $\cos\theta(x,z)$, $V_0$, $\bm X^*$\\[1pt]
      \textbf{equilibrium:} $\min_\phi\mathcal E[\phi]$ s.t.\\ \hspace*{1.2em}$\mathcal V[\phi]=V_0$,\; $\bm X[\phi]=\bm X^*$};
\end{scope}
\end{tikzpicture}}
\caption{The equilibrium problem. (a) Section of the computational box: a liquid drop ($\phi=+1$) in its vapour
($\phi=-1$), separated by a diffuse interface of width $\sim\epsilon$, rests on a wall whose Young angle varies along
the wall (strip: hydrophobic blue, hydrophilic orange; profile below: $\cos\theta(x)$). The box is periodic in $x$ and
$z$ and closed by a no-flux boundary at the top. The centroid constraint holds the drop at $X^*$; its multiplier
$\lambda_x$ is the external force needed to hold it there, equal and opposite to the capillary force $F_x$ exerted by the
pattern. (b) The same problem in three dimensions: the wall field $\cos\theta(x,z)$ and the holding position
$\bm X^*$ are the only inputs; the drop volume $V_0$ is fixed.}\label{fig:setup}
\end{figure}

We consider a liquid drop in contact with its vapour above a flat, chemically heterogeneous wall $y=0$ (\hyperref[fig:setup]{figure~\ref*{fig:setup}}). The two
fluids are described by a phase field $\phi(\bm{x})\in[-1,1]$ ($\phi=+1$ liquid, $\phi=-1$ vapour) and the wall by
its local Young angle $\theta(x,z)$. The equilibrium is the minimiser of the Cahn--Hilliard free energy with the wall energy of Cahn
\citep{Cahn1977,Jacqmin2000}
\begin{equation}
\mathcal{E}[\phi]=\int_\Omega c\Big[\frac{W(\phi)}{\epsilon}+\frac{\epsilon}{2}|\nabla\phi|^2\Big]\,\mathrm{d}V
+\int_{\partial\Omega_w} f_w(\phi;\theta)\,\mathrm{d}A,\qquad
W=\tfrac14(1-\phi^2)^2,\quad c=\frac{3}{2\sqrt2},
\label{eq:energy}
\end{equation}
with the cubic wall energy
\begin{equation}
f_w(\phi;\theta)=-\cos\theta\;g(\phi),\qquad g(\phi)=\tfrac14\,(3\phi-\phi^3).
\label{eq:wall}
\end{equation}
The constant $c$ is chosen so that the planar profile $\phi=\tanh\!\big(n/\sqrt2\epsilon\big)$ carries the
interfacial tension $\sigma=c\int(W/\epsilon+\tfrac{\epsilon}{2}\phi'^2)\,\mathrm{d}n=1$, which is our unit of
energy per unit area. Because $g(\pm1)=\pm\tfrac12$, the wall energy of a wetted minus a dry patch is
$f_w(+1)-f_w(-1)=\gamma_{SL}-\gamma_{SV}=-\sigma\cos\theta$, so that Young's law $\cos\theta_Y=(\gamma_{SV}-\gamma_{SL})/\sigma$
is recovered locally; the cubic form has $g'(\pm1)=0$ and therefore does not alter the bulk values of $\phi$ at
the wall \citep{Jacqmin2000}. The wettability enters the problem \emph{only} through the wall field $\cos\theta(x,z)$.

The minimisation is subject to two constraints,
\begin{equation}
\mathcal{V}[\phi]=\int_\Omega h(\phi)\,\mathrm{d}V=V_0,\qquad
\bm{X}[\phi]=\frac{1}{V_0}\int_\Omega h(\phi)\,\bm{x}_\parallel\,\mathrm{d}V=\bm{X}^*,\qquad
h(\phi)=\tfrac12+g(\phi),
\label{eq:constraints}
\end{equation}
where $\bm{x}_\parallel=(x,z)$ are the wall-parallel coordinates. The volume constraint fixes the drop size,
$V_0=\tfrac43\upi R_0^3$ in three dimensions ($\upi R_0^2$ per unit depth in two). The liquid indicator
$h$ rises from $0$ to $1$ across the interface and, like the wall function $g$, is flat at $\phi=\pm1$; the reason
for this choice, rather than the linear liquid fraction $\tfrac12(1+\phi)$, is explained in \hyperref[sec:gt]{Section~\ref*{sec:gt}}. The centroid constraint pins the
drop at a prescribed position $\bm{X}^*$ on the pattern. On a heterogeneous wall a drop held at an arbitrary
$\bm{X}^*$ is not free: the constraint supplies exactly the external force needed to hold it there, and the
constrained minimiser $\phi^*(\cdot;\bm{X}^*)$ is the static shape of a drop held by that force. Varying
$\bm{X}^*$ therefore traces the quasi-static energy landscape $\mathcal{E}^*(\bm{X}^*)$ along which a drop
migrates, and its gradient is the capillary driving force.

\subsection{The virtual-work force and the contact-line force law}\label{sec:hf}
Translate the wall pattern rigidly by $\bm{\xi}$ relative to the fluid,
$\cos\theta(\bm{x}_\parallel)\to\cos\theta(\bm{x}_\parallel-\bm{\xi})$. The bulk energy and both constraints are
invariant under a simultaneous translation of the fluid, so the constrained minimum energy depends only on the
relative position of drop and pattern. Since $\phi^*$ is stationary with respect to all admissible variations,
the derivative of $\mathcal{E}^*$ with respect to $\bm{\xi}$ is the explicit (partial) derivative alone: this is the
Hellmann--Feynman, or virtual-work, theorem \citep{Feynman1939}:
\begin{equation}
\bm{F}\equiv-\frac{\partial\mathcal{E}^*}{\partial\bm{\xi}_{\rm drop}}
=\int_{\partial\Omega_w}\nabla_\parallel\cos\theta\;g(\phi^*)\,\mathrm{d}A .
\label{eq:hf}
\end{equation}
$\bm{F}$ is the net capillary force exerted by the pattern on the drop (positive towards higher $\cos\theta$). It
requires only the wall trace of the equilibrium field, involves no derivative of $\phi^*$, and is exact at the
constrained minimum. For a straight step $\cos\theta=\cos\theta_1+\Delta\cos\theta\,H(x-x_s)$ crossed by the
drop, $\nabla_\parallel\cos\theta=\Delta\cos\theta\,\delta(x-x_s)\,\hat{\bm{x}}$, and equation~(\ref{eq:hf}) becomes
\begin{equation}
F_x=\sigma\,\Delta\cos\theta\int\big[g(\phi^*(x_s,0,z))+\tfrac12\big]\,\mathrm{d}z\;\equiv\;\sigma\,\Delta\cos\theta\;L_\phi ,
\label{eq:chordlaw}
\end{equation}
where the constant $\tfrac12$ is the contribution of the dry wall (and, in a periodic box, of the returning step),
and $L_\phi$ is the length of the wetted chord of the footprint on the step line, weighted by the diffuse wall
trace. For a sharp interface $L_\phi$ is the geometric chord $L$ and (\ref{eq:chordlaw}) is the classical
contact-line force $F=\sigma\Delta\cos\theta\,L$ of a drop straddling a wettability step
\citep{Brochard1989,Subramanian2005}; in two dimensions $L$ is the unit depth. The derivation shows that the law is an
\emph{identity} of the equilibrium problem whenever the step is straddled, independent of the drop shape, and that
it fails as soon as a contact line sits on the step itself, where $g(\phi^*)$ takes intermediate values: this is
the pinning regime, in which the contact line is held at the step while its local angle sweeps between the two
Young angles \citep{Gibbs1906,Oliver1977}.

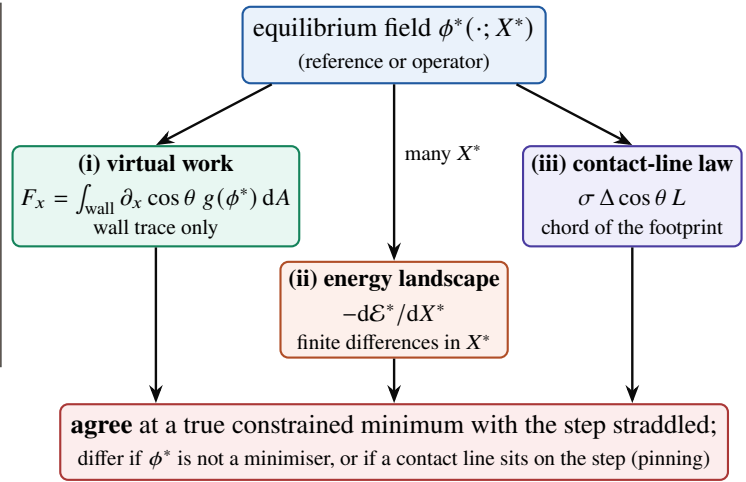
\begin{figure}
\centering
\fitbox{
\begin{tikzpicture}[font=\small]
\node[ttl, anchor=west] at (-0.1,3.1) {(a) drop across a step (top view)};
\fill[cPhob!80] (0,-2.2) rectangle (2.7,2.6);
\fill[cPhil!80] (2.7,-2.2) rectangle (5.6,2.6);
\draw[cInkTwo, line width=0.6pt] (0,-2.2) rectangle (5.6,2.6);
\draw[cInk, line width=1.0pt, dashed] (2.7,-2.2) -- (2.7,2.6);
\node[font=\footnotesize, fill=white, inner sep=1pt] at (2.7,2.3) {step $x_s$};
\node[font=\footnotesize, align=center] at (1.0,-1.75) {$\theta_1$ (phobic)};
\node[font=\footnotesize, align=center] at (4.6,-1.75) {$\theta_2$ (philic)};
\fill[cBlue!55, opacity=0.85] (3.0,0.2) ellipse (1.55 and 1.35);
\draw[cBlue!70!black, line width=1.0pt] (3.0,0.2) ellipse (1.55 and 1.35);
\draw[cRed, line width=2.2pt] (2.7,-1.124) -- (2.7,1.524);
\draw[<->, line width=0.7pt] (2.25,-1.124) -- (2.25,1.524) node[midway, left, font=\footnotesize, fill=cPhob!80, inner sep=1pt] {$L$};
\draw[->, cRed, line width=1.8pt] (3.6,0.2) -- (5.2,0.2);
\node[font=\footnotesize, text=cRed!80!black, fill=white, inner sep=1pt] at (4.95,0.55) {$F_x$};
\node[font=\footnotesize, text=white] at (3.35,-0.45) {footprint};
\node[lbl, anchor=north, align=center] at (2.8,-2.3) {$F_x=\sigma\,\Delta\cos\theta\,L$, \ $\Delta\cos\theta=\cos\theta_2-\cos\theta_1$};
\begin{scope}[xshift=6.2cm]
\node[ttl, anchor=west] at (0.6,3.1) {(b) three routes to the same force};
\node[blk=cBlue] (phi) at (4.6,2.05) {equilibrium field $\phi^*(\cdot;X^*)$\\[-1pt]\scriptsize (reference or operator)};
\node[sblk=cGreen, align=center] (r1) at (1.45,0.05) {\textbf{(i) virtual work}\\[1pt]$F_x=\int_{\rm wall}\partial_x\cos\theta\;g(\phi^*)\,\mathrm dA$\\[-1pt]\scriptsize wall trace only};
\node[sblk=cOrange, align=center] (r2) at (4.6,-1.45) {\textbf{(ii) energy landscape}\\[1pt]$-\mathrm d\mathcal E^*/\mathrm dX^*$\\[-1pt]\scriptsize finite differences in $X^*$};
\node[sblk=cPurple, align=center] (r3) at (7.75,0.05) {\textbf{(iii) contact-line law}\\[1pt]$\sigma\,\Delta\cos\theta\,L$\\[-1pt]\scriptsize chord of the footprint};
\draw[arr] (phi.south west) ++(0.4,0) -- (r1.north);
\draw[arr] (phi.south) -- node[right, font=\scriptsize, pos=0.4] {many $X^*$} (r2.north);
\draw[arr] (phi.south east) ++(-0.4,0) -- (r3.north);
\node[blk=cRed] (cmp) at (4.6,-3.2) {\textbf{agree} at a true constrained minimum with the step straddled;\\[-1pt]
   \scriptsize differ if $\phi^*$ is not a minimiser, or if a contact line sits on the step (pinning)};
\draw[arr] (r1.south) -- (r1.south |- cmp.north);
\draw[arr] (r2.south) -- (cmp.north);
\draw[arr] (r3.south) -- (r3.south |- cmp.north);
\end{scope}
\end{tikzpicture}}
\caption{The virtual-work test of contact-line physics. (a) A drop held across a straight wettability step: the
footprint cuts a chord of length $L$ on the step line, and the pattern pulls the drop towards the more wettable side
with the force $F_x=\sigma\Delta\cos\theta\,L$. (b) For any candidate equilibrium field the force can be obtained in
three independent ways: (i) from the wall trace through the virtual-work integral~(\ref{eq:hf}), (ii) from the slope of
the energy landscape $\mathcal E^*(X^*)$ traced by independent solves, and (iii) from the geometric chord through the
contact-line law~(\ref{eq:chordlaw}). The three coincide at a true constrained minimum when the step is straddled, so
their mismatch measures the quality of an approximate minimiser without reference data.}\label{fig:virtualwork}
\end{figure}

Equation~(\ref{eq:hf}) is the central diagnostic of this paper (\hyperref[fig:virtualwork]{figure~\ref*{fig:virtualwork}}; its discrete form is \hyperref[alg:vw]{algorithm~\ref*{alg:vw}}). An approximate minimiser, in particular a neural
operator, satisfies (\ref{eq:hf}) only to the extent that it is a true constrained minimiser; comparing the
force computed from (\ref{eq:hf}) with the finite-difference slope of its own energy landscape and with
$\sigma\Delta\cos\theta L$ tests the physics of the prediction, not its similarity to a target field.

\subsection{The liquid indicator and the Gibbs--Thomson supersaturation}\label{sec:gt}
The choice of the indicator in~(\ref{eq:constraints}) is not innocuous. In the bulk, stationarity of the Lagrangian
requires $cW'(\phi)/\epsilon=-\lambda_V\,h'(\phi)$, where $\lambda_V$ is the multiplier of the volume constraint, equal
to minus the Laplace pressure $p=\sigma\kappa$ of the drop. With the linear liquid fraction, $h=\tfrac12(1+\phi)$,
the right-hand side is the constant $p/2$, and both bulk phases are shifted off $\phi=\pm1$; under the bound
$|\phi|\le1$ only the vapour can respond, and with $W'(-1+\delta)\simeq2\delta$ it is supersaturated by
\begin{equation}
\delta\simeq\frac{\epsilon\,\sigma\kappa}{4c}.
\label{eq:gt}
\end{equation}
This is the diffuse-interface counterpart of the Kelvin effect. Its consequences grow with the ratio of the box to the
drop volume and with $\epsilon/R_0$: the liquid stored in the vapour is $\tfrac12\delta$ per cell, which is a few per
cent of $V_0$ in our two-dimensional box but about a third of $V_0$ in the three-dimensional one, where a drop at
$\theta_Y=140^\circ$ evaporates completely into the supersaturated vapour (\hyperref[sec:verification]{Section~\ref*{sec:verification}}). The centroid
then includes a uniform background that does not move with the drop, so that the multiplier of the centroid
constraint, $\lambda_x=\partial\mathcal{E}^*/\partial X^*$, exceeds the virtual-work force~(\ref{eq:hf}) by the factor
$V_0/V_{\rm drop}$; and the exchange of mass between drop and vapour is a slow, nearly degenerate mode for
iterative minimisers. All three effects are artefacts of the constraint, not of the physics of the drop. With
$h=\tfrac12+g(\phi)$, which satisfies $h'(\pm1)=0$, the bulk equation is satisfied exactly by $\phi=\pm1$, the vapour
stays at $\phi=-1$, the constraints measure the drop alone, and the centroid translates rigidly with it, so that
$\lambda_x=-F_x$ exactly. We use $h$ throughout and document the linear choice in \hyperref[sec:verification]{Section~\ref*{sec:verification}} for
comparison.

\subsection{Discretisation}
The fields live on a uniform grid of unit spacing with $\epsilon=1$: $N_x\times N_y$ cells in two dimensions and
$N_x\times N_y\times N_z$ in three, periodic in $x$ and $z$, with the wall below the first cell row and a no-flux
boundary at the top. The energy~(\ref{eq:energy}) is evaluated with forward differences,
\begin{equation}
E[\phi]=c\sum_{\rm cells}\Big[\frac{W(\phi_{ijk})}{\epsilon}+\frac{\epsilon}{2}\sum_{d}\big(D_d^+\phi\big)_{ijk}^2\Big]
-\sum_{i,k}\cos\theta_{ik}\;g(\phi_{i0k}),
\label{eq:discrete}
\end{equation}
and the volume and centroid by the corresponding sums. The virtual-work force~(\ref{eq:hf}) is evaluated with a
central difference of the sampled wall field, $F_x=\sum_{i,k}\tfrac12(\cos\theta_{i+1,k}-\cos\theta_{i-1,k})\,g(\phi_{i0k})$,
which telescopes to $\Delta\cos\theta$ across any step. All computations below are carried out with the
discrete problem~(\ref{eq:discrete}); the reference solutions and the neural operators solve exactly the same
problem, so that every difference between them is a difference in the quality of the minimisation.

\section{\dfeno: data-free operator learning of capillary equilibria}\label{sec:methods}

\subsection{The solution operator and the training objective}
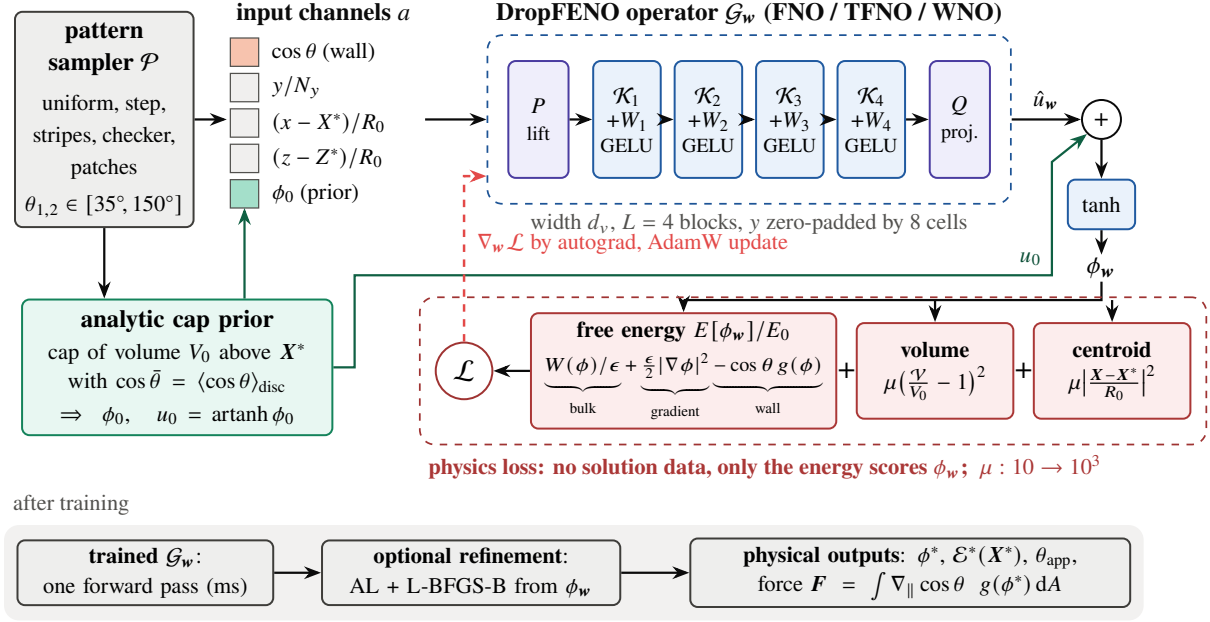
\begin{figure}
\centering
\fitbox{
\begin{tikzpicture}[font=\small]
\node[blk=cInkTwo, minimum width=2.45cm, minimum height=2.75cm] (samp) at (1.22,-1.45)
  {\textbf{pattern}\\\textbf{sampler} $\mathcal P$\\[3pt]\footnotesize uniform, step,\\\footnotesize stripes, checker,\\\footnotesize patches\\[2pt]
   \footnotesize $\theta_{1,2}\in[35^\circ\!,150^\circ]$};
\begin{scope}[shift={(3.0,-0.3)}]
  \foreach \i/\c/\t in {0/cOrange/{$\cos\theta$ (wall)},1/cGrid/{$y/N_y$},2/cGrid/{$(x-X^*)/R_0$},3/cGrid/{$(z-Z^*)/R_0$},4/cGreen/{$\phi_0$ (prior)}} {
    \fill[\c!40, draw=cInkTwo, line width=0.5pt] (0,-\i*0.5) rectangle ++(0.4,-0.4);
    \node[font=\footnotesize, anchor=west] at (0.45,-\i*0.5-0.2) {\t};
  }
  \node[ttl, anchor=south west] at (-0.05,0.05) {input channels $a$};
  \coordinate (phi0in) at (0.2,-2.4);
\end{scope}
\draw[arr] (samp.east) -- (2.95,-1.45);
\foreach \n/\x/\c/\t in {P/7.35/cPurple/{$P$\\[-1pt]\scriptsize lift},L1/8.6/cBlue/{$\mathcal K_1$\\[-2pt]$+W_1$\\[-2pt]\scriptsize GELU},
    L2/9.75/cBlue/{$\mathcal K_2$\\[-2pt]$+W_2$\\[-2pt]\scriptsize GELU},L3/10.9/cBlue/{$\mathcal K_3$\\[-2pt]$+W_3$\\[-2pt]\scriptsize GELU},
    L4/12.05/cBlue/{$\mathcal K_4$\\[-2pt]$+W_4$\\[-2pt]\scriptsize GELU},Q/13.3/cPurple/{$Q$\\[-1pt]\scriptsize proj.}}
  \node[sblk=\c, minimum width=0.85cm, minimum height=1.65cm] (\n) at (\x,-1.45) {\t};
\foreach \a/\b in {P/L1,L1/L2,L2/L3,L3/L4,L4/Q} \draw[arr] (\a) -- (\b);
\node[grp=cBlue, fit=(P)(Q), inner ysep=10pt, inner xsep=8pt] (G) {};
\node[ttl, anchor=south] at (G.north) {DropFENO operator $\mathcal G_{\bm w}$ (FNO / TFNO / WNO)};
\node[lbl, anchor=north] at (G.south) {width $d_v$, $L=4$ blocks, $y$ zero-padded by 8 cells};
\draw[arr] (5.75,-1.45) -- (G.west);
\node[op] (plus) at (15.3,-1.45) {$+$};
\node[sblk=cBlue, minimum width=0.95cm] (tanh) at (15.3,-2.65) {$\tanh$};
\draw[arr] (G.east) -- node[above, font=\footnotesize] {$\hat u_{\bm w}$} (plus);
\draw[arr] (plus) -- (tanh);
\node[font=\small, fill=white, inner sep=1pt] (phi) at (15.3,-3.55) {$\phi_{\bm w}$};
\draw[arr] (tanh) -- (phi);
\node[blk=cGreen, text width=4.1cm, minimum height=1.9cm] (prior) at (2.25,-4.95)
  {\textbf{analytic cap prior}\\[1pt]\footnotesize cap of volume $V_0$ above $\bm X^*$\\ \footnotesize with $\cos\bar\theta=\langle\cos\theta\rangle_{\rm disc}$\\
   \footnotesize $\Rightarrow\;\phi_0,\quad u_0=\operatorname{artanh}\phi_0$};
\draw[arr] (samp.south) -- (samp.south |- prior.north);
\draw[arr, cGreen!55!black] (phi0in |- prior.north) -- (phi0in);
\draw[arr, cGreen!55!black] (prior.east) -- ++(0.3,0) |- (14.6,-3.65) -- (14.6,-2.05) -- (plus.south west);
\node[font=\footnotesize, text=cGreen!45!black, anchor=south] at (14.3,-3.65) {$u_0$};
\node[sblk=cRed, minimum height=1.35cm, text width=4.1cm] (tE) at (9.4,-5.0)
  {\textbf{free energy} $E[\phi_{\bm w}]/E_0$\\[2pt]\scriptsize $\underbrace{W(\phi)/\epsilon}_{\rm bulk}+\underbrace{\tfrac{\epsilon}{2}|\nabla\phi|^2}_{\rm gradient}
   \underbrace{-\cos\theta\,g(\phi)}_{\rm wall}$};
\node[sblk=cRed, minimum height=1.35cm, text width=2.0cm] (tV) at (12.95,-5.0)
  {\textbf{volume}\\[1pt]$\mu\big(\tfrac{\mathcal V}{V_0}-1\big)^2$};
\node[sblk=cRed, minimum height=1.35cm, text width=2.0cm] (tX) at (15.45,-5.0)
  {\textbf{centroid}\\[1pt]$\mu\big|\tfrac{\bm X-\bm X^*}{R_0}\big|^2$};
\node[font=\large] at ($(tE.east)!0.5!(tV.west)$) {$+$};
\node[font=\large] at ($(tV.east)!0.5!(tX.west)$) {$+$};
\node[op, draw=cRed!70!black, minimum size=0.8cm, font=\normalsize] (L) at (6.3,-5.0) {$\mathcal L$};
\draw[arr] (tE.west) -- (L.east);
\draw[cInk, line width=0.9pt] (phi.south) -- (15.3,-4.0) -- (9.4,-4.0);
\foreach \t in {tE,tV} \draw[arr] (\t.north |- 0,-4.0) -- (\t.north);
\draw[arr] (tX.north |- 0,-4.0) -- (tX.north);
\node[grp=cRed, fit=(L)(tE)(tV)(tX), inner sep=6pt] (lossgrp) {};
\node[ttl, text=cRed!75!black, anchor=north west, font=\footnotesize\bfseries] at (lossgrp.south west) {physics loss: no solution data, only the energy scores $\phi_{\bm w}$; \ $\mu:10\to10^3$};
\draw[garr] (L.north) -- (6.3,-2.25) -- (G.west |- 0,-2.25);
\node[font=\footnotesize, text=cRed, anchor=west] at (6.35,-3.2) {$\nabla_{\bm w}\mathcal L$ by autograd, AdamW update};
\node[sblk=cInkTwo, text width=3.4cm] (inf) at (1.8,-7.85) {\textbf{trained} $\mathcal G_{\bm w}$:\\ one forward pass (ms)};
\node[sblk=cInkTwo, text width=4.0cm] (ref) at (6.4,-7.85) {\textbf{optional refinement}:\\ AL + L-BFGS-B from $\phi_{\bm w}$};
\node[sblk=cInkTwo, text width=6.0cm] (outp) at (12.6,-7.85) {\textbf{physical outputs}: $\phi^*$, $\mathcal E^*(\bm X^*)$, $\theta_{\rm app}$,\\
  force $\bm F=\int\nabla_\parallel\cos\theta\;g(\phi^*)\,\mathrm dA$};
\draw[arr] (inf) -- (ref); \draw[arr] (ref) -- (outp);
\begin{scope}[on background layer]
  \node[fill=cGrid!35, rounded corners=6pt, fit=(inf)(ref)(outp), inner sep=5pt] (strip) {};
\end{scope}
\node[lbl, anchor=south west] at (strip.north west) {after training};
\end{tikzpicture}}
\caption{Overall architecture of \dfeno, the data-free operator, and its physics loss. Wall patterns and holding positions are
drawn afresh at every iteration from the distribution $\mathcal P$ (\hyperref[fig:patterns]{figure~\ref*{fig:patterns}}) and turned into the input
channels $a$, which include the analytic cap prior $\phi_0$. The neural operator $\mathcal G_{\bm w}$ (lifting $P$, four
operator blocks, projection $Q$; \hyperref[fig:layer]{figures~\ref*{fig:layer}} and~\ref{fig:kernels}) returns a correction $\hat u_{\bm w}$
that is added to $u_0=\operatorname{artanh}\phi_0$ and mapped through $\tanh$ to the field $\phi_{\bm w}$, which
satisfies $|\phi_{\bm w}|<1$ by construction. The physics loss scores $\phi_{\bm w}$ by its free energy~(\ref{eq:discrete})
and by penalties on the volume and centroid constraints, equation~(\ref{eq:loss}); its gradient with respect to the
weights, obtained by automatic differentiation through the energy, is the only training signal (dashed). No equilibrium
field is used at any stage of training. After training, a prediction costs one forward pass and can optionally be refined by the
reference minimiser (\hyperref[fig:refsolver]{figure~\ref*{fig:refsolver}}).}\label{fig:overview}
\end{figure}

For a fixed drop volume the equilibrium defines a solution operator
\begin{equation}
\mathcal{G}:\;\big(\cos\theta(x,z),\,\bm{X}^*\big)\;\longmapsto\;\phi^*(x,y,z),
\label{eq:operator}
\end{equation}
from the wall wettability field and the holding position to the constrained energy minimiser. We approximate
$\mathcal{G}$ by a neural operator $\mathcal{G}_{\bm{w}}$ with weights $\bm{w}$ and train it by minimising the
expected value of the physical objective over a distribution $\mathcal{P}$ of wall patterns and positions,
\begin{equation}
\mathcal{L}(\bm{w})=\mathbb{E}_{(a,\bm X^*)\sim\mathcal{P}}\Big[\frac{E[\phi_{\bm w}]}{E_0}
+\mu\Big\{\Big(\frac{\mathcal{V}[\phi_{\bm w}]}{V_0}-1\Big)^2+\Big|\frac{\bm{X}[\phi_{\bm w}]-\bm{X}^*}{R_0}\Big|^2\Big\}\Big],
\qquad \phi_{\bm w}=\mathcal{G}_{\bm w}(a,\bm X^*),
\label{eq:loss}
\end{equation}
with $E_0=2\upi R_0\sigma$ (two dimensions) or $4\upi R_0^2\sigma$ (three dimensions), the energy of a free drop.
No equilibrium field enters (\ref{eq:loss}): the inputs $a=\cos\theta(x,z)$ are drawn afresh at every iteration
from $\mathcal{P}$ and the only training signal is the energy of the predicted field, which is the same energy
that defines the physics. Because the minimiser of (\ref{eq:loss}) over all maps is the pointwise constrained
minimiser of $E$ (for $\mu\to\infty$), training is an amortised optimisation: the cost of solving the
variational problem is paid once, during training, for the whole family $\mathcal{P}$
\citep{Amos2023}. The pipeline is drawn in \hyperref[fig:overview]{figure~\ref*{fig:overview}} and listed as \hyperref[alg:train]{algorithm~\ref*{alg:train}}. The penalty weight is increased geometrically from $\mu=10$ to $\mu=10^3$ over the first
$60\%$ of training, a continuation that lets the network first learn the shape and then enforce the constraints.

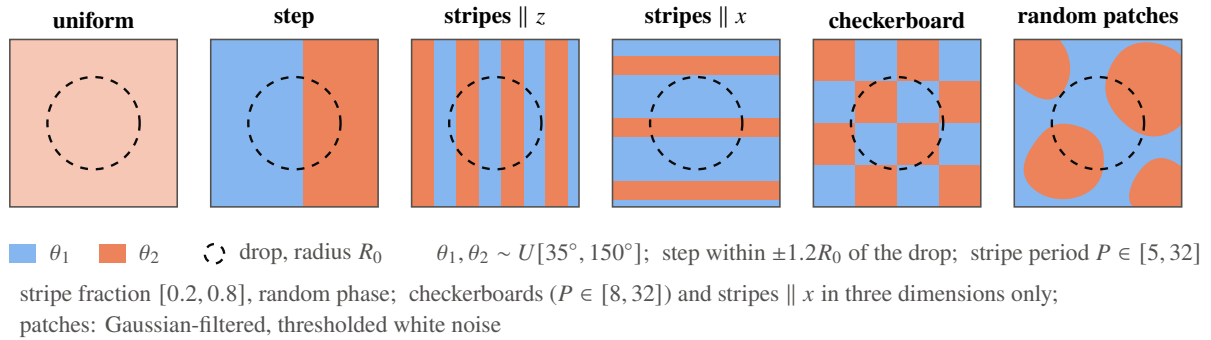
\begin{figure}
\centering
\fitbox{
\begin{tikzpicture}[font=\small]
\def\s{2.25}
\newcommand{\tile}[3]{
  \begin{scope}[xshift=#1 cm]
    \fill[cPhob] (0,0) rectangle (\s,\s);
    \begin{scope}\clip (0,0) rectangle (\s,\s); #3 \end{scope}
    \draw[cInkTwo, line width=0.6pt] (0,0) rectangle (\s,\s);
    \node[font=\footnotesize\bfseries, anchor=south] at (\s/2,\s+0.02) {#2};
  \end{scope}}
\tile{0}{uniform}{\fill[cPhil!45] (0,0) rectangle (\s,\s);}
\tile{2.7}{step}{\fill[cPhil] (1.25,0) rectangle (\s,\s);}
\tile{5.4}{stripes $\parallel z$}{\foreach \i in {0,2,4,6} \fill[cPhil] (\i*0.3,0) rectangle ++(0.3,\s);}
\tile{8.1}{stripes $\parallel x$}{\foreach \i in {0,2,4} \fill[cPhil] (0,\i*0.42+0.1) rectangle ++(\s,0.25);}
\tile{10.8}{checkerboard}{\foreach \i in {0,...,3} \foreach \j in {0,...,3} {\pgfmathparse{int(mod(\i+\j,2))}\ifnum\pgfmathresult=1 \fill[cPhil] (\i*0.5625,\j*0.5625) rectangle ++(0.5625,0.5625);\fi}}
\tile{13.5}{random patches}{
  \fill[cPhil] plot[smooth cycle, tension=0.9] coordinates {(0.2,0.3) (0.9,0.15) (1.2,0.7) (0.8,1.1) (0.3,0.9)};
  \fill[cPhil] plot[smooth cycle, tension=0.9] coordinates {(1.4,1.2) (2.1,1.0) (2.4,1.7) (1.9,2.2) (1.3,1.9)};
  \fill[cPhil] plot[smooth cycle, tension=0.9] coordinates {(0.1,1.6) (0.6,1.5) (0.7,2.1) (0.2,2.4) (-0.1,2.0)};
  \fill[cPhil] plot[smooth cycle, tension=0.9] coordinates {(1.6,0.0) (2.4,0.1) (2.3,0.5) (1.8,0.6)};}
\foreach \x in {0,2.7,5.4,8.1,10.8,13.5} \draw[cInk, line width=0.9pt, dashed] (\x+1.125,1.125) circle (0.62);
\fill[cPhob] (0,-0.75) rectangle ++(0.35,0.25); \node[lbl, anchor=west] at (0.4,-0.63) {$\theta_1$};
\fill[cPhil] (1.2,-0.75) rectangle ++(0.35,0.25); \node[lbl, anchor=west] at (1.6,-0.63) {$\theta_2$};
\draw[cInk, dashed, line width=0.9pt] (2.75,-0.63) circle (0.13); \node[lbl, anchor=west] at (2.95,-0.63) {drop, radius $R_0$};
\node[lbl, anchor=west] at (5.6,-0.63) {$\theta_{1},\theta_2\sim U[35^\circ,150^\circ]$;\; step within $\pm1.2R_0$ of the drop;\; stripe period $P\in[5,32]$};
\node[lbl, anchor=north west, align=left] at (0,-0.9) {stripe fraction $[0.2,0.8]$, random phase;\; checkerboards ($P\in[8,32]$) and stripes $\parallel x$ in three dimensions only;\\ patches: Gaussian-filtered, thresholded white noise};
\end{tikzpicture}}
\caption{The six families of wall patterns of the training distribution $\mathcal P$ (top view of the wall,
schematic). Each training instance draws a family, two Young angles and the parameters of the family, and a holding
position within $\pm R_0/2$ of the box centre (dashed circle: drop footprint of radius $R_0$). The test patterns of
\hyperref[sec:verification]{Sections~\ref*{sec:verification}}--\ref{sec:physics} are never drawn exactly.}\label{fig:patterns}
\end{figure}

\paragraph{The pattern distribution.} $\mathcal{P}$ is a mixture of six families of wall fields (four in two dimensions) (\hyperref[fig:patterns]{figure~\ref*{fig:patterns}}), with the two
Young angles of each pattern drawn uniformly in $[35^\circ,150^\circ]$: uniform walls; a single smoothed step
$\tfrac12[1+\tanh(x/0.75)]$ placed within $\pm1.2R_0$ of the drop; stripes of period $P\in[5,32]$, area
fraction $[0.2,0.8]$ and random phase, running along $z$ or (in three dimensions) along $x$; checkerboards
($P\in[8,32]$, three dimensions only); and smooth random patches obtained by Gaussian filtering and thresholding
white noise. The drop position is drawn within $\pm R_0/2$ of the box centre. The distribution is generic: it
contains no information on any of the test configurations of \hyperref[sec:verification]{Sections~\ref*{sec:verification}}--\ref{sec:physics}
beyond their pattern family, and the test patterns (e.g.\ the specific Young angles of the lattice Boltzmann walls
used for comparison) are never drawn exactly.

\subsection{Output parametrisation and the analytic prior}
The network does not output $\phi$ directly. It outputs a correction $\hat{u}$ to the field
$u_0=\operatorname{artanh}\phi_0$ of a diffuse spherical cap,
\begin{equation}
\phi_{\bm w}=\tanh\big(u_0+\hat{u}_{\bm w}\big),\qquad
\phi_0=-\tanh\!\Big(\frac{|\bm{x}-\bm{x}_c|-R_c}{\sqrt2\,\epsilon}\Big),
\label{eq:prior}
\end{equation}
where the cap of volume $V_0$, centred above $\bm{X}^*$, has the Cassie-like mean angle
$\cos\bar\theta=\langle\cos\theta\rangle$ over a disc of radius $R_0$ under the drop. The $\tanh$ enforces the
bound $|\phi|<1$ and the prior supplies the trivial part of the answer (the size and the mean shape), so that the
network learns the response to the pattern; the last layer is initialised to zero, so that training starts from
the prior. The prior is itself an approximation built from the inputs alone and contains no solution data; an
ablation without it (\hyperref[sec:ablation]{Section~\ref*{sec:ablation}}) quantifies its role.

\subsection{Architectures}\label{sec:arch}
All three operators share the structure of \citet{Li2021}: a pointwise lifting $P$ of the input channels to a
width $d_v$, $L=4$ blocks
\begin{equation}
v_{l+1}=\operatorname{GELU}\big(\mathcal{K}_lv_l+W_lv_l\big),
\label{eq:block}
\end{equation}
with $W_l$ a pointwise linear map, and a pointwise projection $Q$ to the scalar $\hat u$. The inputs are the
wall field $\cos\theta$ extended uniformly in $y$, the normalised coordinates $y/N_y$, $(x-X^*)/R_0$ and, in
three dimensions, $(z-Z^*)/R_0$, and the prior $\phi_0$. Because $y$ is not periodic, the latent fields are
zero-padded by eight cells at the top before the global operators and cropped afterwards. The three operators
differ only in the global kernel $\mathcal{K}_l$. \hyperref[fig:layer]{Figure~\ref*{fig:layer}} shows the forward pass with its tensor shapes,
\hyperref[fig:kernels]{figure~\ref*{fig:kernels}} the three kernels, and \hyperref[alg:forward]{algorithm~\ref*{alg:forward}} lists one prediction.

\paragraph{FNO.} $\mathcal{K}_l$ is a convolution evaluated in Fourier space,
$(\mathcal{K}v)(\bm x)=\mathcal{F}^{-1}\big[R(\bm k)\,\mathcal{F}v(\bm k)\big]$, with a dense complex weight
$R\in\mathbb{C}^{d_v\times d_v}$ on each retained mode $|k_d|\le k_{\max,d}$ and zero elsewhere \citep{Li2021}. Because
the field is real, only non-negative wavenumbers are kept in the last direction, so that $m_d=2k_{\max,d}$ modes are
retained in every other direction and $m_d=k_{\max,d}$ in the last one; one kernel has $d_v^2\prod_d m_d$ complex
weights.

\paragraph{TFNO.} The weight tensor $R_{io\bm k}$ of the FNO is replaced by a complex canonical-polyadic (CP)
factorisation of rank $r$ \citep{Kossaifi2023},
\begin{equation}
R_{i o k_x k_y k_z}=\sum_{q=1}^{r}\lambda_q\,A_{iq}\,B_{oq}\,K^{(x)}_{k_xq}K^{(y)}_{k_yq}K^{(z)}_{k_zq},
\label{eq:cp}
\end{equation}
which reduces the number of complex weights of one kernel from $d_v^2\prod_dm_d$ to $r(1+2d_v+\sum_dm_d)$ and
is applied without forming $R$, by contracting the factors in sequence.

\paragraph{WNO.} $\mathcal{K}_l$ acts in a wavelet basis \citep{Tripura2023}: a $J=3$-level orthonormal Haar
transform of $v_l$, a learned channel mixing, location by location, of the approximation and the detail
coefficients at the coarsest level, and the inverse transform with the finer detail coefficients set to zero. The kernel is local in space and in scale, in
contrast to the global Fourier modes of the FNO, which makes it a candidate for a field that contains
a thin interface and a pinned contact line.

We refer to the three variants as \dfeno-FNO, \dfeno-TFNO and \dfeno-WNO, and to them
collectively as \dfeno; in figures and tables they are labelled FNO, TFNO and WNO. They are implemented in PyTorch from these definitions; \hyperref[tab:arch]{table~\ref*{tab:arch}} lists their sizes.

\subsection{Training}
\begin{figure}
\centering
\includegraphics[width=\textwidth]{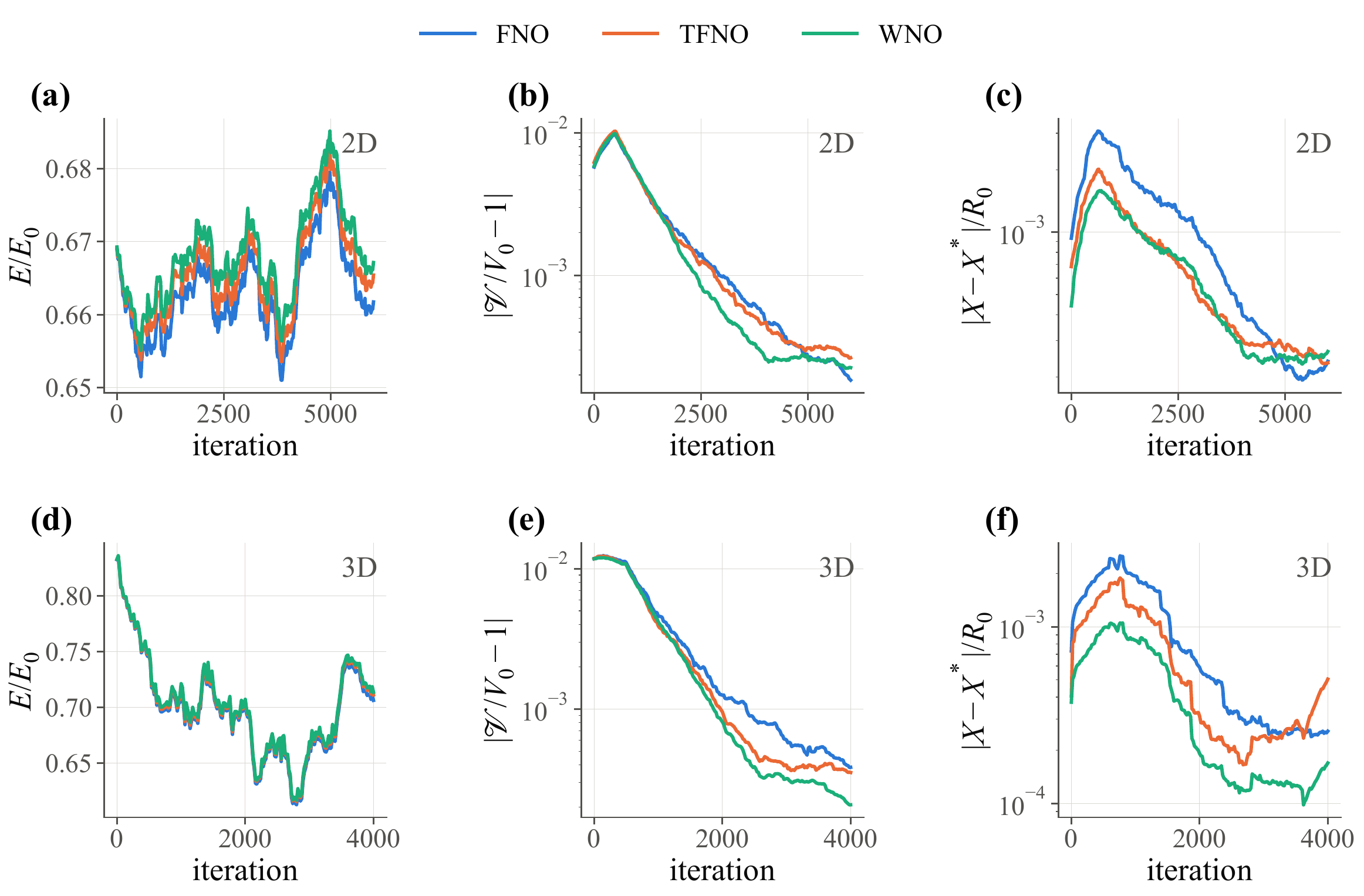}
\caption{Data-free training of the three operators in two (a--c) and three (d--f) dimensions. (a,d) Energy of the
predicted fields, averaged over each batch of freshly drawn patterns (running mean); the scatter reflects the
variety of the patterns, not the optimisation. (b,e) Volume and (c,f) centroid residuals, enforced only through the
penalty whose weight grows from $10$ to $10^3$ over the first $60\%$ of training.}\label{fig:training}
\end{figure}

Training uses AdamW (weight decay $10^{-5}$) with a one-cycle learning-rate schedule (peak $10^{-3}$),
gradient clipping at norm one, batches of eight (two dimensions) or two (three dimensions) independently drawn
patterns, $6000$ (two dimensions) and $4000$ (three dimensions) iterations, i.e.\ $4.8\times10^4$ and
$8\times10^3$ distinct problem instances, each seen once. The training histories of the energy and of the
constraint residuals are shown in \hyperref[fig:training]{figure~\ref*{fig:training}}. All computations run on CPUs: training on four threads per job (two for the extended-budget runs of \hyperref[sec:ablation]{Section~\ref*{sec:ablation}}), reference solves and evaluation on two. The configuration, cost and training history of every run, including
the ablations, are given in the Supplementary Information ({\color{linkblue}table~}\xref{tab:si_runs}, {\color{linkblue}figures~}\xref{fig:si_hist}
and~\xref{fig:si_sched}).

\subsection{Reference solutions}\label{sec:reference}
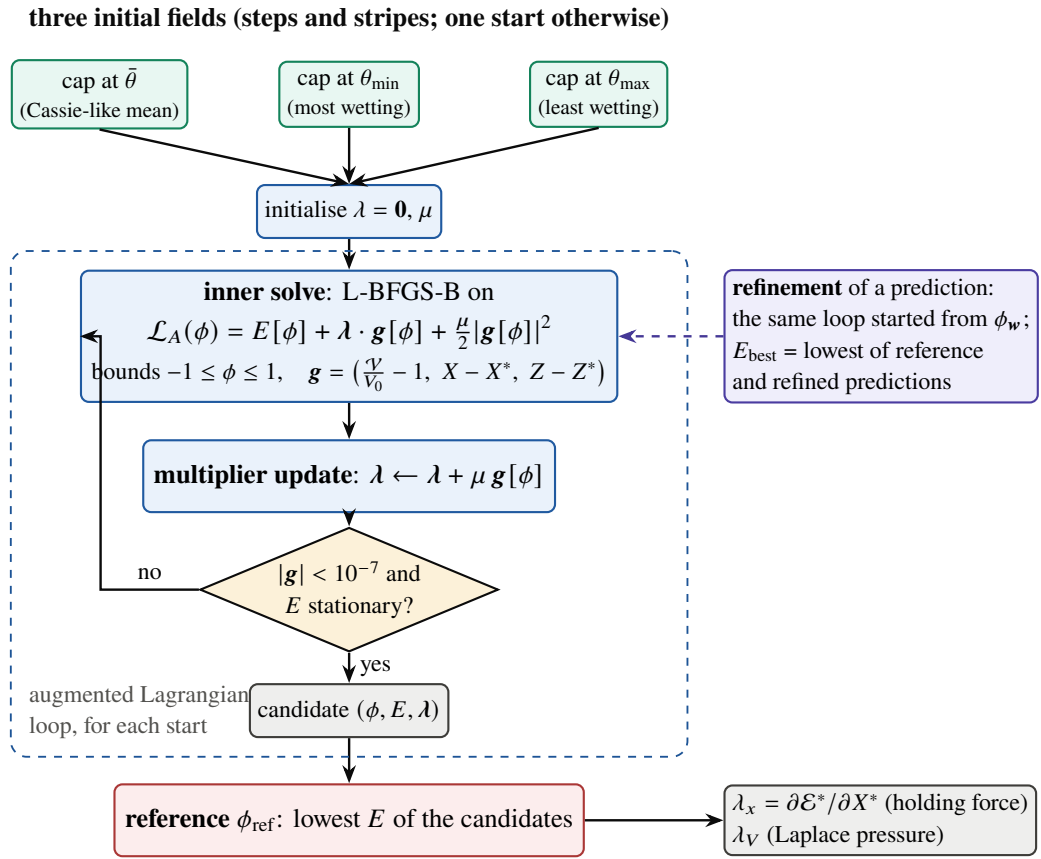
\begin{figure}
\centering
\fitbox{
\begin{tikzpicture}[font=\small, node distance=0.55cm]
\node[sblk=cGreen] (s1) at (0,0) {cap at $\bar\theta$\\[-1pt]\scriptsize (Cassie-like mean)};
\node[sblk=cGreen] (s2) at (3.3,0) {cap at $\theta_{\min}$\\[-1pt]\scriptsize (most wetting)};
\node[sblk=cGreen] (s3) at (6.6,0) {cap at $\theta_{\max}$\\[-1pt]\scriptsize (least wetting)};
\node[ttl, anchor=south] at (3.3,0.7) {three initial fields (steps and stripes; one start otherwise)};
\node[sblk=cBlue, minimum width=1.6cm] (init) at (3.3,-1.55) {initialise $\lambda=\bm 0$, $\mu$};
\foreach \s in {s1,s2,s3} \draw[arr] (\s.south) -- (init.north);
\node[blk=cBlue] (inner) at (3.3,-3.2) {\textbf{inner solve}: L-BFGS-B on\\[2pt]
   $\displaystyle\mathcal L_A(\phi)=E[\phi]+\bm\lambda\cdot\bm g[\phi]+\tfrac{\mu}{2}|\bm g[\phi]|^2$\\[2pt]
   \footnotesize bounds $-1\le\phi\le1$, \;\; $\bm g=\big(\tfrac{\mathcal V}{V_0}-1,\;X-X^*,\;Z-Z^*\big)$};
\node[blk=cBlue] (upd) at (3.3,-5.05) {\textbf{multiplier update}: $\bm\lambda\leftarrow\bm\lambda+\mu\,\bm g[\phi]$};
\node[diamond, aspect=2.4, draw=cInk, fill=cYellow!15, line width=0.8pt, inner sep=1pt, font=\footnotesize, align=center] (test) at (3.3,-6.55)
   {$|\bm g|<10^{-7}$ and\\ $E$ stationary?};
\draw[arr] (init) -- (inner); \draw[arr] (inner) -- (upd); \draw[arr] (upd) -- (test);
\draw[arr] (test.west) -- ++(-1.3,0) node[above, font=\footnotesize, pos=0.5] {no} |- (inner.west);
\node[sblk=cInkTwo] (cand) at (3.3,-8.15) {candidate $(\phi,E,\bm\lambda)$};
\draw[arr] (test.south) -- node[right, font=\footnotesize] {yes} (cand.north);
\node[blk=cRed] (best) at (3.3,-9.6) {\textbf{reference} $\phi_{\rm ref}$: lowest $E$ of the candidates};
\draw[arr] (cand) -- (best);
\node[grp=cBlue, fit=(inner)(upd)(test)(cand), inner xsep=26pt] (loop) {};
\node[lbl, anchor=south west, align=left] at ($(loop.south west)+(0.12,0.12)$) {augmented Lagrangian\\loop, for each start};
\node[sblk=cInkTwo, align=left, anchor=west] (out) at (8.25,-9.6)
  {$\lambda_x=\partial\mathcal E^*/\partial X^*$ (holding force)\\ $\lambda_V$ (Laplace pressure)};
\draw[arr] (best) -- (out);
\node[sblk=cPurple, align=left, anchor=west] (refi) at (8.25,-3.2)
  {\textbf{refinement} of a prediction:\\ the same loop started from $\phi_{\bm w}$;\\ $E_{\rm best}$ = lowest of reference\\ and refined predictions};
\draw[arr, cPurple!70!black, dashed] (refi.west) -- (inner.east);
\end{tikzpicture}}
\caption{The reference minimiser used to judge the operators (\hyperref[sec:reference]{Section~\ref*{sec:reference}}). An augmented-Lagrangian loop
enforces the volume and centroid constraints around a bound-constrained quasi-Newton (L-BFGS-B) minimisation of the
discrete energy directly in $\phi\in[-1,1]$. For steps and stripes the loop is run from three caps and the lowest
minimum is kept. Started from an operator prediction instead of a cap, the same loop is the ``refinement'' of
\hyperref[sec:accuracy]{Section~\ref*{sec:accuracy}}. The converged multipliers are the holding force $\lambda_x$ and the Laplace pressure.}\label{fig:refsolver}
\end{figure}

To judge the operators we solve each test instance individually by a classical method: an augmented-Lagrangian
loop for the constraints around a bound-constrained quasi-Newton minimisation (L-BFGS-B) of the discrete energy
directly in $\phi\in[-1,1]$, started from the analytic cap, until the constraint residuals are below $10^{-7}$ and
the energy is stationary. Working directly in $\phi$ with bounds, rather than in the $\tanh$ parametrisation that is
natural for the network, avoids the vanishing gradients of $\tanh$ at the bulk values $\phi=\pm1$. Because steps
and stripes admit several metastable minima, the reference for those problems is the lowest of three solves
started from caps with the Cassie-like mean angle and with the two extreme angles of the pattern. The reference
solutions are never used in training. The procedure is shown in \hyperref[fig:refsolver]{figure~\ref*{fig:refsolver}} and listed as
\hyperref[alg:ref]{algorithm~\ref*{alg:ref}}.

\begin{figure}
\centering
\fitbox{
\begin{tikzpicture}[font=\small]
\node[ttl, anchor=west] at (-0.1,1.75) {(a) forward pass (two dimensions; shapes channels $\times N_x\times N_y$)};
\node[sblk=cGrid, minimum width=1.3cm, minimum height=1.25cm] (a) at (0.7,0) {$a$\\[-1pt]\scriptsize $4\times128\times64$};
\node[sblk=cPurple, minimum height=1.25cm] (P) at (3.05,0) {\textbf{lift} $P$\\[-1pt]\scriptsize $1{\times}1$, GELU, $1{\times}1$};
\node[sblk=cInkTwo, minimum height=1.25cm] (pad) at (5.4,0) {pad $y$\\[-1pt]\scriptsize $+8$ cells};
\node[sblk=cBlue, minimum height=1.25cm] (B) at (7.75,0) {\textbf{block} $l$\\[-1pt]\scriptsize $d_v\times128\times72$};
\node[sblk=cInkTwo, minimum height=1.25cm] (crop) at (9.95,0) {crop\\[-1pt]\scriptsize $y$};
\node[sblk=cPurple, minimum height=1.25cm] (Q) at (12.3,0) {\textbf{project} $Q$\\[-1pt]\scriptsize $1{\times}1$ to $4d_v$, GELU,\\[-2pt]\scriptsize $1{\times}1$ to $1$};
\node[sblk=cGrid, minimum width=1.3cm, minimum height=1.25cm] (u) at (14.85,0) {$\hat u$\\[-1pt]\scriptsize $1\times128\times64$};
\foreach \x/\y in {a/P,P/pad,pad/B,B/crop,crop/Q,Q/u} \draw[arr] (\x) -- (\y);
\draw[arr, line width=0.8pt] ($(B.north)+(0.35,0)$) arc (-20:200:0.37 and 0.32);
\node[font=\footnotesize, anchor=west] at ($(B.north)+(0.45,0.38)$) {$\times\,L=4$};
\begin{scope}[yshift=-3.1cm]
\node[ttl, anchor=west] at (-0.1,1.5) {(b) operator block $v_{l+1}=\operatorname{GELU}\!\big(\mathcal K_l v_l+W_lv_l\big)$};
\node[font=\normalsize] (vl) at (0.3,-0.35) {$v_l$};
\coordinate (sp) at (1.1,-0.35);
\node[sblk=cBlue, minimum height=0.95cm] (K) at (4.9,0.35) {\textbf{global kernel} $\mathcal K_l$\\[-1pt]\scriptsize Fourier (FNO), CP-Fourier (TFNO)\\[-2pt]\scriptsize or Haar wavelet (WNO)};
\node[sblk=cPurple, minimum height=0.75cm] (W) at (4.9,-1.2) {\textbf{local map} $W_l$: $1{\times}1$ convolution};
\node[op] (sum) at (8.15,-0.35) {$+$};
\node[sblk=cInkTwo, minimum width=1.2cm] (act) at (9.6,-0.35) {GELU};
\node[font=\normalsize] (vl1) at (11.1,-0.35) {$v_{l+1}$};
\draw[line width=0.9pt] (vl) -- (sp);
\draw[arr] (sp) |- (K.west); \draw[arr] (sp) |- (W.west);
\draw[arr] (K.east) -| (sum.north); \draw[arr] (W.east) -| (sum.south);
\draw[arr] (sum) -- (act); \draw[arr] (act) -- (vl1);
\node[lbl, align=left, anchor=west] at (11.8,-0.35)
  {$\mathcal K_l$: couples all cells (long range)\\ $W_l$: keeps local detail\\ Fourier weights act on modes,\\ not on cells};
\end{scope}
\end{tikzpicture}}
\caption{Block diagram of the operator. (a) Forward pass in two dimensions with the tensor shapes: pointwise lifting
$P$ of the four input channels to the latent width $d_v$, zero padding of the non-periodic $y$ direction, $L=4$
operator blocks, cropping and pointwise projection $Q$ to the scalar correction $\hat u$. In three dimensions the
input has five channels and the grid is $64\times32\times64$. (b) One operator block, equation~(\ref{eq:block}):
a global kernel $\mathcal K_l$ and a pointwise linear map $W_l$ act in parallel, followed by a GELU activation. The three
operators differ only in $\mathcal K_l$ (\hyperref[fig:kernels]{figure~\ref*{fig:kernels}}).}\label{fig:layer}
\end{figure}

\begin{figure}
\centering
\fitbox{
\begin{tikzpicture}[font=\small]
\newcommand{\fieldicon}[3]{
  \begin{scope}[shift={#1}]
    \fill[#3!18, draw=#3!70!black, line width=0.6pt] (-#2/2,-0.45) rectangle (#2/2,0.45);
    \draw[#3!50, line width=0.3pt, step=0.15] (-#2/2,-0.45) grid (#2/2,0.45);
  \end{scope}}
\node[ttl, anchor=west] at (-0.1,1.05) {(a) FNO: dense spectral convolution};
\node[font=\normalsize] (va) at (0.25,0) {$v$};
\node[sblk=cBlue, minimum width=1.25cm] (fft) at (1.65,0) {FFT};
\begin{scope}[shift={(3.55,0)}]
  \fill[cGrid!40, draw=cInkTwo, line width=0.6pt] (-0.75,-0.6) rectangle (0.75,0.6);
  \fill[cBlue!45, draw=cBlue!70!black] (-0.75,-0.6) rectangle (-0.15,-0.05);
  \fill[cBlue!45, draw=cBlue!70!black] (-0.75,0.6) rectangle (-0.15,0.05);
  \node[font=\scriptsize] at (0.3,0) {$\hat v(\bm k)$};
  \node[lbl, font=\scriptsize, anchor=north] at (0,-0.62) {keep $|k_d|\le k_{\max,d}$};
  \coordinate (specE) at (0.75,0); \coordinate (specW) at (-0.75,0);
\end{scope}
\node[sblk=cBlue] (R) at (6.95,0) {multiply each mode by\\ $R(\bm k)\in\mathbb C^{d_v\times d_v}$};
\node[sblk=cBlue] (zp) at (10.0,0) {zero the\\ other modes};
\node[sblk=cBlue, minimum width=1.25cm] (ifft) at (11.8,0) {iFFT};
\node[font=\normalsize] (Ka) at (12.95,0) {$\mathcal K v$};
\draw[arr] (va) -- (fft); \draw[arr] (fft) -- (specW); \draw[arr] (specE) -- (R); \draw[arr] (R) -- (zp); \draw[arr] (zp) -- (ifft); \draw[arr] (ifft) -- (Ka);
\node[lbl, anchor=west, align=left] at (13.5,0) {params per block\\ $d_v^2\prod_d m_d$ complex\\ (2D: $1.57$\,M real)};
\begin{scope}[yshift=-3.15cm]
\node[ttl, anchor=west] at (-0.1,1.7) {(b) TFNO: the same convolution with a CP-factorised weight tensor};
\node[circle, draw=cOrange!80!black, fill=cOrange!20, minimum size=0.75cm, font=\small] (lam) at (2.4,0) {$\lambda$};
\foreach \n/\a/\t/\e in {A/150/{$A$}/{$i$},B/30/{$B$}/{$o$},Kx/210/{$K^{(x)}$}/{$k_x$},Ky/270/{$K^{(y)}$}/{$k_y$},Kz/330/{$K^{(z)}$}/{$k_z$}} {
  \node[rectangle, rounded corners=2pt, draw=cOrange!80!black, fill=cOrange!10, minimum size=0.6cm, font=\scriptsize] (\n) at ($(lam)+(\a:1.25)$) {\t};
  \draw[cOrange!70!black, line width=0.9pt] (lam) -- (\n);
  \draw[cInkTwo, line width=0.7pt] (\n) -- ++(\a:0.55) node[pos=1.35, font=\scriptsize, text=cInk] {\e};
}
\node[lbl, anchor=west] at (4.35,0.55) {$R_{io\bm k}=\displaystyle\sum_{q=1}^{r}\lambda_q\,A_{iq}\,B_{oq}\,K^{(x)}_{k_xq}K^{(y)}_{k_yq}K^{(z)}_{k_zq}$};
\node[lbl, anchor=west, align=left] at (4.35,-0.65) {$R$ is never formed: $\hat v$ is contracted with $A$\\ ($d_v\to r$), scaled mode-wise by $\lambda_qK^{(x)}K^{(y)}K^{(z)}$,\\ and expanded with $B$ ($r\to d_v$)};
\node[lbl, anchor=west, align=left] at (11.75,0) {params per block\\ $r\big(1+2d_v+\sum_dm_d\big)$ complex\\ (2D, $r=32$: $8.3$\,k real)};
\end{scope}
\begin{scope}[yshift=-7.0cm]
\node[ttl, anchor=west] at (-0.1,1.35) {(c) WNO: Haar wavelet kernel, $J=3$ levels};
\node[font=\normalsize] (vc) at (0.25,0) {$v$};
\foreach \j/\w/\x in {0/1.6/1.55,1/1.1/3.3,2/0.75/4.75,3/0.5/5.95} {
  \fill[cGreen!15, draw=cGreen!60!black, line width=0.6pt] (\x-\w/2,-\w/3.2) rectangle (\x+\w/2,\w/3.2);
}
\node[font=\scriptsize] at (1.55,0) {$128\times72$};
\node[font=\scriptsize] at (3.3,0) {$64{\times}36$};
\node[font=\tiny] at (4.75,0) {$32{\times}18$};
\node[font=\scriptsize, anchor=north] at (5.95,-0.2) {$16{\times}9$};
\foreach \a/\b in {0.45/0.72,2.38/2.72,3.88/4.35,5.15/5.68} \draw[arr] (\a,0) -- (\b,0);
\node[lbl, anchor=south] at (3.9,0.55) {Haar analysis: $a_{j+1},d_{j+1}=\mathrm{H}(a_j)$};
\node[sblk=cGreen] (mix) at (8.5,0) {learned channel mixing\\ of $a_3$ and $d_3$ at each\\ of the $16\times9$ locations};
\draw[arr] (6.25,0) -- (mix);
\node[sblk=cGreen] (inv) at (11.75,0) {inverse Haar\\ ($d_1,d_2$ set to $0$)};
\node[font=\normalsize] (Kc) at (13.6,0) {$\mathcal K v$};
\draw[arr] (mix) -- (inv); \draw[arr] (inv) -- (Kc);
\node[lbl, anchor=north] at (6.6,-0.85) {local in space and in scale: the kernel acts on the coarsest coefficients only, and the\\
  pointwise path $W_l$ carries the finer detail. Params per block $(1+n_d)\,d_v^2N_J$, with $n_d=3$ (2D) or $7$ (3D) detail bands\\ and $N_J$ coarsest cells (2D: $0.59$\,M).};
\end{scope}
\end{tikzpicture}}
\caption{The three global kernels $\mathcal K_l$. (a) FNO: Fourier transform, multiplication of the retained modes
by dense complex matrices, inverse transform. (b) TFNO: the same convolution with the weight tensor replaced by its rank-$r$
canonical-polyadic factorisation, equation~(\ref{eq:cp}), drawn as a tensor network (one leg per index) and applied
factor by factor. (c) WNO: a three-level Haar analysis of the padded latent field, learned channel mixing of the
coarsest approximation and detail coefficients at every coarse location, and the inverse transform with the finer
detail bands set to zero. Parameter counts are per block for the two-dimensional configurations of
\hyperref[tab:arch]{table~\ref*{tab:arch}}.}\label{fig:kernels}
\end{figure}

\begin{table}
\centering
\caption{The three operators. Retained Fourier modes (FNO, TFNO) or coarsest-level wavelet grid (WNO), latent width $d_v$ and CP rank $r$ are given as two-dimensional / three-dimensional values; all operators have $L=4$ blocks and the same lifting and projection.}\label{tab:arch}
\footnotesize
\begin{tabular}{@{}llcccrr@{}}
\toprule
operator & global kernel $\mathcal{K}_l$ & modes or grid & $d_v$ & $r$ & params (2D) & params (3D)\\
\midrule
FNO & dense spectral & $24\times16$ / $12\times8\times12$ & 32 / 20 & -- & 6.30\,M & 14.75\,M \\
TFNO & CP-factorised spectral & $24\times16$ / $12\times8\times12$ & 32 / 20 & 32 / 48 & 42.8\,k & 39.7\,k \\
WNO & Haar wavelet, $J=3$ & $16\times9$ / $8\times5\times8$ & 32 / 20 & -- & 2.37\,M & 4.10\,M \\
\bottomrule
\end{tabular}
\end{table}

\begin{algorithm}[t]
\caption{Virtual-work test of a predicted or computed equilibrium (\hyperref[fig:virtualwork]{figure~\ref*{fig:virtualwork}})}\label{alg:vw}
\begin{algorithmic}[1]
\Require step pattern $\cos\theta=\cos\theta_1+\Delta\cos\theta\,H(x-x_s)$; held positions $X^*_1<\dots<X^*_M$; a map $X^*\mapsto\phi^*$ (operator or minimiser)
\For{$m=1,\dots,M$}
  \State $\phi^*_m\gets\phi^*(\cdot;X^*_m)$;\quad $\mathcal E^*_m\gets E[\phi^*_m]$
  \State $F_m\gets\sum_{i,k}\tfrac12\big(\cos\theta_{i+1,k}-\cos\theta_{i-1,k}\big)\,g(\phi^*_{m,i0k})$ \Comment{virtual-work force, (\ref{eq:hf})}
  \State $L_m\gets$ wetted chord of the footprint $\phi^*_m(x_s,0,z)>0$ on the step line
\EndFor
\State $S_m\gets-(\mathcal E^*_{m+1}-\mathcal E^*_{m-1})/(X^*_{m+1}-X^*_{m-1})$ \Comment{slope of the landscape}
\State \Return $F_m$, $S_m$ and $\sigma\Delta\cos\theta\,L_m$; agreement of the three for straddled steps tests the physics of $\phi^*$
\end{algorithmic}
\end{algorithm}
\begin{algorithm}[t]
\caption{Data-free training of the solution operator (\hyperref[fig:overview]{figure~\ref*{fig:overview}})}\label{alg:train}
\begin{algorithmic}[1]
\Require pattern distribution $\mathcal P$; operator $\mathcal G_{\bm w}$; iterations $N$; batch size $B$; penalty range $\mu_0=10$, $\mu_1=10^3$; peak learning rate $\eta=10^{-3}$
\State initialise $\bm w$, with the last layer of $Q$ set to zero so that $\phi_{\bm w}=\phi_0$ at the start
\For{$n=0,\dots,N-1$}
  \State $\mu\gets\mu_0\,(\mu_1/\mu_0)^{\min(1,\;n/0.6N)}$ \Comment{penalty continuation}
  \For{$b=1,\dots,B$} \Comment{fresh problems, never repeated}
    \State draw $\big(\cos\theta_b(x,z),\,\bm X^*_b\big)\sim\mathcal P$
    \State $(a_b,u_{0,b})\gets\textrm{Inputs}(\cos\theta_b,\bm X^*_b)$ \Comment{\hyperref[alg:forward]{algorithm~\ref*{alg:forward}}, lines 1--4}
    \State $\phi_b\gets\tanh\!\big(u_{0,b}+\mathcal G_{\bm w}(a_b)\big)$
  \EndFor
  \State $\mathcal L\gets\frac1B\sum_b\Big[\frac{E[\phi_b]}{E_0}+\mu\Big\{\Big(\frac{\mathcal V[\phi_b]}{V_0}-1\Big)^2+\Big|\frac{\bm X[\phi_b]-\bm X^*_b}{R_0}\Big|^2\Big\}\Big]$ \Comment{equation~(\ref{eq:loss})}
  \State $\bm g\gets\nabla_{\bm w}\mathcal L$ by automatic differentiation through $E$, $\mathcal V$, $\bm X$; clip $\|\bm g\|\le1$
  \State $\bm w\gets\textrm{AdamW}(\bm w,\bm g)$ with one-cycle learning rate (peak $\eta$), weight decay $10^{-5}$
\EndFor
\State \Return $\bm w$ \Comment{no equilibrium field has been used}
\end{algorithmic}
\end{algorithm}
\begin{algorithm}[t]
\caption{Prediction of an equilibrium field (one forward pass; \hyperref[fig:layer]{figures~\ref*{fig:layer}} and~\ref{fig:kernels})}\label{alg:forward}
\begin{algorithmic}[1]
\Require wall field $\cos\theta(x,z)$, holding position $\bm X^*$, drop volume $V_0$, trained weights $\bm w$
\State $\cos\bar\theta\gets$ mean of $\cos\theta$ over the disc $|\bm x_\parallel-\bm X^*|<R_0$ \Comment{Cassie-like angle}
\State $\phi_0\gets$ diffuse spherical cap of volume $V_0$ and angle $\bar\theta$ above $\bm X^*$, equation~(\ref{eq:prior})
\State $u_0\gets\operatorname{artanh}\phi_0$
\State $a\gets\big[\cos\theta,\;y/N_y,\;(x-X^*)/R_0,\;(z-Z^*)/R_0,\;\phi_0\big]$ \Comment{$\cos\theta$ extended in $y$; $z$ channel in 3D only}
\State $v_0\gets P(a)$; pad $v_0$ by 8 cells at the top in $y$
\For{$l=0,\dots,L-1$}
  \State $v_{l+1}\gets\operatorname{GELU}\big(\mathcal K_lv_l+W_lv_l\big)$ \Comment{FNO, TFNO or WNO kernel $\mathcal K_l$}
\EndFor
\State crop $v_L$ in $y$; $\hat u\gets Q(v_L)$
\State \Return $\phi_{\bm w}=\tanh(u_0+\hat u)$
\end{algorithmic}
\end{algorithm}
\begin{algorithm}[t]
\caption{Reference minimiser and refinement (\hyperref[fig:refsolver]{figure~\ref*{fig:refsolver}})}\label{alg:ref}
\begin{algorithmic}[1]
\Require wall field, $\bm X^*$, $V_0$; set of initial fields $\mathcal S$ (three caps for steps and stripes, one cap otherwise, or an operator prediction $\phi_{\bm w}$ for refinement)
\For{each $\phi^{(0)}\in\mathcal S$}
  \State $\phi\gets\phi^{(0)}$; $\bm\lambda\gets\bm 0$; choose $\mu$
  \Repeat
    \State $\phi\gets\arg\min_{-1\le\phi\le1}\;E[\phi]+\bm\lambda\cdot\bm g[\phi]+\frac{\mu}{2}|\bm g[\phi]|^2$ by L-BFGS-B, warm-started at $\phi$
    \State $\bm\lambda\gets\bm\lambda+\mu\,\bm g[\phi]$ \Comment{$\bm g=(\mathcal V/V_0-1,\,X-X^*,\,Z-Z^*)$}
  \Until{$|\bm g[\phi]|<10^{-7}$ and $E$ stationary}
  \State store the candidate $(\phi,E[\phi],\bm\lambda)$
\EndFor
\State \Return the candidate of lowest energy; its multiplier $\lambda_x=\partial\mathcal E^*/\partial X^*$ is the holding force
\end{algorithmic}
\end{algorithm}

\begin{table}
\centering
\caption{Held-out test problems. None of them is used in training. The reference minimum of the step and stripe
problems is the lowest of three starts; for all problems it is further compared with the refined operator
predictions (\hyperref[sec:accuracy]{Section~\ref*{sec:accuracy}}). Angles in degrees; $R_0=16$ (2D) and $12$ (3D) lattice units.}\label{tab:testsets}
\small
\begin{tabular}{lp{7.6cm}cc}
\toprule
set & content & 2D & 3D\\
\midrule
uniform & $\theta_Y=40$--$150$ (2D), $60,90,120,140$ (3D) & 12 & 4\\
step & $128.8\to79.7$, $128.8\to104.5$, $143.7\to74.5$ (2D; first only in 3D); $-1.6\le X^*/R_0\le1.6$ & 51 & 9\\
stripes & mixed $143.7\,|\,74.5$, $f=\tfrac12$, half-period $w=2$--$12$; area-fraction $128.8\,|\,79.7$, $P=6,12$ (3D: $6$), $f=\tfrac16$--$\tfrac56$; sharp edges & 16 & 11\\
random & training distribution, independent seed & 100 & 20\\
\bottomrule
\end{tabular}
\end{table}

\section{Verification of the reference and of the virtual-work force}\label{sec:verification}

\begin{figure}
\centering
\includegraphics[width=\textwidth]{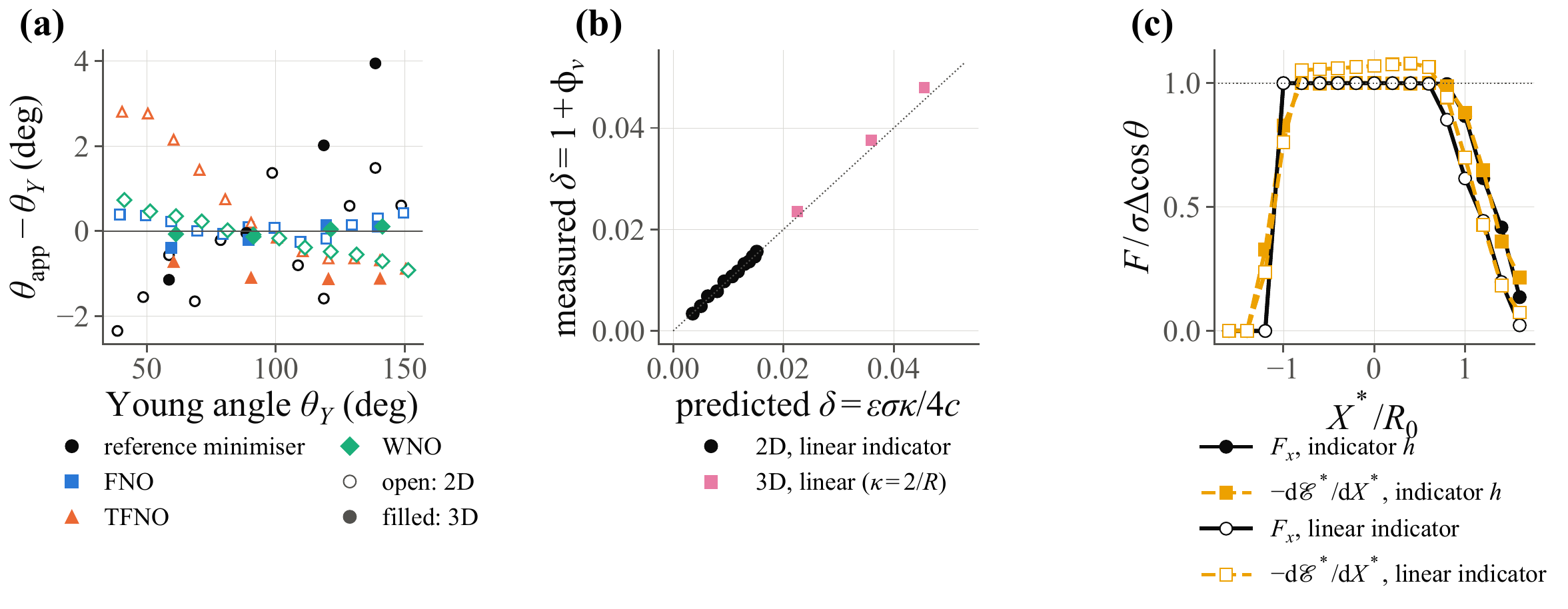}
\caption{Verification of the reference minimiser and of the constraint formulation. (a) Apparent contact angle
minus Young angle on uniform walls, for the reference and the operators, in two (open symbols) and three (filled)
dimensions. (b) Supersaturation $\delta=1+\phi_v$ of the vapour with the linear liquid fraction as indicator,
against the Gibbs--Thomson prediction~(\ref{eq:gt}); with the indicator $h$ used in this work $\delta$ vanishes to
machine precision. (c) Force on a drop held across the $128.8^\circ\!\to\!79.7^\circ$ step: virtual-work force
(\ref{eq:hf}) and slope of the energy landscape, with the indicator $h$ (filled) and with the linear fraction (open),
for which the slope exceeds the force by $V_0/V_{\rm drop}$.}\label{fig:verify}
\end{figure}

Before the operators are assessed, the reference solutions must be shown to be minimisers of the discrete
problem~(\ref{eq:discrete}), and the virtual-work force~(\ref{eq:hf}) must be shown to be the physical force. Three
independent checks are made; their results are collected in \hyperref[fig:verify]{figure~\ref*{fig:verify}}. Each check is also run with
the linear indicator $\tfrac12(1+\phi)$, to make concrete the constraint artefact of \hyperref[sec:gt]{Section~\ref*{sec:gt}}.

\paragraph{Young's law.} On uniform walls with $40^\circ\le\theta_Y\le150^\circ$ the apparent contact angle,
obtained from a least-squares circle fitted to the $\phi=0$ contour above the first three cell rows, departs from
the prescribed Young angle by at most $\NYoungDevMaxTwo^\circ$ in two dimensions (r.m.s.\ $\NYoungDevRmsTwo^\circ$)
and by at most $\NYoungDevMaxThree^\circ$ in the two orthogonal sections of the three-dimensional caps
(\hyperref[fig:verify]{figure~\ref*{fig:verify}}a). The departure is not a convergence error: restarting the minimiser from caps of angle
$\theta_Y\pm25^\circ$ returns the same state to $10^{-9}E_0$ (the one exception, a start at $15^\circ$ on the
$40^\circ$ wall, ends in a distinct state $0.027E_0$ higher). It is a property of the discrete Cahn wall at an interface width
of about three cells, and it is the problem the operators are trained to solve. All comparisons below are therefore
made with the discrete minimiser, not with the continuum angle.

\paragraph{The vapour and the drop volume.} With the indicator $h$ the bulk vapour of every reference solution sits exactly at
the bound $\phi=-1$ (median deviation zero in both dimensions; the only departures are the exponential tails of the
interface), and the volume constraint measures the drop alone. With the
linear indicator it does not (\hyperref[fig:verify]{figure~\ref*{fig:verify}}b). For the two-dimensional drop at $\theta_Y=90^\circ$ the vapour
sits at $\phi_v=-1+\NDeltaVapNinety$, and over the twelve uniform walls the measured shift $\delta=1+\phi_v$, divided
by the Gibbs--Thomson prediction~(\ref{eq:gt}) with $\kappa=1/R$ the curvature of the fitted circle, is
$\NGTratioMean$ on average (largest deviation $\NGTratioSpread$). In three dimensions, where $\kappa=2/R$ and the box
is larger relative to the drop, the shift at $90^\circ$ is $\NDeltaVapNinetyThree$, a fraction $\NVapFracNinetyThree$
of the prescribed volume is held in the vapour, and the drop at $140^\circ$ evaporates entirely (largest value of
$\phi$ in the box, $\NPhiMaxFortyThree$). The supersaturation is quantitatively the Kelvin effect of the diffuse
model, and with the linear indicator the `drop' that a solver returns is smaller than the drop that was asked for.

\paragraph{The force on a drop straddling a step.} For a drop held across a step, the virtual-work force must equal
$\sigma\Delta\cos\theta$ per unit depth independently of the drop shape and of the held position, as long as the step
is straddled (equation~\ref{eq:chordlaw}). Over the $\NNstraddleTwo$ straddling states of the three two-dimensional
step walls of \hyperref[tab:testsets]{table~\ref*{tab:testsets}} it does so to within $\NHFstraddleDev$. The slope of the energy landscape,
$-\mathrm{d}\mathcal{E}^*/\mathrm{d}X^*$, a quantity computed independently from the energies alone, is
$(\NSlopeRatioHMin\textrm{--}\NSlopeRatioHMax)\,\sigma\Delta\cos\theta$, and the drop itself, measured by the midpoint
of its contact lines, moves by $\NKinRatioHMin\textrm{--}\NKinRatioHMax$ times the displacement of the centroid
(\hyperref[fig:verify]{figure~\ref*{fig:verify}}c, filled symbols). The force, the energy landscape and the kinematics are thus mutually
consistent, and the contact-line law $F=\sigma\Delta\cos\theta\,L$ is recovered as an identity of the minimiser. With
the linear indicator the force is unchanged, but the slope is $(\NSlopeRatioMin\textrm{--}\NSlopeRatioMax)\,
\sigma\Delta\cos\theta$ and the drop moves by $\NKinRatioMin\textrm{--}\NKinRatioMax$ times the centroid (open
symbols): the centroid includes the vapour excess, which does not move with the drop, and the multiplier of the
centroid constraint overestimates the capillary force by the factor $V_0/V_{\rm drop}$. The same consideration
applies to any diffuse-interface or pseudopotential calculation of the force on a drop from the response of its
energy to a displacement of its centre of mass, and to any physics-informed loss built on a linear volume fraction.

\section{Accuracy of the data-free operators}\label{sec:accuracy}

\subsection{Test problems and error measures}
The operators are tested on problems that were never used in training (\hyperref[tab:testsets]{table~\ref*{tab:testsets}}): (i) twelve uniform
walls with $40^\circ\le\theta_Y\le150^\circ$; (ii) a drop held at seventeen positions $-1.6\le X^*/R_0\le1.6$ across
three sharp-edged steps whose Young angles are those of the lattice Boltzmann walls used for comparison
($128.8^\circ\!\to\!79.7^\circ$, $128.8^\circ\!\to\!104.5^\circ$, $143.7^\circ\!\to\!74.5^\circ$); (iii) striped walls with
the same angles, of half-period $w=2$--$12$ and area fraction $f=1/6$--$5/6$, with sharp (unsmoothed) stripe edges and
periods down to $P=4$, below the range sampled in training; and (iv) $100$ patterns drawn from the training
distribution with an independent random seed. The structured sets (ii) and (iii) contain configurations that are
deliberately away from the training distribution (held positions up to $1.6R_0$ from the box centre, sharper and
narrower stripes), and they are the ones of physical interest.

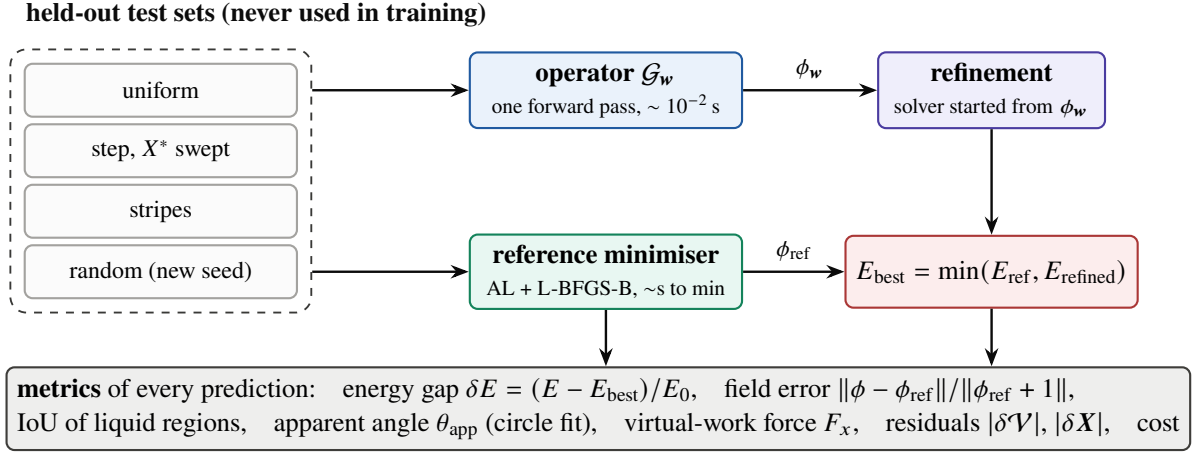
\begin{figure}
\centering
\fitbox{
\begin{tikzpicture}[font=\small]
\node[ttl, anchor=west] at (-0.1,1.0) {held-out test sets (never used in training)};
\foreach \n/\y/\t in {ts1/0/{uniform},ts2/-0.8/{step, $X^*$ swept},ts3/-1.6/{stripes},ts4/-2.4/{random (new seed)}}
  \node[sblk=cGrid, minimum width=3.6cm, anchor=west] (\n) at (0,\y) {\t};
\node[grp, fit=(ts1)(ts4), inner sep=5pt] (ts) {};
\node[blk=cBlue, minimum width=3.6cm] (op) at (7.7,0.0) {\textbf{operator} $\mathcal G_{\bm w}$\\[-1pt]\scriptsize one forward pass, $\sim10^{-2}$\,s};
\node[blk=cGreen, minimum width=3.6cm] (rf) at (7.7,-2.4) {\textbf{reference minimiser}\\[-1pt]\scriptsize AL + L-BFGS-B, $\sim$s to min};
\draw[arr] (ts.east |- op) -- (op);
\draw[arr] (ts.east |- rf) -- (rf);
\node[blk=cPurple, minimum width=3.0cm] (pol) at (12.8,0.0) {\textbf{refinement}\\[-1pt]\scriptsize solver started from $\phi_{\bm w}$};
\draw[arr] (op) -- node[above, font=\footnotesize] {$\phi_{\bm w}$} (pol);
\node[blk=cRed, minimum width=3.0cm] (best) at (12.8,-2.4) {$E_{\rm best}=\min(E_{\rm ref},E_{\rm refined})$};
\draw[arr] (pol) -- (best);
\draw[arr] (rf) -- node[above, font=\footnotesize] {$\phi_{\rm ref}$} (best);
\node[blk=cInkTwo, align=left, anchor=north] (met) at (7.6,-3.65)
  {\textbf{metrics} of every prediction:\quad
   energy gap $\delta E=(E-E_{\rm best})/E_0$,\quad field error $\|\phi-\phi_{\rm ref}\|/\|\phi_{\rm ref}+1\|$,\\
   IoU of liquid regions,\quad apparent angle $\theta_{\rm app}$ (circle fit),\quad virtual-work force $F_x$,\quad
   residuals $|\delta\mathcal V|$, $|\delta\bm X|$,\quad cost};
\draw[arr] (rf.south) -- (rf.south |- met.north);
\draw[arr] (best.south) -- (best.south |- met.north);
\end{tikzpicture}}
\caption{Evaluation workflow. Every held-out problem is predicted by the operator in one forward pass and solved
independently by the reference minimiser; the prediction is also refined by the same minimiser. The best known minimum
$E_{\rm best}$ is the lowest of the reference and refined energies, and every prediction is scored against it and
against the reference field by the metrics listed.}\label{fig:workflow}
\end{figure}

For each problem we compare the energy of the prediction with the best minimum known for that problem. Because
stripes and steps admit several metastable minima, the reference solver is started from three caps (with the
Cassie-like mean angle and with the two extreme angles of the pattern) and the lowest minimum is kept; in addition,
every operator prediction is used as the starting point of a reference solve (``refined'' prediction), and the
lowest of all these minima defines $E_{\rm best}$ (\hyperref[fig:workflow]{figure~\ref*{fig:workflow}}). The energy gap $(E-E_{\rm best})/E_0$ is a strict measure: it is
non-negative for any field that satisfies the constraints, it vanishes only at the global minimum among those found,
and it does not depend on the choice of norm, which affects pointwise field errors at a diffuse interface. We also
report the relative $L^2$ error of the field, the intersection-over-union (IoU) of the liquid regions $\phi>0$, the
constraint residuals and the apparent angles. Per-run and per-family statistics, the distributions of the
constraint residuals and the forces of every random problem are tabulated in the Supplementary Information
({\color{linkblue}tables~}\xref{tab:si_acc}--\xref{tab:si_pol}, {\color{linkblue}figures~}\xref{fig:si_constraints} and~\xref{fig:si_force}).

\subsection{Energy gap, field error and basin selection}
\begin{figure}
\centering
\includegraphics[width=\textwidth]{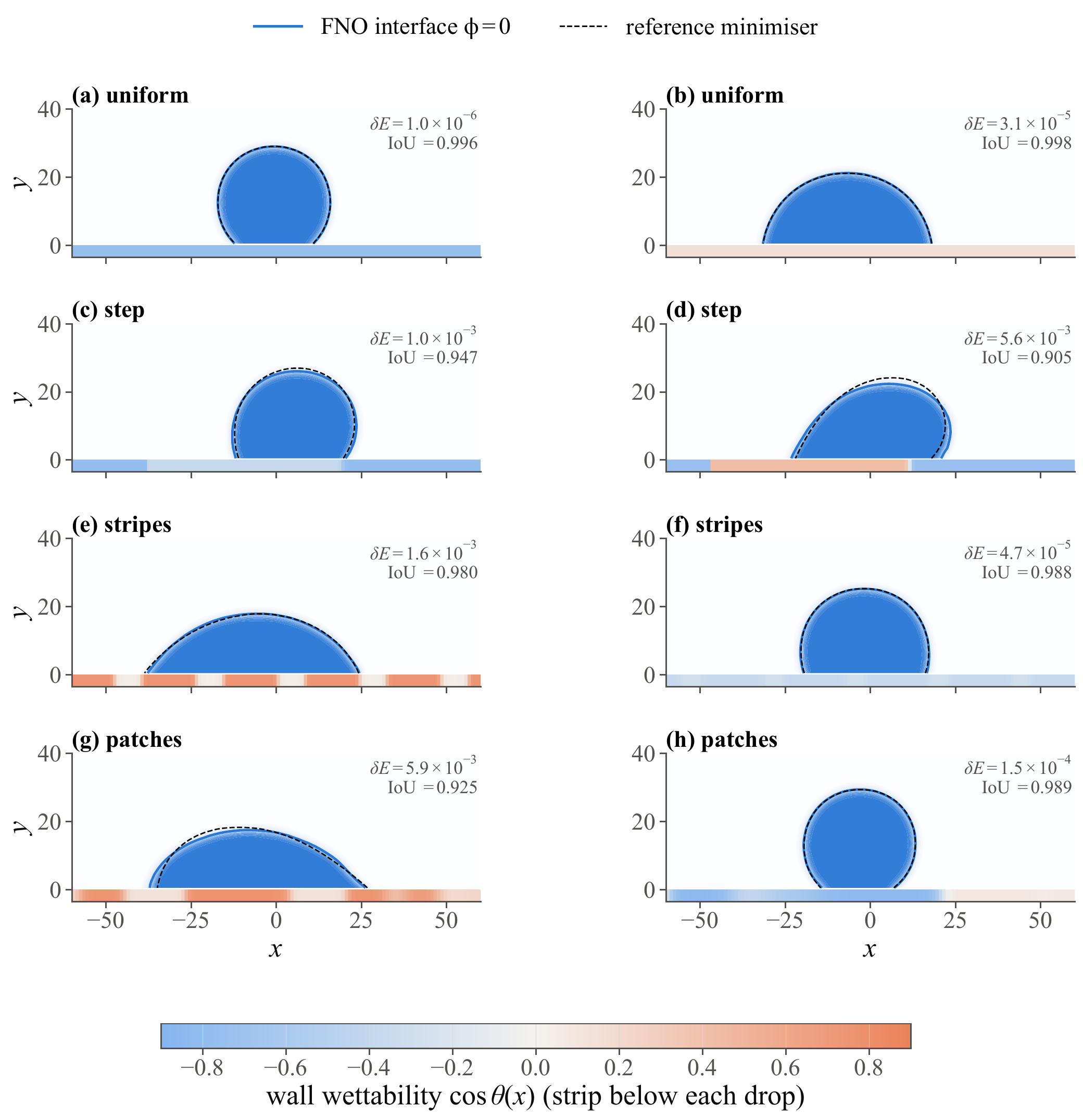}
\caption{Held-out two-dimensional problems, two of each pattern family, never seen in training: field predicted by
the data-free FNO (shading, with its interface $\phi=0$ as a solid line) and interface of the reference minimiser
(dashed). The strip below each drop shows the wall wettability $\cos\theta(x)$, from hydrophobic (blue) to
hydrophilic (orange). The energy gap to the best known minimum, $\delta E=(E-E_{\rm best})/E_0$, and the intersection-over-union of the liquid
regions are given in each panel. The same problems predicted by the TFNO and the WNO are shown in the Supplementary
Information.}\label{fig:gallery2d}
\end{figure}

\hyperref[fig:gallery2d]{Figure~\ref*{fig:gallery2d}} shows predictions for eight held-out two-dimensional problems, and
\hyperref[fig:accuracy]{figure~\ref*{fig:accuracy}} summarises the two-dimensional results. On uniform walls the energy gap of the FNO has a
median of $\NdEmedUniFnoTwo$ (TFNO $\NdEmedUniTfnoTwo$, WNO $\NdEmedUniWnoTwo$), and the apparent angles agree with the
reference to a fraction of a degree. On the held-out random patterns the median gaps are $\NdEmedRandFnoTwo$,
$\NdEmedRandTfnoTwo$ and $\NdEmedRandWnoTwo$ for the FNO, TFNO and WNO, and the median field errors are
$\NLtwoRandFnoTwo$, $\NLtwoRandTfnoTwo$ and $\NLtwoRandWnoTwo$. To put the energy gap in perspective: in two dimensions a gap of
$10^{-3}E_0$ equals the tension energy of $0.1$ lattice units of additional interface, and the median field error of
the FNO corresponds to a mean displacement of the $\phi=0$ contour by about half a lattice cell. The volume constraint, enforced only through the penalty, is satisfied to a median
$\NVrelRandFnoTwo$ (FNO).

The difficulty of a pattern grows with the number of contact-line positions it makes available: the gaps on the
uniform walls are an order of magnitude smaller than on steps, stripes and random patches, on which the contact
line must be placed at, or pinned on, a wettability edge. The largest gaps (up to $\NdEmaxRandFnoTwo$) occur for random
patches whose many edges create closely spaced metastable states.

A central question for a data-free operator is whether it selects the right state among several metastable minima.
Refining each prediction with the reference solver answers it. On the $100$ random problems the refined FNO prediction
returns to the reference minimum (same energy to $10^{-7}E_0$, median $L^2$ distance below $10^{-6}$) in all but
$\NNotBestRandFnoTwo$ case, and in that case it reaches a minimum \emph{lower} than the best of the three reference
starts: the operator, trained to minimise the energy over a whole distribution of problems, found a better state than
the classical solver started from the natural initial guesses. The gap on the random patterns is therefore an
approximation error inside the correct basin, not a wrong basin. On the structured sets, which lie partly outside the
training distribution, the results are less uniform: $\NHigherStepFnoTwo$ of the $51$ step problems and
$\NHigherStrFnoTwo$ of the $16$ stripe problems are refined into a minimum \emph{higher} than the reference: the
prediction sits in the basin of a neighbouring metastable state, with a different number of stripes wetted or with
the front line on the other side of the step, and these cases account for the largest gaps of
\hyperref[fig:accuracy]{figure~\ref*{fig:accuracy}}(a). The two other operators behave in the same way: every refined TFNO and WNO
prediction on the uniform and random walls returns to the reference minimum (with one exception for the TFNO), and on
the structured walls the same few step and stripe problems ($\NHigherStepTfnoTwo$ and $\NHigherStrTfnoTwo$ for the
TFNO, $\NHigherStepWnoTwo$ and $\NHigherStrWnoTwo$ for the WNO) end in a neighbouring metastable state. Basin
selection is therefore a property of the problem (of how close in energy the competing states are) more than of
the architecture.

\paragraph{The three architectures.} The FNO is the most accurate operator on every family: its median energy gap
is three to eight times smaller than that of the TFNO and the WNO on uniform walls, steps and random patterns, but
only $1.2$--$1.6$ times smaller on the stripes, where all three operators return similar
Cassie-like states (\hyperref[sec:stripes]{Section~\ref*{sec:stripes}}; \hyperref[fig:accuracy]{figure~\ref*{fig:accuracy}}a). The TFNO achieves this with $\NParamsTfnoTwo$\,k parameters, against
$\NParamsFnoTwo$\,M for the FNO: a factor of about $150$ fewer, at the cost of at most a factor of about three in the energy gap. The WNO,
whose local wavelet kernel might be expected to suit a thin interface, does not outperform the global Fourier
kernels; its errors are largest on the step walls (median $\NdEmedStepWnoTwo$, against $\NdEmedStepFnoTwo$ for the FNO), plausibly because there the response of the drop to the pattern is non-local:
the position of the far contact line is set by the volume constraint, not by the wall beneath it.

\subsection{Three dimensions}
\hyperref[tab:accuracy]{Table~\ref*{tab:accuracy}} collects the results in both dimensions, and \hyperref[fig:accuracy3d]{figure~\ref*{fig:accuracy3d}} shows the
three-dimensional ones by family. In three dimensions the FNO reaches median energy
gaps of $\NdEmedUniFnoThree$ on uniform walls, $\NdEmedStrFnoThree$ on stripes and $\NdEmedRandFnoThree$ on the
random patterns, comparable to two dimensions, although it was trained on six times fewer problem instances. The step
problems are the hardest ($\NdEmedStepFnoThree$, IoU $\NIoUStepFnoThree$): the drop is held up to $1.6R_0$ from the
step, where in the periodic box it also meets the returning step, and the contact line is pinned on an edge in about
half of the held positions (\hyperref[sec:step]{Section~\ref*{sec:step}}). Refinement of the uniform and step predictions returns the reference
minimum in all but $\NHigherStepFnoThree$ case. The ranking of the architectures is that of two dimensions but the
spread is smaller: on the random patterns the TFNO reaches $\NdEmedRandTfnoThree$ and the WNO $\NdEmedRandWnoThree$,
on the steps $\NdEmedStepTfnoThree$ and $\NdEmedStepWnoThree$, and each of them, after refinement, ends in a higher
minimum on one step problem. A three-dimensional prediction costs $\NTinfRandFnoThree$\,ms (TFNO
$\NTinfRandTfnoThree$\,ms, WNO $\NTinfRandWnoThree$\,ms) against a median of $\NTrefRandFnoThree$\,s for the reference
solve, a factor of about $\NSpeedRandFnoThree$.

\subsection{Cost}
\begin{figure}
\centering
\includegraphics[width=\textwidth]{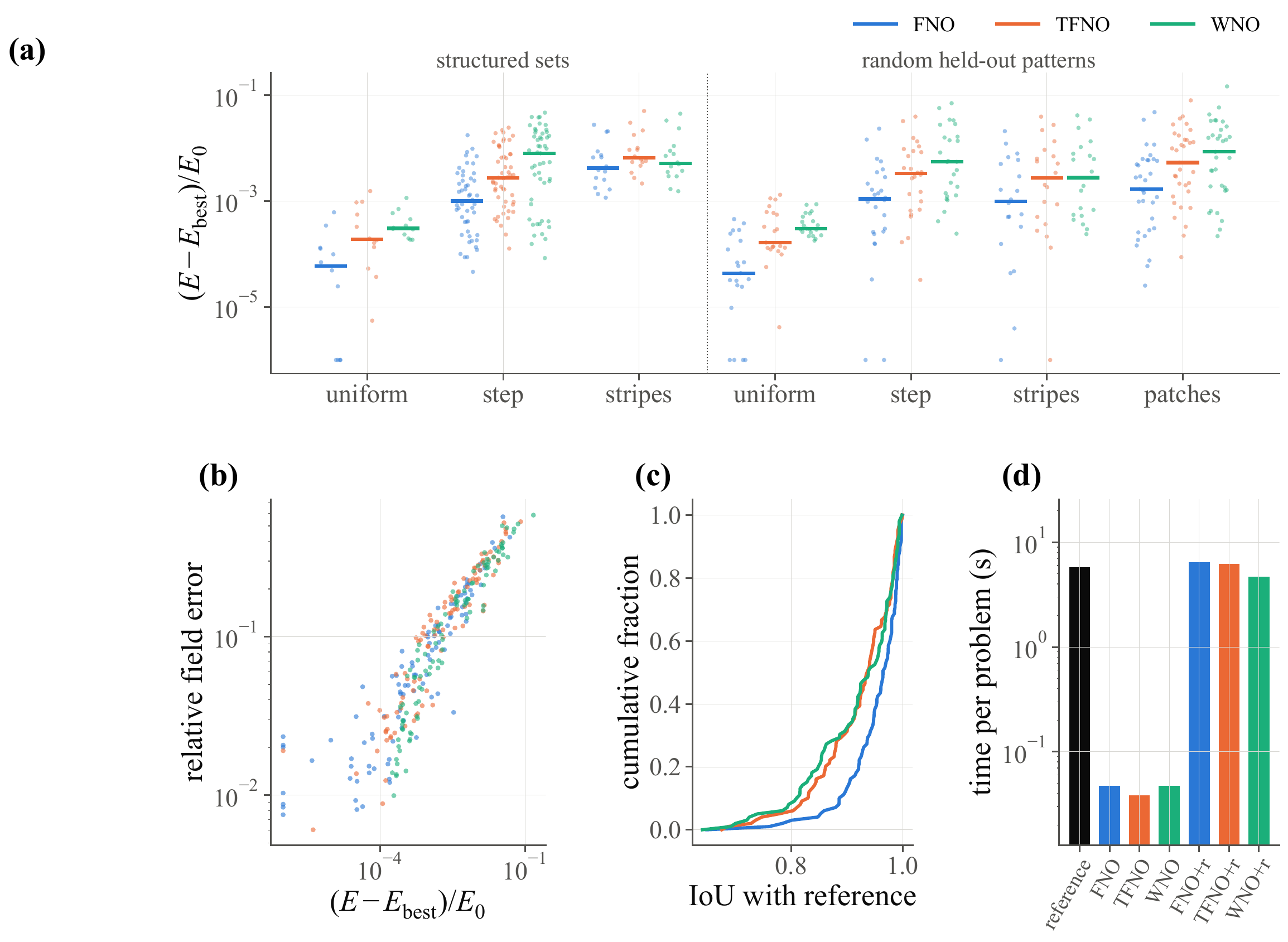}
\caption{Accuracy and cost in two dimensions. (a) Energy gap to the best known minimum for every test problem
(dots) and its median (bars), by pattern family, for the three operators; structured sets on the left, random
held-out patterns on the right. (b) Field error against energy gap. (c) Cumulative distribution of the
intersection-over-union of the predicted and reference liquid regions. (d) Wall-clock time per problem on two CPU
threads: reference solve from the analytic cap, operator prediction, and reference solve started from the operator
prediction. Gaps below $10^{-6}$ are shown at $10^{-6}$.}\label{fig:accuracy}
\end{figure}

\begin{figure}
\centering
\includegraphics[width=\textwidth]{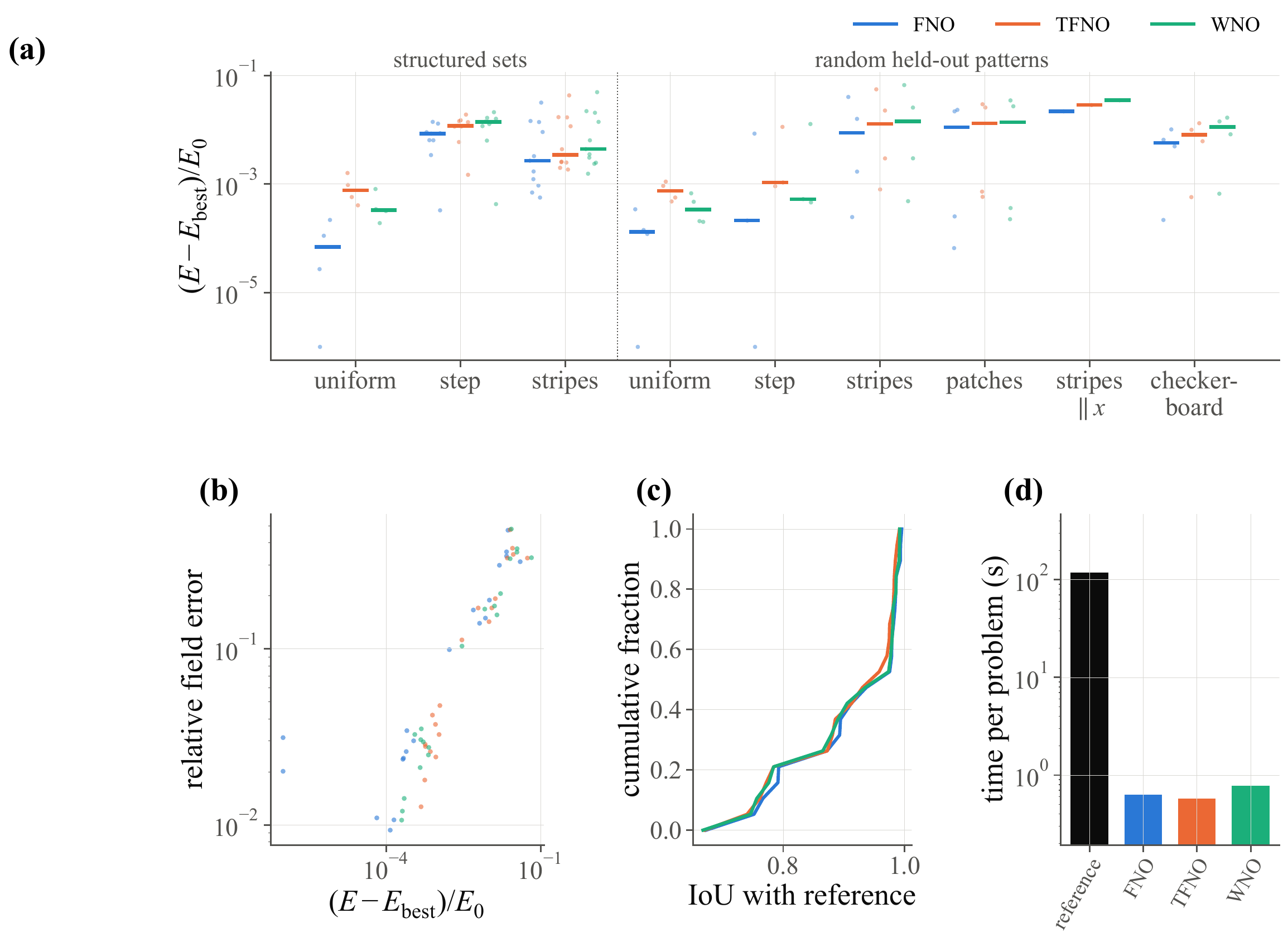}
\caption{Accuracy and cost in three dimensions, as in \hyperref[fig:accuracy]{figure~\ref*{fig:accuracy}}. The random held-out set contains
twenty patterns, so that some families have a single member. (d) Only the reference solve from the analytic
cap and the predictions are timed, since the random three-dimensional problems were not refined. Gaps below
$10^{-6}$ are shown at $10^{-6}$.}\label{fig:accuracy3d}
\end{figure}

A prediction costs $\NTinfRandFnoTwo$\,ms (FNO) on two CPU threads, against a median of $\NTrefRandFnoTwo$\,s for
one reference solve on the same two threads, a factor of about $\NSpeedRandFnoTwo$ (\hyperref[fig:accuracy]{figure~\ref*{fig:accuracy}}d); training costs
$40$--$55$ minutes per operator in two dimensions and $1.2$--$2.3$ hours in three, on four CPU threads. The training cost of
the FNO is recovered after about $570$ two-dimensional or $60$ three-dimensional problems, and a parametric study such as the
force--position curves of \hyperref[sec:step]{Section~\ref*{sec:step}}, which with three starts per position takes the classical solver about a
quarter of an hour in two dimensions and forty minutes in three, costs a few seconds. Using the prediction as the
initial guess of the reference solver does not shorten the solve, however (median $\NTpolRandFnoTwo$\,s): the
multipliers of the augmented Lagrangian start from zero, so that the first subproblem pulls even an accurate initial
field away from the constrained minimum, and the quasi-Newton solver retraces most of the path (about $200$
iterations in the first outer cycle from either start, for the five random problems that we profiled). Warm-starting
the multipliers as well, by fitting them to the predicted field, removes this penalty but helps only at coarse
tolerance (\hyperref[sec:discussion]{Section~\ref*{sec:discussion}}). The operator and the
classical solver are complementary: the first gives an approximate equilibrium in the correct basin in milliseconds,
the second makes it exact when that is needed.

\begin{table}
\centering
\caption{Accuracy and cost of the data-free operators on the held-out sets of \hyperref[tab:testsets]{table~\ref*{tab:testsets}}: median energy gap $(E-E_{\rm best})/E_0$ per set, median intersection-over-union of the liquid regions on the random set, wall-clock time of one prediction $t_{\rm op}$ and median time of one reference solve $t_{\rm ref}$ on the random set. In three dimensions only the uniform and step predictions were refined, so that $E_{\rm best}$ is the reference minimum for the stripe and random sets.}\label{tab:accuracy}
\footnotesize
\begin{tabular}{@{}llccccccc@{}}
\toprule
 & & \multicolumn{4}{c}{median $(E-E_{\rm best})/E_0$} & IoU & $t_{\rm op}$ & $t_{\rm ref}$\\
\cmidrule(lr){3-6}
 & operator & uniform & step & stripes & random & random & (ms) & (s)\\
\midrule
2D & FNO & $6.0\times10^{-5}$ & $1.0\times10^{-3}$ & $4.2\times10^{-3}$ & $6.7\times10^{-4}$ & 0.965 & 47 & 5.8 \\
2D & TFNO & $1.9\times10^{-4}$ & $2.8\times10^{-3}$ & $6.5\times10^{-3}$ & $2.0\times10^{-3}$ & 0.936 & 38 & 5.8 \\
2D & WNO & $3.1\times10^{-4}$ & $8.0\times10^{-3}$ & $5.2\times10^{-3}$ & $1.9\times10^{-3}$ & 0.939 & 47 & 5.8 \\
\midrule
3D & FNO & $6.9\times10^{-5}$ & $8.5\times10^{-3}$ & $2.7\times10^{-3}$ & $1.0\times10^{-3}$ & 0.956 & 628 & 117.8 \\
3D & TFNO & $7.7\times10^{-4}$ & $1.2\times10^{-2}$ & $3.4\times10^{-3}$ & $2.0\times10^{-3}$ & 0.944 & 575 & 117.8 \\
3D & WNO & $3.3\times10^{-4}$ & $1.4\times10^{-2}$ & $4.4\times10^{-3}$ & $1.8\times10^{-3}$ & 0.954 & 778 & 117.8 \\
\bottomrule
\end{tabular}
\end{table}

\section{Capillary physics recovered by \dfeno without data}\label{sec:physics}

\subsection{A drop held across a wettability step}\label{sec:step}
\begin{figure}
\centering
\includegraphics[width=\textwidth]{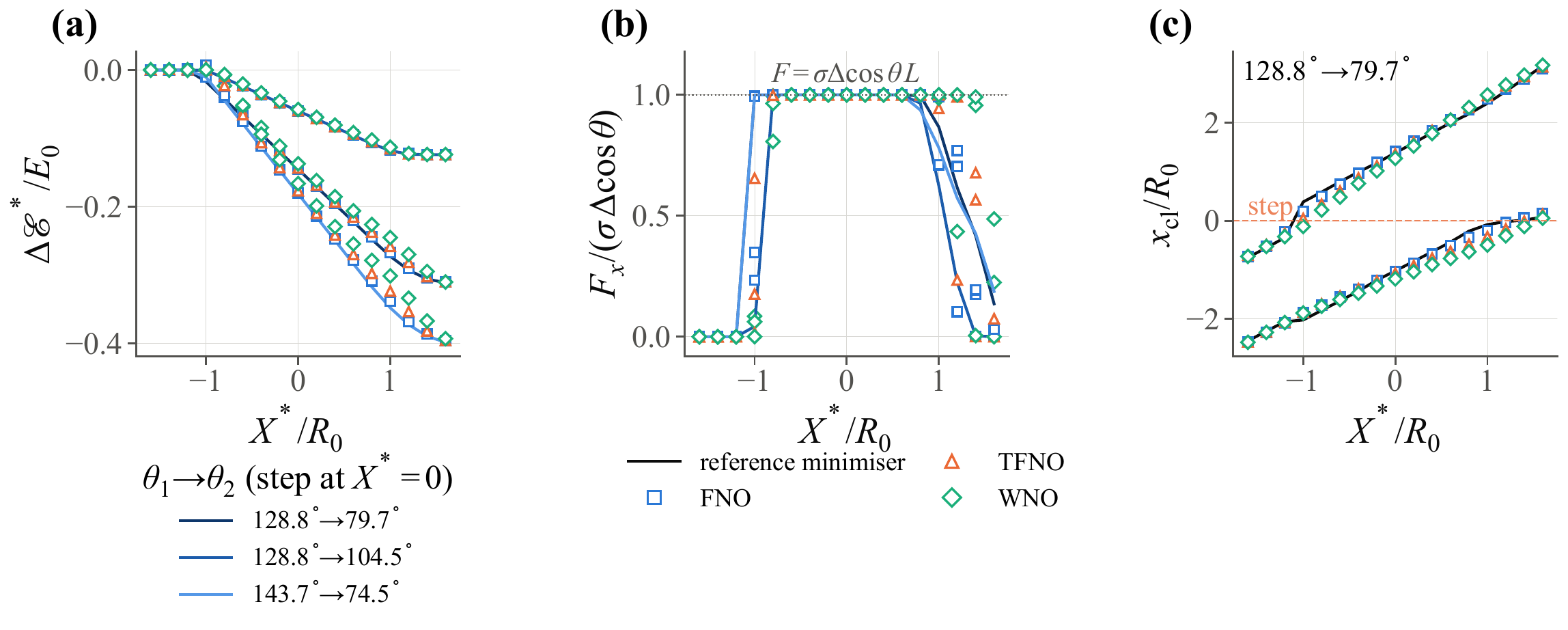}
\caption{A two-dimensional drop held at position $X^*$ across a wettability step (step at $x=0$, hydrophobic for
$x<0$): reference minimiser (lines) and operators (symbols), each point an independent solve or prediction.
(a) Equilibrium energy landscape for three steps. (b) Virtual-work force~(\ref{eq:hf}) in units of
$\sigma\Delta\cos\theta$; the plateau at unity is the contact-line law~(\ref{eq:chordlaw}). (c) Positions of the two
contact lines for the $128.8^\circ\!\to\!79.7^\circ$ step: the front line snaps onto the hydrophilic side at
$X^*\simeq\NXsnapA R_0$, the rear line is pinned at the step for $\NXpinA\lesssim X^*/R_0\lesssim\NXrelA$.}\label{fig:step2d}
\end{figure}

\hyperref[fig:step2d]{Figure~\ref*{fig:step2d}} follows a two-dimensional drop that is held at successive positions $X^*$ across a step from
a hydrophobic ($x<0$) to a hydrophilic ($x>0$) wall. Each point is an independent constrained minimisation (reference)
or an independent forward pass of an operator; the operators were never shown a step at most of these positions
relative to the box, nor these Young angles.

\paragraph{Energy landscape.} The equilibrium energy $\mathcal{E}^*(X^*)$ is flat while the drop rests entirely on
one wall, falls steeply while it straddles the step, and flattens again once it rests on the hydrophilic wall
(\hyperref[fig:step2d]{figure~\ref*{fig:step2d}}a). The total decrease of the energy, $\NEdropA E_0$, $\NEdropB E_0$ and $\NEdropC E_0$ for the three steps, is
the work that the step can do on a drop that crosses it quasi-statically, and an upper bound on the energy available to
a released drop. The operators reproduce the landscape, including
the positions of its two kinks, without having been trained on any landscape: each point is a separate prediction,
and the landscape emerges from the consistency of the predictions across positions.

\paragraph{Force and the contact-line law.} The virtual-work force (\hyperref[fig:step2d]{figure~\ref*{fig:step2d}}b) makes the structure
explicit. It vanishes while the drop is on one wall, jumps to $F=\sigma\Delta\cos\theta$ when the advancing contact
line reaches the step ($X^*\simeq\NXsnapA R_0$ for the $128.8^\circ\!\to\!79.7^\circ$ step), and keeps exactly this
value, independently of the deformation of the drop, for as long as both contact lines are on opposite sides of the
step: this is the contact-line force law of equation~(\ref{eq:chordlaw}), here an identity of the equilibrium problem. The
FNO reproduces the plateau to a median relative error of $\NFstrErrFnoTwo$ in $\NFstrAgreeFnoTwo$ of the
$\NNstrTwo$ straddling states (TFNO $\NFstrAgreeTfnoTwo$, WNO $\NFstrAgreeWnoTwo$). The states it misses lie at the
snap position itself, where the drop with its front line just before the step and the drop with the front line just
past it are nearly degenerate.

\paragraph{Pinning of the receding contact line.} Beyond $X^*\simeq\NXpinA R_0$ the rear contact line reaches the step
and stays there (\hyperref[fig:step2d]{figure~\ref*{fig:step2d}}c): instead of crossing onto the hydrophilic wall it is pinned at the edge,
while the drop is pulled further. Gibbs' criterion \citep{Gibbs1906,Oliver1977} allows a contact line on an edge
to take any angle between the two Young angles; as $X^*$ increases the local angle at the rear decreases from the
hydrophobic towards the hydrophilic value, the wall trace $g(\phi^*)$ at the step decreases from $+\tfrac12$, and the
force falls continuously to zero, reached when the rear line is released at $X^*\simeq\NXrelA R_0$. The response of
the drop to the step is therefore strongly asymmetric: the advancing line \emph{snaps} onto the hydrophilic side, the
receding line is \emph{pinned} at the edge. In the pinning regime the force is a functional of the field at the
contact line itself, which is the most difficult output of the problem, and the operators are least accurate there (mean
absolute error of $F/\sigma\Delta\cos\theta$ of $\NFpinErrFnoTwo$ for the FNO over the $\NNpinTwo$ pinned states): their
rear contact line reaches the step too late, so that the predicted drop still straddles the step and carries the
full force over part of the range in which the reference drop is already pinned.

\begin{figure}
\centering
\includegraphics[width=\textwidth]{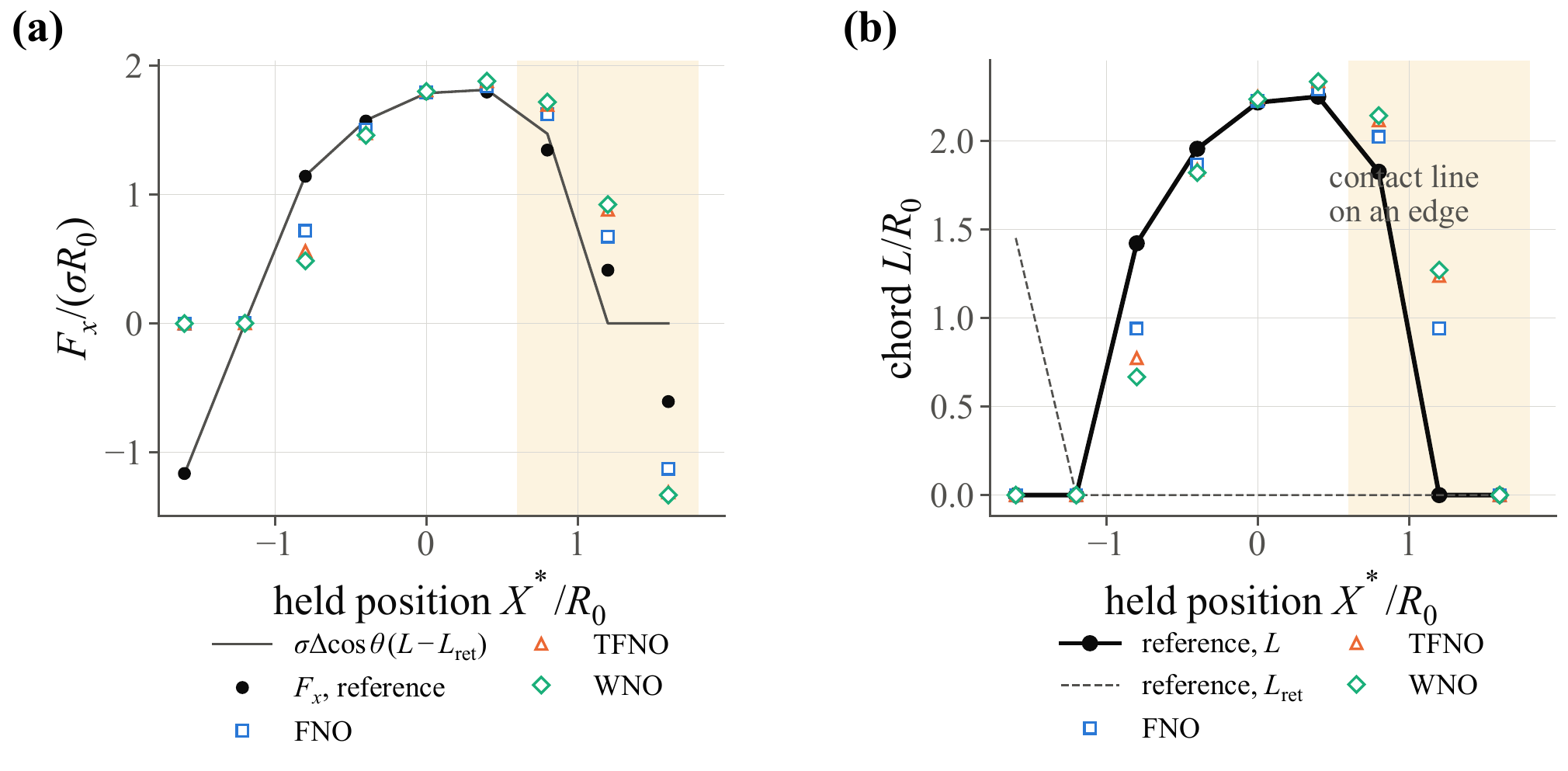}
\caption{A three-dimensional drop ($R_0=12$, $128.8^\circ\!\to\!79.7^\circ$) held at nine positions across a
step in a periodic box. (a) Virtual-work force~(\ref{eq:hf}) of the reference minimiser (filled circles) and of the
operators (open symbols) against the contact-line law $\sigma\Delta\cos\theta\,(L-L_{\rm ret})$ evaluated with the
reference chords (line). (b) The chord $L$ cut by the footprint on the step line and, dashed, the chord $L_{\rm ret}$
on the returning step of the periodic box. Shaded: states in which a contact line sits on an edge, where the force
departs from the chord law.}\label{fig:step3d}
\end{figure}

\paragraph{The three-dimensional step: the chord law.} In three dimensions the force on a straddling drop is no
longer a constant: by (\ref{eq:chordlaw}) it is $\sigma\Delta\cos\theta$ times the chord $L$ that the footprint cuts
on the step line, which grows from zero as the advancing line reaches the step, passes through a maximum close to
the footprint diameter, and falls again as the rear line approaches (\hyperref[fig:step3d]{figure~\ref*{fig:step3d}}). In the periodic box
a drop held far from the step also touches the returning step, and the law then reads
$F_x=\sigma\Delta\cos\theta\,(L-L_{\rm ret})$. The reference force, computed from the wall trace alone, follows
this law with the geometric chords measured on the $\phi=0$ contour to within $\NStrThreeDev$ per cent in all
$\NStrThreeN$ straddling states, with a maximum $F_x=\NFmaxThree\,\sigma R_0$. In the remaining states a contact line
sits on an edge (the rear line on the step, or the advancing line on the returning, hydrophilic-to-hydrophobic
edge), and the force departs from the chord law exactly as the pinning argument of \hyperref[sec:hf]{Section~\ref*{sec:hf}} predicts:
the chord on the line vanishes while $g(\phi^*)$ on the edge, and with it the force, takes intermediate values.
This is the statement, here derived and verified for the equilibrium problem, that our lattice Boltzmann
simulations found to hold for the force that drives a three-dimensional drop across a step.
The FNO reproduces the force of the straddling states to a median relative error of $\NFerrStrThreeFno$ (TFNO
$\NFerrStrThreeTfno$, WNO $\NFerrStrThreeWno$); over all held positions, including the pinned ones, its median error
is $\NFerrThreeFno$ of the largest force (TFNO $\NFerrThreeTfno$, WNO $\NFerrThreeWno$). As in two dimensions, the
operators err where a contact line sits on an edge (\hyperref[fig:step3d]{figure~\ref*{fig:step3d}}).

\subsection{Apparent angles on striped walls}\label{sec:stripes}
\begin{figure}
\centering
\includegraphics[width=\textwidth]{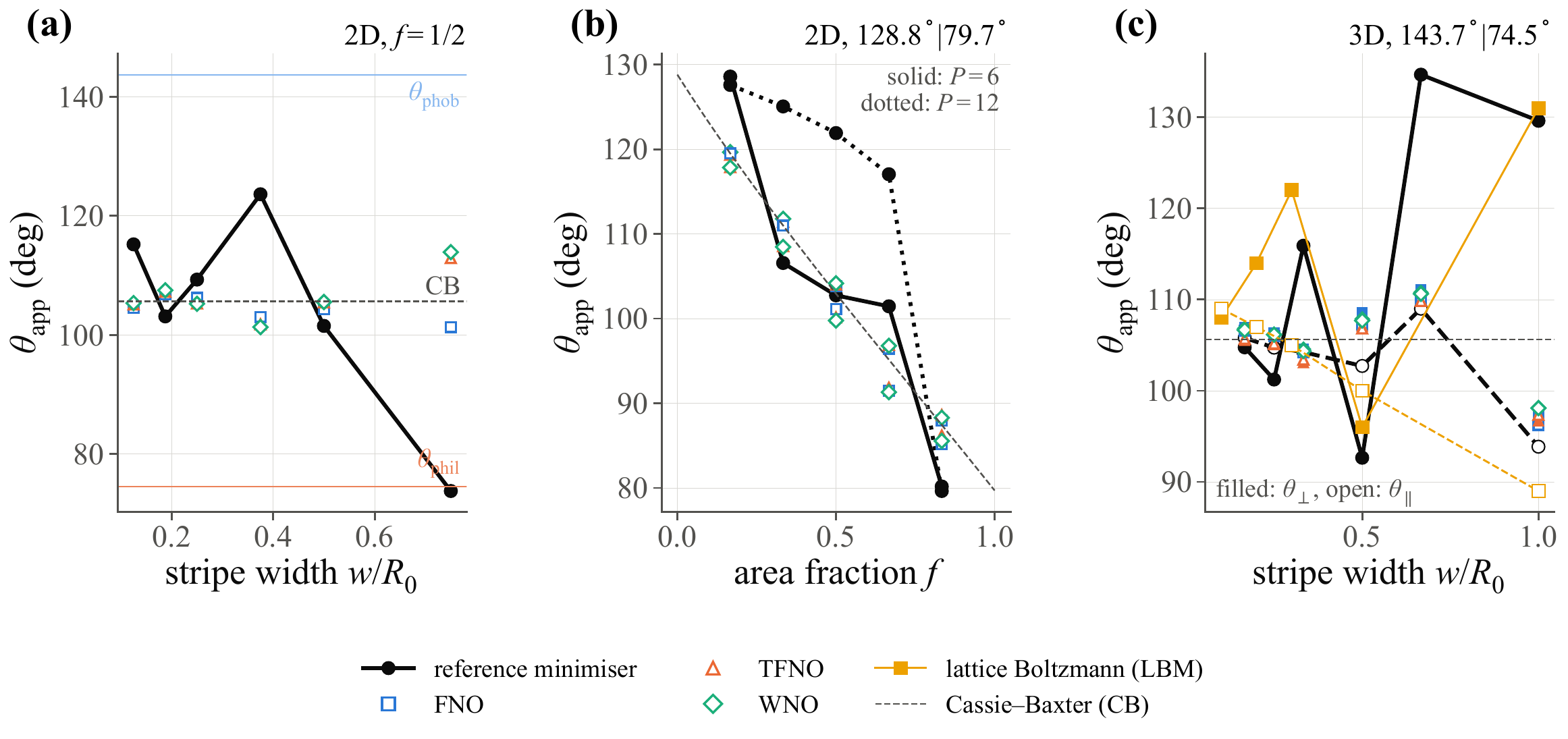}
\caption{Apparent contact angles on striped walls. (a) Two dimensions, mixed $143.7^\circ|74.5^\circ$ stripes of
width $w$; the Cassie--Baxter angle and the two Young angles are shown as horizontal lines. (b) Two dimensions,
$128.8^\circ|79.7^\circ$ stripes of period $P$ and hydrophilic fraction $f$, with the Cassie--Baxter angle.
(c) Three dimensions, mixed stripes: angle across ($\theta_\perp$, filled) and along ($\theta_\parallel$, open) the
stripes, compared with the drops of our three-dimensional lattice Boltzmann simulations at the same $w/R_0$.}\label{fig:stripes}
\end{figure}

\paragraph{Two dimensions: pinning on the stripe edges.} On a striped wall the equilibrium drop has its contact lines
on stripe edges whenever it can, and its apparent angle is then not the Cassie--Baxter angle but whatever angle the
volume and the positions of the two pinned lines impose \citep{Brandon2003,Kusumaatmaja2007}. The reference minimisers
show this clearly (\hyperref[fig:stripes]{figure~\ref*{fig:stripes}}a,b): for the $143.7^\circ\,|\,74.5^\circ$ stripes of equal width the
apparent angle varies non-monotonically between $\NThRefRangeLoTwo^\circ$ and $\NThRefRangeHiTwo^\circ$ as the half-period
$w$ grows from $2$ to $12$ cells, about the Cassie--Baxter value of $105.6^\circ$; on the $128.8^\circ\,|\,79.7^\circ$
walls of varying area fraction it stays within $9^\circ$ of the Cassie--Baxter curve for the shorter period $P=6$ but departs from
it by more than $20^\circ$ for $P=12$, for which the drop spans only about two periods. The operators do not resolve this
selection. Their apparent angles stay within $\NThOpRangeLoFnoTwo^\circ$--$\NThOpRangeHiFnoTwo^\circ$ (FNO) on the
equal-width stripes, close to the Cassie--Baxter angle, with an r.m.s.\ deviation from the reference of
$\NThRmsStrFnoTwo^\circ$ (TFNO $\NThRmsStrTfnoTwo^\circ$, WNO $\NThRmsStrWnoTwo^\circ$) over all stripe problems: the
predicted contact lines sit between edges rather than on them, and the predicted drop is an average over the
pattern. Because a contact-line displacement along a stripe is one of the softest modes of the energy
(\hyperref[sec:discussion]{Section~\ref*{sec:discussion}}), these predictions are nevertheless close in energy to the pinned state (median gap
$\NdEmedStrFnoTwo E_0$), and refinement carries all but $\NHigherStrFnoTwo$ of them into the reference minimum. The
apparent angle on a striped wall is thus the quantity most sensitive to the soft-mode error of an energy-trained
operator, and the one for which a refinement step is indispensable.

\begin{figure}
\centering
\includegraphics[width=\textwidth]{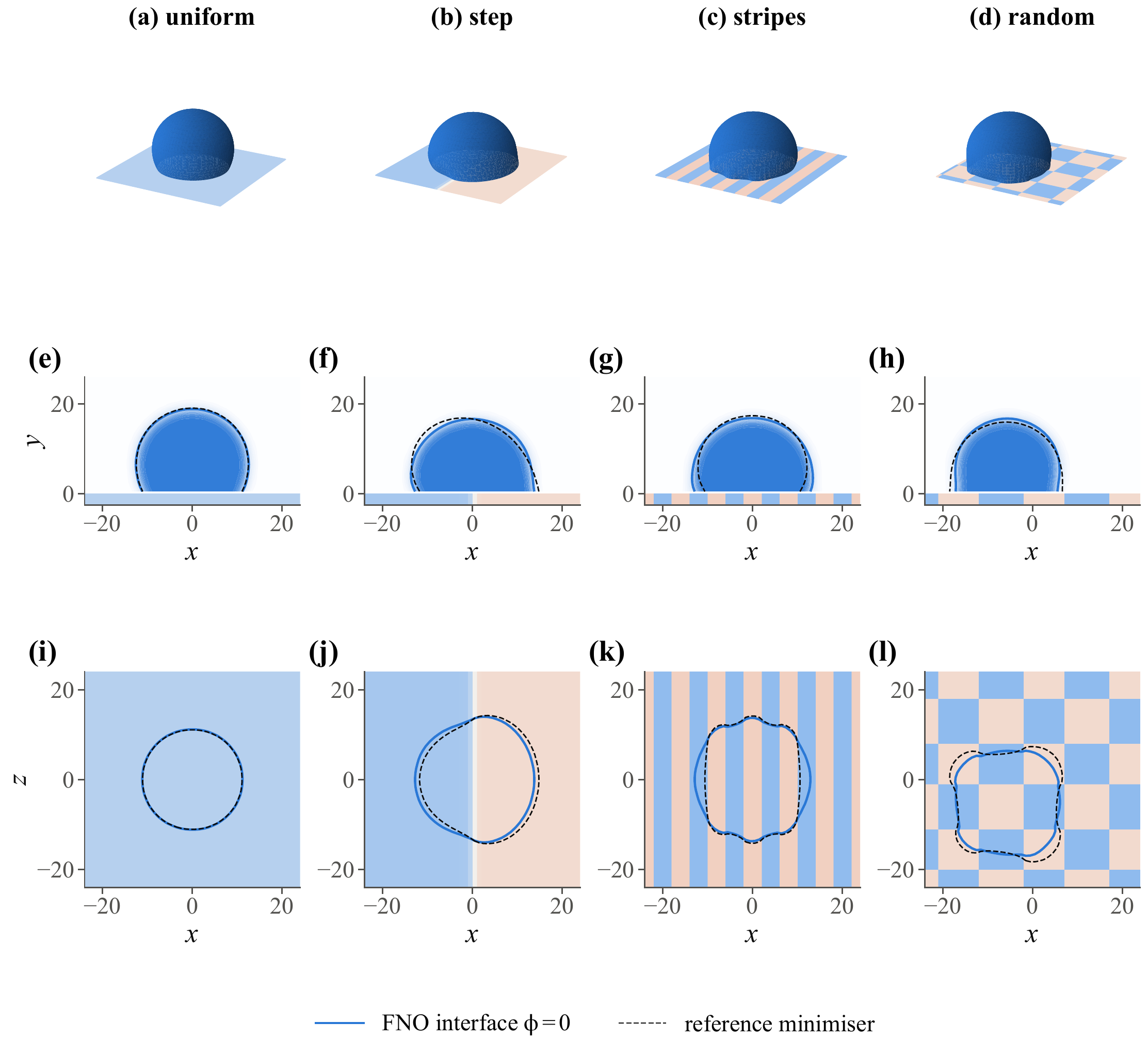}
\caption{Held-out three-dimensional problems ($R_0=12$, $64\times32\times64$ cells): a uniform wall at
$\theta_Y=120^\circ$, a drop held centrally on the $128.8^\circ\!\to\!79.7^\circ$ step, $143.7^\circ\,|\,74.5^\circ$ stripes
of half-period $w=4$ and a checkerboard. (a--d) Drop surface $\phi=0$ predicted by the data-free FNO above the
patterned wall (colour: $\cos\theta$, blue hydrophobic, orange hydrophilic), shown over the central $44\times44$
cells. (e--h) Section through the centre of the drop perpendicular to $z$: predicted field (shading), its interface
(solid) and the interface of the reference minimiser (dashed). (i--l) Wetted footprint $\phi(x,0,z)=0$ of the FNO
(solid) and of the reference (dashed) over the wall pattern.}\label{fig:gallery3d}
\end{figure}

\paragraph{Three dimensions: the anisotropy of the apparent angle.} On a three-dimensional striped wall the
pinning acts only across the stripes: the contact line can slide along a stripe edge but not across it, so that the
footprint is shaped by the stripes (\hyperref[fig:gallery3d]{figure~\ref*{fig:gallery3d}}k). The apparent angle measured in the section
across the stripes, $\theta_\perp$, and along them, $\theta_\parallel$, then differ. The reference minimisers on the
$143.7^\circ\,|\,74.5^\circ$ stripes show an anisotropy of up to $\NAnisoRefMaxThree^\circ$
(\hyperref[fig:stripes]{figure~\ref*{fig:stripes}}c): for $w/R_0=1$, $\theta_\perp=\NThPerpRefWTwelve^\circ$ and
$\theta_\parallel=\NThParRefWTwelve^\circ$; for $w/R_0=1/2$, $\theta_\perp=\NThPerpRefWSix^\circ$ and
$\theta_\parallel=\NThParRefWSix^\circ$; for $w/R_0=1/3$, $\theta_\perp=\NThPerpRefWFour^\circ$ and
$\theta_\parallel=\NThParRefWFour^\circ$. These are, to within six degrees, the angles of the drops in our three-dimensional lattice Boltzmann
simulations at comparable $w/R_0$ ($131^\circ$ and $89^\circ$ at $w/R_0=1$, $96^\circ$ and $100^\circ$ at
$1/2$, $122^\circ$ and $105^\circ$ at $0.3$), obtained there at a $2.5$ times larger drop radius and as the end state of
an impact and relaxation. The agreement shows that the anisotropy found in the dynamic
simulations is an equilibrium property, set by which stripe edges the contact line is pinned on, and that the
non-monotonic dependence on $w/R_0$ reflects the discrete number of stripes spanned by the footprint. The
operators, in contrast, predict nearly isotropic drops: $\theta_\perp$ and $\theta_\parallel$ differ by at most
$\NAnisoMaxFnoThree^\circ$ for the FNO (TFNO $\NAnisoMaxTfnoThree^\circ$, WNO $\NAnisoMaxWnoThree^\circ$) and stay near the Cassie--Baxter value, as in two dimensions, while the energy gap on these
walls remains small (median $\NdEmedStrFnoThree E_0$). The anisotropy of the apparent angle is, with the pinned
receding line on a step, the clearest example of a physically important quantity that lies in the soft modes of the
energy, and therefore of a quantity that an energy-trained operator resolves least accurately.

\section{Gravity and inclined walls}\label{sec:gravity}

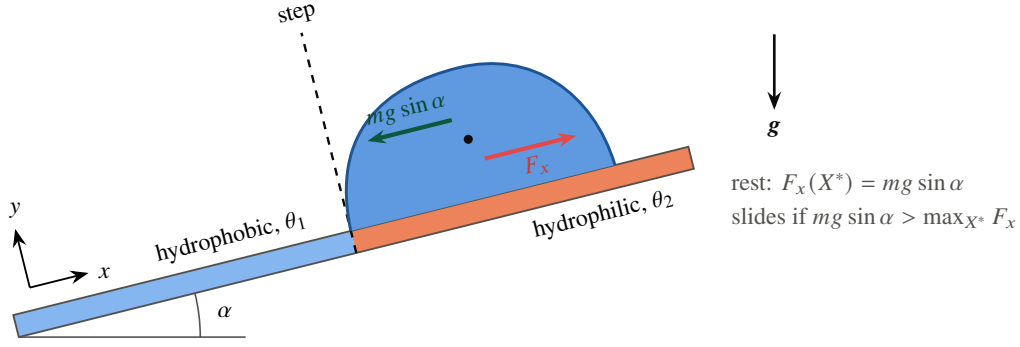
\begin{figure}
\centering
\fitbox{
\begin{tikzpicture}[font=\small]
\begin{scope}[rotate=14]
  \fill[cPhob] (-4.6,-0.3) rectangle (0,0);
  \fill[cPhil] (0,-0.3) rectangle (4.6,0);
  \draw[cInkTwo, line width=0.7pt] (-4.6,-0.3) rectangle (4.6,0);
  \draw[cInk, dashed, line width=0.9pt] (0,-0.3) -- (0,2.7);
  \node[font=\footnotesize, fill=white, inner sep=1pt, rotate=14] at (0,2.95) {step};
  \node[font=\footnotesize, rotate=14] at (-1.55,0.3) {hydrophobic, $\theta_1$};
  \node[font=\footnotesize, rotate=14] at (3.3,-0.6) {hydrophilic, $\theta_2$};
  \fill[cBlue!75] (0,0) .. controls (0.05,1.35) and (0.9,1.75) .. (1.9,1.75) .. controls (2.9,1.75) and (3.55,1.0) .. (3.6,0) -- cycle;
  \draw[cBlue!70!black, line width=1.1pt] (0,0) .. controls (0.05,1.35) and (0.9,1.75) .. (1.9,1.75) .. controls (2.9,1.75) and (3.55,1.0) .. (3.6,0);
  \fill[cInk] (1.8,0.8) circle (0.06);
  \draw[->, cRed, line width=1.6pt] (1.95,0.5) -- (3.2,0.5);
  \node[font=\footnotesize, text=cRed!85!black, rotate=14] at (2.6,0.25) {$F_x$};
  \draw[->, cGreen!50!black, line width=1.6pt] (1.65,1.1) -- (0.5,1.1);
  \node[font=\footnotesize, text=cGreen!40!black, rotate=14] at (1.1,1.38) {$mg\sin\alpha$};
  \draw[->, line width=0.7pt] (-4.3,0.3) -- (-3.5,0.3) node[right, font=\footnotesize] {$x$};
  \draw[->, line width=0.7pt] (-4.3,0.3) -- (-4.3,1.1) node[above, font=\footnotesize] {$y$};
\end{scope}
\begin{scope}[rotate=14]\coordinate (C) at (-4.6,-0.3);\end{scope}
\draw[cInkTwo, line width=0.6pt] (C) -- ++(3.0,0);
\draw[cInkTwo, line width=0.6pt] (C) ++(2.4,0) arc (0:14:2.4);
\node[font=\footnotesize] at ($(C)+(7:2.75)$) {$\alpha$};
\draw[->, cInk, line width=1.0pt] (5.6,2.6) -- (5.6,1.6) node[below, font=\footnotesize] {$\bm g$};
\node[lbl, anchor=north west, align=left] at (4.9,0.9) {rest: $F_x(X^*)=mg\sin\alpha$\\[1pt] slides if $mg\sin\alpha>\max_{X^*}F_x$};
\end{tikzpicture}}
\caption{A drop on a wall inclined by $\alpha$, with a wettability step and the hydrophobic side downhill. The weight
component $mg\sin\alpha$ pulls the drop downhill and the pattern force $F_x$ of the step pulls it towards the hydrophilic
side; the drop rests where the two balance, equation~(\ref{eq:balance}), and slides off when the weight exceeds the
largest force the step can exert. The normal component $mg\cos\alpha$ flattens the drop.}\label{fig:incline}
\end{figure}

The results above are for drops much smaller than the capillary length. For larger drops gravity flattens the drop,
and on an inclined wall its component along the wall must be balanced by the wall pattern for the drop to remain at
rest. In this section we extend \dfeno to gravity in two dimensions, verify the reference solutions against the
Young--Laplace equation, and use the virtual-work force to predict where a drop comes to rest on an inclined wall with
a wettability step and at which inclination it slides off (\hyperref[fig:incline]{figure~\ref*{fig:incline}}).

\subsection{Formulation}\label{sec:gformulation}
With the Bond number $Bo=\Delta\rho\,g R_0^2/\sigma$ and a wall inclined by an angle $\alpha$, gravity has a component
normal to the wall, $Bo_n=Bo\cos\alpha$, and a component along it, $Bo_t=Bo\sin\alpha$, directed downhill (towards
$-x$). The gravitational energy of the liquid is
\begin{equation}
\mathcal{E}_g[\phi]=\frac{\sigma}{R_0^2}\int_\Omega h(\phi)\,\big(Bo_n\,y+Bo_t\,x\big)\,\mathrm{d}V,
\label{eq:gravity}
\end{equation}
with $h$ the liquid indicator of the constraints~(\ref{eq:constraints}) and $y$ the height above the wall. The normal
part does not depend on the position of the drop along the wall, so that the bulk energy remains invariant under a
translation of the fluid and the virtual-work force~(\ref{eq:hf}) is unchanged: it is still the wall integral of the
equilibrium field, and the contact-line law~(\ref{eq:chordlaw}) still holds for a straddled step, independently of the
deformation of the drop by gravity. The tangential part is linear in $x$ and equals $\sigma Bo_t V_0X^*/R_0^2$ for a drop
held at its centroid $X^*$; it therefore does not change the held shape, and only adds the weight
$W=\sigma Bo_tV_0/R_0^2=mg\sin\alpha$ to the force balance. A free drop on the inclined wall rests where the pattern force
balances its weight,
\begin{equation}
F_x(X_{\rm eq};Bo_n)=mg\sin\alpha,\qquad \frac{\mathrm{d}F_x}{\mathrm{d}X^*}\Big|_{X_{\rm eq}}<0,
\label{eq:balance}
\end{equation}
where $F_x(X^*;Bo_n)$ is the virtual-work force of the held drop at the normal Bond number $Bo_n$. The largest force a
step can exert is the plateau value $\sigma\Delta\cos\theta$ of a straddling drop, so that in two dimensions the drop
slides off the step when
\begin{equation}
\sin\alpha>\sin\alpha_c=\frac{\Delta\cos\theta\,R_0^2}{Bo\,V_0}=\frac{\Delta\cos\theta}{\upi\,Bo}.
\label{eq:alphac}
\end{equation}
This is the analogue, for a chemical step, of the retention force $\sigma w(\cos\theta_R-\cos\theta_A)$ of a drop of
width $w$ held by contact-angle hysteresis \citep{Furmidge1962}: the difference of the cosines of the two Young angles
takes the place of the hysteresis. For the $128.8^\circ\!\to\!79.7^\circ$ step, $\Delta\cos\theta=\NGDeltaCos$, and
equation~(\ref{eq:alphac}) gives $\alpha_c=\NGAlphaCOne^\circ$ for $Bo=1$ and $\NGAlphaCHalf^\circ$ for $Bo=0.5$; for
$Bo<\NGBoMin$ the step holds the drop at any inclination.

\paragraph{Training and reference.} Because the tangential part does not change the held shape, the operator needs
only $Bo_n$ as an additional input. We train an FNO and a TFNO with the architectures and budget of
\hyperref[sec:methods]{Section~\ref*{sec:methods}}, with a constant input channel $Bo_n$, drawn uniformly in
$[0,1.2]$ for every training instance, and with $\mathcal{E}_g$ added to the energy in the loss~(\ref{eq:loss}); the cap
prior is unchanged, so that the network learns the deformation by gravity. Training takes $\NGTrainFno$\,h (FNO) and
$\NGTrainTfno$\,h (TFNO) on four CPU threads, with the reference solves running at the same time. The reference minimiser of
\hyperref[sec:reference]{Section~\ref*{sec:reference}} is used with the same energy. For the inclined wall it is run
without the centroid constraint and with both components of gravity, starting from a cap on the hydrophilic side
($X^*=R_0$); it then finds the rest position of a free drop directly, without using equation~(\ref{eq:balance}).

\subsection{Drop shapes and accuracy}
\begin{figure}
\centering
\includegraphics[width=\textwidth]{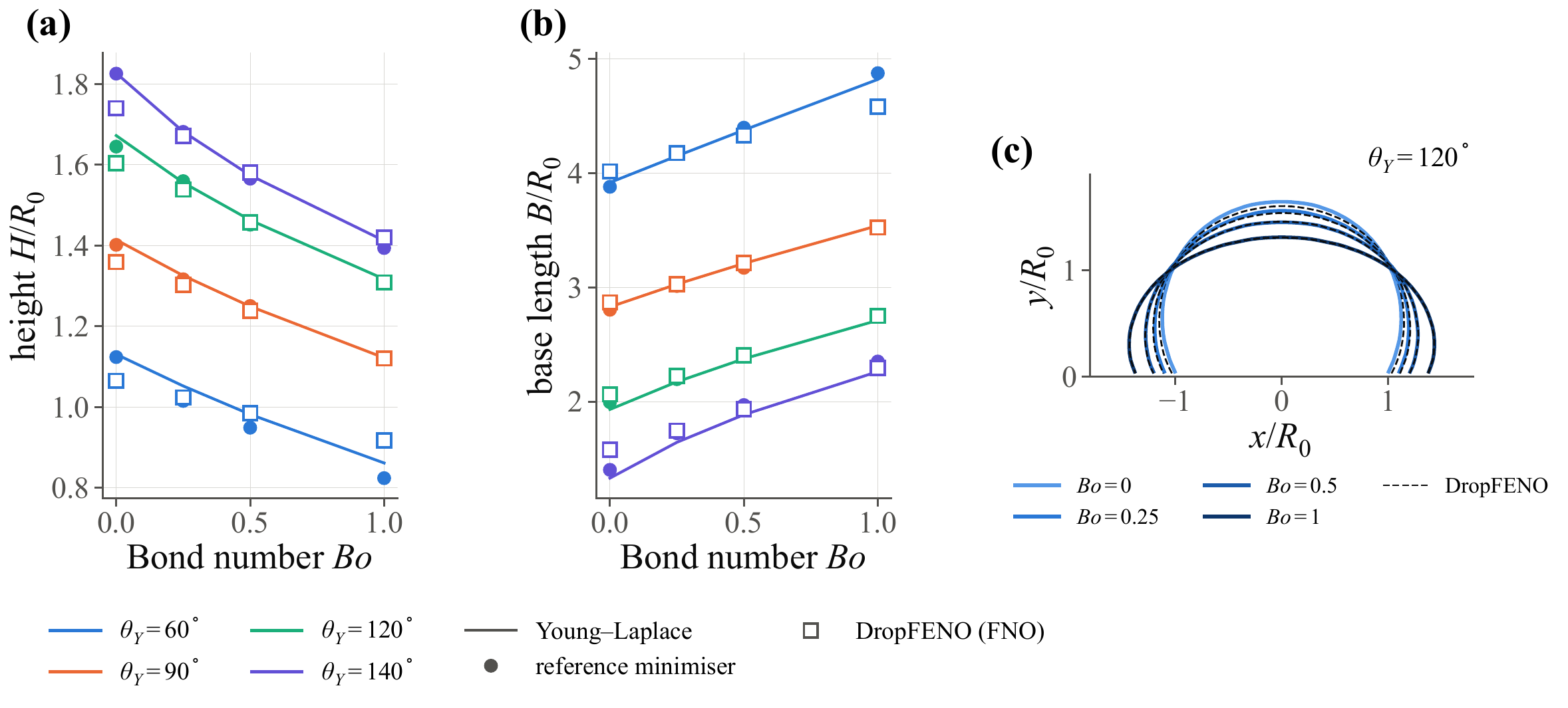}
\caption{Drops on uniform walls under gravity (two dimensions). (a) Height and (b) base length against the Bond number
for four Young angles: sharp-interface Young--Laplace solution (lines), reference minimiser (filled circles) and \dfeno
(FNO, open squares). (c) Interfaces $\phi=0$ at $\theta_Y=120^\circ$ for $Bo=0$, $0.25$, $0.5$ and $1$: reference
(solid) and \dfeno (dashed).}\label{fig:gshapes}
\end{figure}

On uniform walls the diffuse-interface reference minima agree with the sharp-interface solution of the Young--Laplace
equation with gravity for the same volume and Young angle: for $Bo>0$ the drop height and base length differ from it by
at most $\NGYLdevH$ and $\NGYLdevB$ per cent (\hyperref[fig:gshapes]{figure~\ref*{fig:gshapes}}a,b). At $Bo=0$ the
differences are up to $\NGYLdevHzero$ and $\NGYLdevBzero$ per cent; they are the lattice effect on the contact angle
discussed in \hyperref[sec:verification]{Section~\ref*{sec:verification}}. The largest difference in height under gravity
occurs for the flattest drops ($\theta_Y=60^\circ$, a height of $\NGHsixtyOne$ cells at $Bo=1$), for which the
interface width is no longer small compared with the drop height. At $Bo=1$ the drop is $\NGFlatHSixty$ per cent lower and its base
$\NGFlatBSixty$ per cent wider than at $Bo=0$ for $\theta_Y=60^\circ$, and $\NGFlatHOneForty$ per cent lower and
$\NGFlatBOneForty$ per cent wider for $\theta_Y=140^\circ$. \dfeno reproduces this deformation with a median error of
$\NGOpHerr$ cells in the height and $\NGOpBerr$ cells in the base length (\hyperref[fig:gshapes]{figure~\ref*{fig:gshapes}}c).
\hyperref[tab:gravity]{Table~\ref*{tab:gravity}} lists the energy gaps: the FNO reaches a median of $\NGdEUniformFno$ on
uniform walls, $\NGdEStepFno$ on the step and $\NGdERandomFno$ on random patterns with random $Bo_n$. These gaps are
larger than without gravity ($\NdEmedUniFnoTwo$, $\NdEmedStepFnoTwo$ and $\NdEmedRandFnoTwo$), most on uniform walls,
because the same training budget now has to cover the additional parameter $Bo_n$. The TFNO is less accurate than the FNO
by a factor of four to nine, a larger factor than without gravity. A prediction takes $\NGTinfFno$\,ms (FNO), against
a median of $\NGTref$\,s for a reference solve.

\subsection{The step under gravity}
\begin{figure}
\centering
\includegraphics[width=\textwidth]{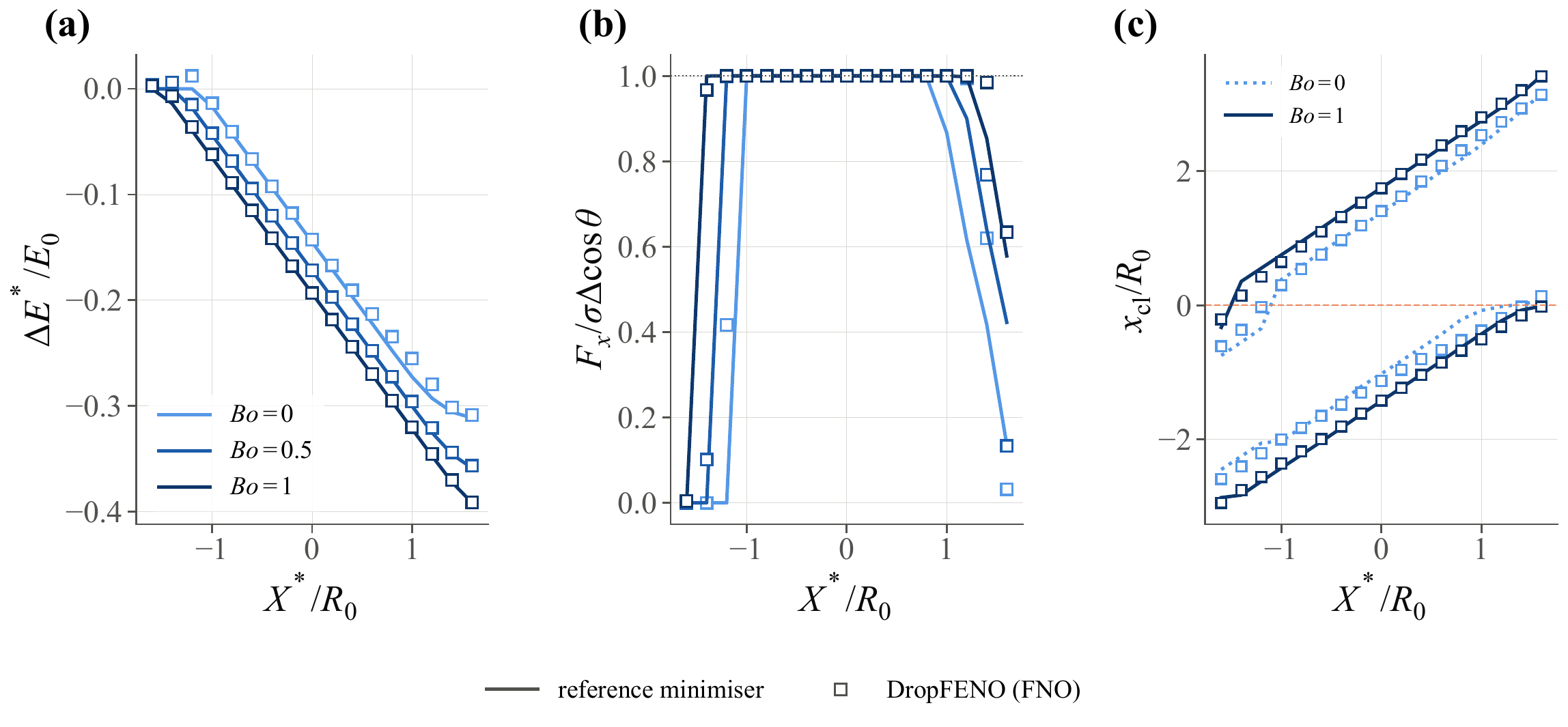}
\caption{A drop held across the $128.8^\circ\!\to\!79.7^\circ$ step on a horizontal wall at $Bo=0$, $0.5$ and $1$:
reference minimiser (lines) and \dfeno (FNO, open squares). (a) Energy landscape, including the gravitational energy.
(b) Virtual-work force; the plateau $F=\sigma\Delta\cos\theta$ does not depend on $Bo$. (c) Positions of the two contact
lines at $Bo=0$ (dotted) and $Bo=1$ (solid).}\label{fig:gstep}
\end{figure}

\hyperref[fig:gstep]{Figure~\ref*{fig:gstep}} repeats the step of \hyperref[sec:step]{Section~\ref*{sec:step}} at
$Bo=0$, $0.5$ and $1$ on a horizontal wall. The plateau of the force is unchanged by gravity, as the virtual-work argument
predicts: for the straddling states of the reference at $Bo=1$ it equals $\sigma\Delta\cos\theta$ to within $\NGPlatOne$.
Gravity changes where the plateau begins and ends, because a flatter drop has a longer base: the plateau extends from
$X^*\simeq\NGSnapZero R_0$ to $\NGPinZero R_0$ at $Bo=0$ and from $\NGSnapOne R_0$ to $\NGPinOne R_0$ at $Bo=1$, where
the first value is the position at which the advancing line reaches the step and the second the position at which the
receding line becomes pinned at the step and the force begins to fall. \dfeno follows these shifts. As without gravity, its largest
errors are in the pinning range: over the held positions at which the reference force lies between $5$ and $95$ per cent of
the plateau, the mean absolute error of $F_x/\sigma\Delta\cos\theta$ of the FNO is $\NGPinErrFno$.

\subsection{Rest position and sliding on an inclined wall}
\begin{figure}
\centering
\includegraphics[width=\textwidth]{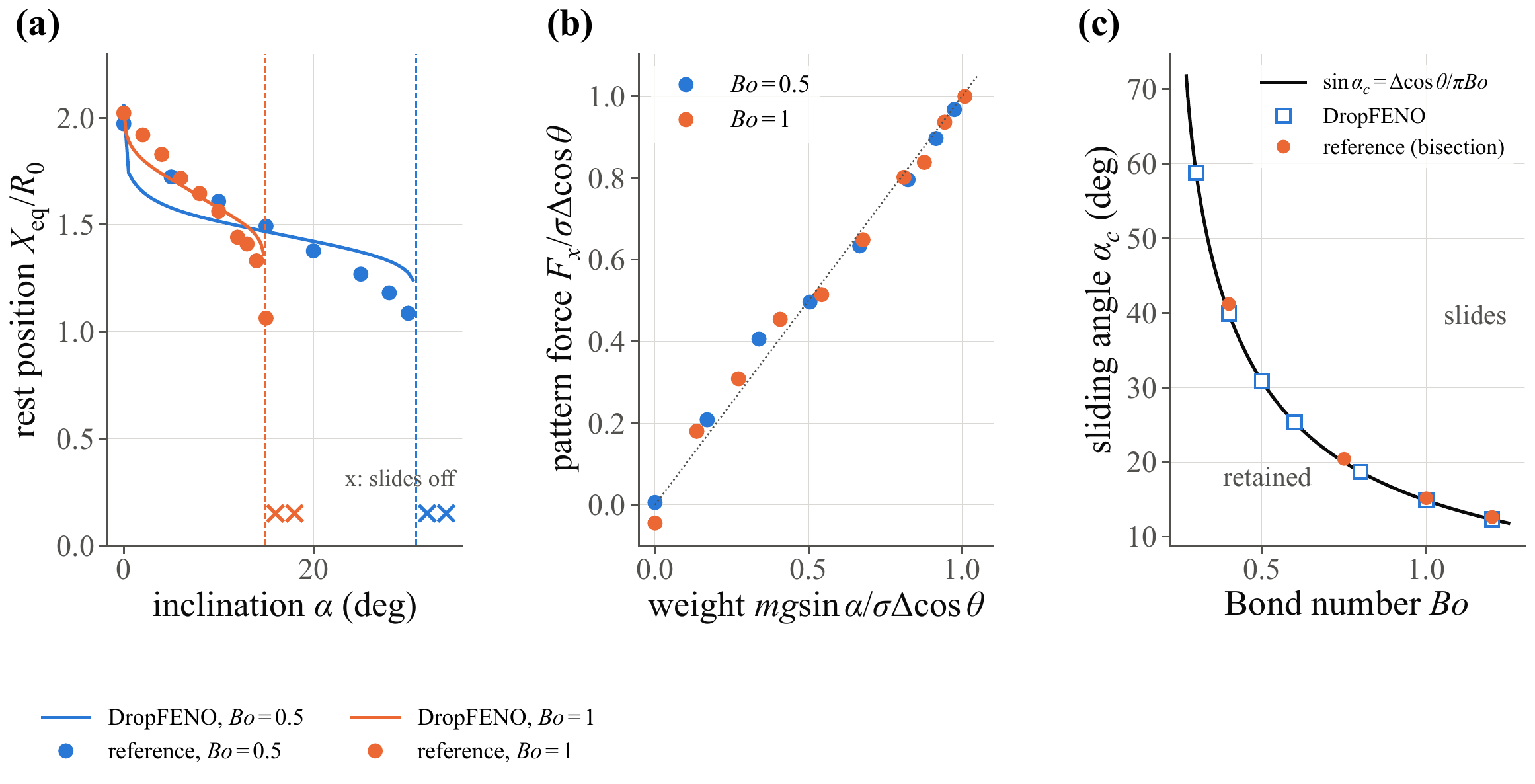}
\caption{Free drops on the inclined wall with the $128.8^\circ\!\to\!79.7^\circ$ step. (a) Rest position against the
inclination for $Bo=0.5$ and $1$: free drops of the reference minimiser (circles; crosses: the drop slides off) and
\dfeno from the force balance~(\ref{eq:balance}) on held drops (lines); dashed lines mark the threshold of
equation~(\ref{eq:alphac}). (b) Pattern force against weight for the retained reference drops; the dotted line is
equality. (c) Sliding angle against the Bond number: equation~(\ref{eq:alphac}) (line), \dfeno (squares) and reference
bisection (circles with the final bracket).}\label{fig:gincline}
\end{figure}

On the inclined wall the free drops of the reference come to rest with the downhill contact line held at the step, in the
pinning range of the force (\hyperref[fig:gincline]{figure~\ref*{fig:gincline}}a). As $\alpha$ increases the weight grows and
the drop rests closer to the step, until it slides off between $\alpha=\NGHeldMaxOne^\circ$ and $\NGSlidMinOne^\circ$ for
$Bo=1$ and between $\NGHeldMaxHalf^\circ$ and $\NGSlidMinHalf^\circ$ for $Bo=0.5$, close to the values
$\NGAlphaCOne^\circ$ and $\NGAlphaCHalf^\circ$ of equation~(\ref{eq:alphac}). For the $\NGNheld$ retained drops the pattern force computed from the wall trace balances the
weight to within $\NGForceBal\,\sigma\Delta\cos\theta$ (\hyperref[fig:gincline]{figure~\ref*{fig:gincline}}b); the small
residual reflects the finite tolerance of the free minimisation, in which a translation of the drop is a very soft
mode, and the discrete lattice. The reference thresholds obtained by bisection at $\NGCritN$ Bond numbers lie within
$\NGCritRefDev^\circ$ of equation~(\ref{eq:alphac}), all slightly above it, which is consistent with the same small
residual resistance of the free drops (\hyperref[fig:gincline]{figure~\ref*{fig:gincline}}c).

\dfeno predicts the same quantities from held drops only. For each inclination we evaluate the FNO force
$F_x(X^*;Bo\cos\alpha)$ on $241$ held positions, which takes a few seconds, and apply the force
balance~(\ref{eq:balance}) to a drop released at $X^*=R_0$. For the retained drops the predicted rest positions differ
from those of the free reference drops by $\NGXeqErr R_0$ on average (at most $\NGXeqErrMax R_0$); the one exception is
$\alpha=15^\circ$ at $Bo=1$, just above the threshold of equation~(\ref{eq:alphac}), at which the reference drop is
still retained and \dfeno predicts that it slides. This error is consistent with the error of the
operator in the pinning range, where the rest position is set. The sliding angles predicted by the FNO, from its maximum
force, lie within $\NGCritOpDev^\circ$ of equation~(\ref{eq:alphac}) for $0.3\le Bo\le1.2$. The retention threshold, which
depends only on the plateau of the force, is therefore predicted accurately, whereas the rest position, which depends on
the pinned contact line, inherits the soft-mode error of the operator identified in
\hyperref[sec:discussion]{Section~\ref*{sec:discussion}}.

\begin{table}
\centering
\caption{Accuracy of \dfeno with gravity on held-out problems: median energy gap $(E-E_{\rm ref})/E_0$ (capillary plus gravitational energy) per set, median intersection-over-union of the liquid regions on the random set, median error of the drop height over all sets (lattice units) and wall-clock time of one prediction. The median reference solve takes 5.8\,s.}\label{tab:gravity}
\small
\begin{tabular}{@{}lcccccc@{}}
\toprule
 & \multicolumn{3}{c}{median $(E-E_{\rm ref})/E_0$} & IoU & $|\delta H|$ & $t_{\rm op}$\\
\cmidrule(lr){2-4}
operator & uniform & step & random & random & (cells) & (ms)\\
\midrule
FNO & $2.7\times10^{-4}$ & $2.1\times10^{-3}$ & $9.0\times10^{-4}$ & 0.959 & 0.63 & 14\\
TFNO & $2.5\times10^{-3}$ & $9.4\times10^{-3}$ & $6.2\times10^{-3}$ & 0.927 & 1.04 & 11\\
\bottomrule
\end{tabular}
\end{table}

\section{Ablations: factorisation, prior, budget and resolution}\label{sec:ablation}

\begin{figure}
\centering
\includegraphics[width=\textwidth]{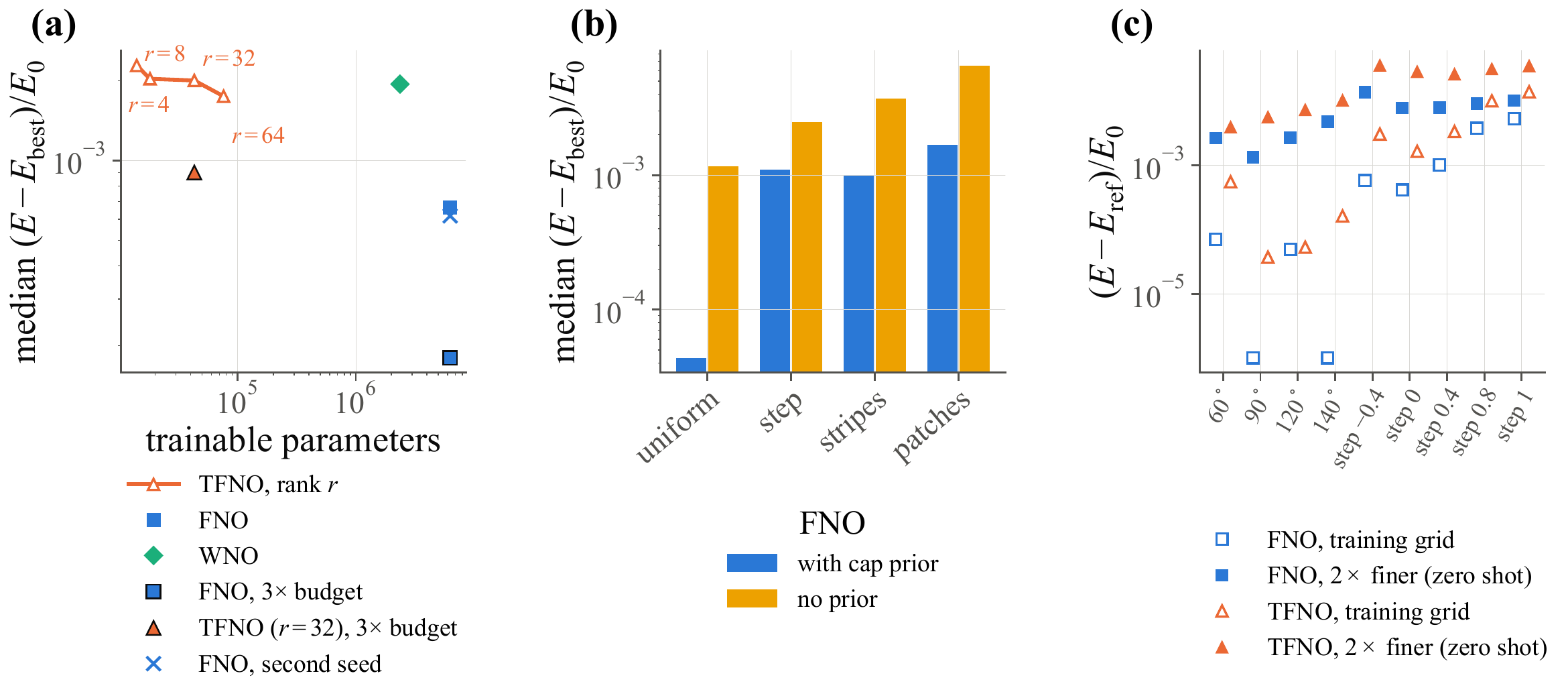}
\caption{Ablations in two dimensions. (a) Median energy gap on the random held-out set against the number of
trainable parameters: TFNO for CP ranks $r=4$--$64$, FNO, WNO, a second FNO seed and FNO and TFNO trained with three
times the budget. (b) FNO with and without the analytic cap prior, by pattern family. (c) Zero-shot transfer to a
grid refined by a factor of two (same physical problem, $\epsilon$ and $R_0$ doubled in lattice units).}\label{fig:ablation}
\end{figure}

\paragraph{Rank of the factorised kernel.} \hyperref[fig:ablation]{Figure~\ref*{fig:ablation}}(a) places every two-dimensional operator in the
plane of parameter count and median energy gap on the random held-out set. The CP rank of the TFNO matters little:
rank $4$ ($\NAblParRankFour$\,k parameters) gives $\NAblRankFour$, rank $8$ gives $\NAblRankEight$, rank $32$ (the
default, $\NAblParRankThirtyTwo$\,k) $\NAblRankThirtyTwo$ and rank $64$ ($\NAblParRankSixtyFour$\,k)
$\NAblRankSixtyFour$, against $\NAblFnoBase$ for the dense FNO with $\NParamsFnoTwo$\,M parameters. A sixteen-fold
increase of the rank thus improves the gap by less than a factor of $1.4$, and the remaining factor of about $2.5$ to
the dense FNO is not closed within the range of ranks tested: the dense spectral weights of the FNO are
compressible to below one per cent of their size at a moderate, not a negligible, loss of accuracy.

\paragraph{Seed and training budget.} A second FNO trained from a different initialisation and with a different
stream of training patterns reaches $\NAblSeedOne$ (first seed $\NAblFnoBase$), and its median gaps on the individual
pattern families differ from those of the first by at most about $20\%$, except on the uniform members of the set, where the
gaps are smallest and differ by $40\%$. With this single pair of seeds as an
indication of the spread, the differences between architectures reported in \hyperref[sec:accuracy]{Section~\ref*{sec:accuracy}}, which are factors
of two or more, are significant, whereas differences of a few tens of per cent are not. Tripling the number of
training iterations, and with it the number of distinct training patterns, reduces the median gap of the FNO from
$\NAblFnoBase$ to $\NAblFnoLong$, a factor of nearly four, and that of the TFNO from $\NAblRankThirtyTwo$ to
$\NAblTfnoLong$ (\hyperref[fig:ablation]{figure~\ref*{fig:ablation}}a). The operators are therefore far from saturated at the budget used in the
rest of the paper, whose accuracies are conservative; in the data-free setting additional training instances cost
only the compute to draw and evaluate them. Even with three times the budget, however, the TFNO does not reach the
FNO trained with the base budget, so that the advantage of the dense kernel is not an artefact of under-training.

\paragraph{The analytic prior.} Without the cap prior of equation~(\ref{eq:prior}) (the network then outputs
$\phi=\tanh\hat u$ directly, with the same inputs other than $\phi_0$), the FNO still learns the constrained
equilibrium, but its median gap on the random set rises from $\NAblFnoBase$ to $\NAblNoPrior$, and on every pattern
family it is two to four times larger, on uniform walls more than twenty times (\hyperref[fig:ablation]{figure~\ref*{fig:ablation}}b). The prior
is not required, but it is worth more than tripling the training budget: it supplies the drop size and the mean
shape, which the network otherwise has to learn from the volume and centroid penalties alone, and leaves the network to
learn only the response of the drop to the pattern.

\paragraph{Resolution.} Neural operators are often credited with discretisation invariance. For the present problem the
claim fails, for a physical reason. Refining the grid by a factor of two at fixed physical problem (drop radius and
interface width both doubled in lattice units, step width scaled accordingly) and evaluating the trained operators
without retraining increases the median energy gap on uniform walls from $\NSrUniFnoOne$ to $\NSrUniFnoTwo$ for the FNO
and on the step problems from $\NSrStepFnoOne$ to $\NSrStepFnoTwo$ (\hyperref[fig:ablation]{figure~\ref*{fig:ablation}}c). The discrete energy
depends on the interface width in lattice units, which sets both the contact angle actually realised by the Cahn
wall (\hyperref[sec:verification]{Section~\ref*{sec:verification}}) and the cost of the interface, and an operator trained at $\epsilon=1$ has learnt the
minimiser of that discrete problem, not of its continuum limit. Retraining at the finer resolution costs no data,
only compute.

\begin{figure}
\centering
\includegraphics[width=\textwidth]{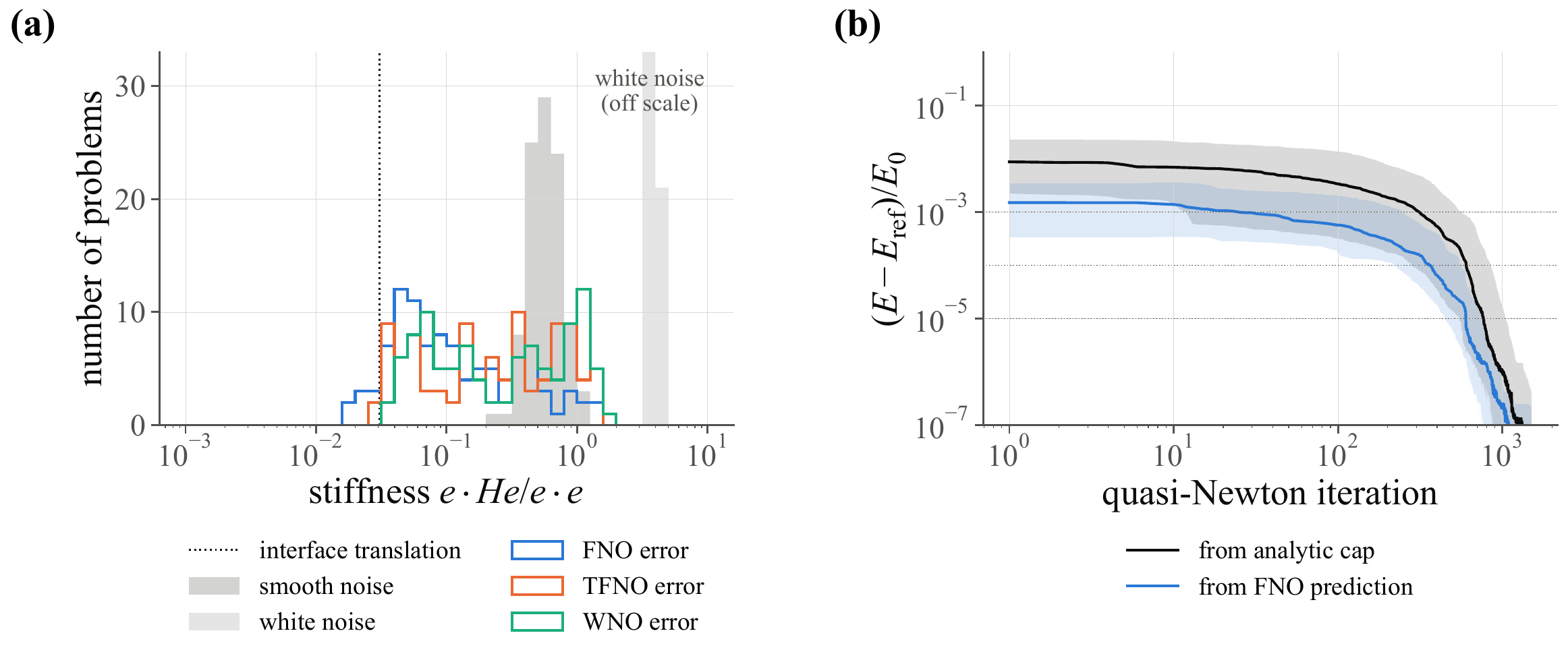}
\caption{Where the error of an energy-trained operator lives. (a) Stiffness $e\cdot He/e\cdot e$ of the operator error
$e=\phi_{\bm w}-\phi^*$ on the $100$ random held-out problems, with $H$ the Hessian of the discrete energy at the
reference minimiser, against interface-localised smooth and white noise and the translation mode of the interface
(dotted). (b) Energy gap along a quasi-Newton solve started from the analytic cap and from the FNO prediction
(median and quartiles over $30$ random problems); dotted lines mark the tolerances of \hyperref[sec:discussion]{Section~\ref*{sec:discussion}}.}\label{fig:diagnostics}
\end{figure}

\section{Discussion}\label{sec:discussion}

\paragraph{Where the error of an energy-trained operator lives.} An operator trained on the energy is penalised for
an error $e=\phi_{\bm w}-\phi^*$ in proportion to $\tfrac12e\cdot He$, with $H$ the Hessian of the energy at the
minimiser, and not in proportion to $|e|^2$. It is therefore free to leave error in the soft directions of $H$
and is driven hard to remove it from the stiff ones. The Rayleigh quotient $e\cdot He/e\cdot e$ of the actual FNO
error on the random patterns has a median of $\NRayOpFno$ (TFNO $\NRayOpTfno$, WNO $\NRayOpWno$), against $\NRaySmoothFno$ for smooth random perturbations
of the same interfacial support and $\NRayWhiteFno$ for white ones (\hyperref[fig:diagnostics]{figure~\ref*{fig:diagnostics}}a); the translation mode of the interface, the
softest physically meaningful direction, has $\NRayTranslFno$. The residual error is thus a small displacement or
tilt of the interface and of the contact line, not a distortion of the interface profile. Three observations of
\hyperref[sec:accuracy]{Sections~\ref*{sec:accuracy}}--\ref{sec:physics} follow from this. First, the energy gap is small while the field error is
not, because the second is dominated by modes that cost almost no energy. Second, the quantities that are
\emph{first} order in these soft modes (the position of a contact line on an edge, and with it the force in the
pinning regime) are the least accurate outputs, whereas the force on a straddling drop, which depends only on
which side of the step each line lies, is reproduced to a few parts per million. Third, the prediction accelerates a classical solver only for a coarse tolerance. In a single stiff-penalty
quasi-Newton pass with multipliers fitted to the initial field, the FNO prediction reaches a gap of $10^{-3}E_0$ after a
median of $\NWsOpThree$ iterations, against $\NWsCapThree$ from the analytic cap (median ratio $\NWsGainThree$ over
$\NWsN$ random problems); but to reach $10^{-4}E_0$ and $10^{-5}E_0$ it still needs $\NWsOpFour$ and $\NWsOpFive$
iterations, a gain of only $\NWsGainFour$ and $\NWsGainFive$, because the remaining error lies precisely in the
ill-conditioned directions that limit the convergence of the solver (\hyperref[fig:diagnostics]{figure~\ref*{fig:diagnostics}}b). The operator and the solver are, in this sense, complementary
rather than redundant.

\paragraph{The virtual-work test as a verification principle.} Without data, the usual measures of a surrogate
(agreement with a held-out solution) presuppose the classical solver that the surrogate is meant to replace.
The virtual-work identity~(\ref{eq:hf}) supplies a test that does not: an operator that has learnt the constrained
minimiser must return a force that equals the slope of its own energy landscape and, on a straddled step, the
contact-line value $\sigma\Delta\cos\theta\,L$. Both checks require only forward passes of the operator. The same
construction applies to any operator trained on a variational principle with a parametric external field
(the Hellmann--Feynman force on a nucleus in electronic-structure surrogates, the configurational force on a defect
in elasticity, the force on an inclusion in a membrane), and it tests the physics of a prediction rather than its similarity to a reference field.

\paragraph{Constraints in physics-informed losses.} The Gibbs--Thomson artefact of \hyperref[sec:gt]{Section~\ref*{sec:gt}} is not specific to
neural operators. It affects any diffuse-interface calculation that imposes the volume or the centre of mass through
the linear liquid fraction (a common choice in phase-field energy minimisation and in physics-informed losses for
two-phase problems), and its size grows with the ratio of the box to the drop volume and with the interface width relative to the drop: the
vapour is shifted by $\delta\simeq\epsilon\sigma\kappa/4c$, and each unit volume of vapour holds a liquid volume $\delta/2$. In our three-dimensional box it holds a third of the
liquid in the vapour and evaporates a drop at $140^\circ$. An indicator $h$ with $h'(\pm1)=0$ adds no cost and
removes it exactly; we recommend it for any such loss.

\paragraph{Limitations.} The diffuse interface is three cells wide and the drops are $16$ (two dimensions) and $12$
(three dimensions) cells in radius. The operators solve the discrete problem at this resolution, including its
lattice-induced departures from Young's law, and do not transfer zero-shot to a finer grid
(\hyperref[sec:ablation]{Section~\ref*{sec:ablation}}). Gravity is treated in two dimensions only (\hyperref[sec:gravity]{Section~\ref*{sec:gravity}}); in three dimensions the
retention force is $\sigma\Delta\cos\theta$ times the chord, which itself depends on the Bond number. The problem is
static: the operators predict equilibria and the landscape
$\mathcal{E}^*(\bm X^*)$, not the dissipative dynamics along it, and the hysteresis of a moving contact line enters
only through the multiplicity of metastable states. Which of these states an operator returns is decided by its
training and is not controlled; on the structured walls a few predictions lie in the basin of a neighbouring state.
Finally, the pattern distribution is generic but finite; patterns far outside it, such as stripes much narrower than
the interface, are outside the scope of the present operators.

\section{Conclusions}\label{sec:conclusions}

\dfeno, a set of neural operators trained on the free energy alone without any solution field, learns the capillary
equilibrium of drops on chemically patterned walls in two and three dimensions to within a small fraction of the
free-drop energy, at a cost two orders of magnitude below that of a per-instance minimisation, and almost always in the
basin of the reference state.
The constrained-minimisation structure of the problem supplies the means to verify them without data. The
virtual-work theorem makes the capillary force an explicit functional of the equilibrium field; on a straddled step it
reduces identically to the contact-line law $F=\sigma\Delta\cos\theta\,L$, which the reference minimiser satisfies to
better than one part in a thousand in two dimensions and to about one per cent with the geometric chord in three, and which the
operators reproduce as an emergent property of predictions made independently at each position. The comparison
identifies the limits of energy-based training: the error lives in the soft modes of the energy, so that
contact-line positions on edges (the pinned receding line at a step, the edge-pinned lines that set the apparent
angle on stripes) are the least accurate outputs, while the energy and the force on a straddling drop are the most
accurate. For these outputs a short refinement by a classical solver, started from the prediction, recovers the
pinned state in nearly all cases. These accuracies are conservative: tripling the training budget, which without data costs only compute,
reduces the median gap of the FNO by a further factor of nearly four. Among the architectures the dense FNO is the most accurate, the CP-factorised TFNO trades a factor of three
in accuracy for a factor of $150$ in size, and the local wavelet kernel of the WNO brings no advantage. With gravity,
which leaves the virtual-work force unchanged, the same force balance gives the rest position of a drop on an inclined
patterned wall and the inclination at which it slides off; the threshold depends only on the plateau of the force and
is predicted accurately, whereas the rest position depends on the pinned contact line. Finally, the
choice of the liquid indicator in the constraints matters: the linear liquid fraction creates a Gibbs--Thomson
supersaturation of the vapour that can hold a third of the liquid and biases forces inferred from energy landscapes,
and an indicator flat in the bulk phases removes it. Both the virtual-work test and the constraint remedy carry over to
any operator or physics-informed network trained on a variational principle with constraints.

\section*{Supplementary Information}
The Supplementary Information contains a notation table, the configuration and cost of every training run, per-run,
per-family and per-angle accuracy tables, refinement statistics, training histories and schedules, constraint
residuals, virtual-work forces of every random problem, galleries of the TFNO and WNO predictions, schematics of the
wavelet decomposition and of the input layout, and the pattern sampler and stiffness diagnostic as algorithms.

\section*{Data and code availability}
The \dfeno source code (the free energy and constraints, the FNO, TFNO and WNO operators, the data-free training, the
reference minimiser, and the scripts that reproduce every figure and table) is available at
\url{https://github.com/Ganesh1591/DropFENO} and will be made public upon publication. No training data exist: the
operators are trained on the free energy alone.

\section*{Author contributions (CRediT)}
\textbf{G.S. Meshram}: conceptualisation, methodology, software, validation, formal analysis, investigation,
visualisation, writing (original draft). \textbf{P.P. Chakrabarti}: supervision, methodology,
writing (review and editing). \textbf{S. Chakraborty}: supervision, conceptualisation, writing (review and
editing).

\section*{Acknowledgements}
The authors acknowledge the Computational Facility of the Indian Institute of Technology Kharagpur for providing the
computing resources used in this work.

\section*{Declaration of interests}
The authors report no conflict of interest.


\end{document}


\maketitle

\noindent This document supplements the main paper on \dfeno with the notation, the architecture and training details of every
run, additional accuracy statistics, training histories, constraint residuals, the virtual-work forces of the random
held-out problems, galleries of the TFNO and WNO predictions, and two auxiliary algorithms. No additional training or
solves were performed for it: every figure and table is produced from the result files of the main study by
the script make\_si.py and the gallery scripts. Figures, tables and algorithms are numbered with the prefix S;
references without it point to the main paper.

\tableofcontents

\section{Notation}\label{sec:notation}
Table~\ref{tab:notation} lists the symbols used in the main paper and in this document.
\begin{table}[htbp]
\centering
\caption{Notation. Lattice units throughout: unit cell spacing, $\epsilon=1$, $\sigma=1$.}\label{tab:notation}
\footnotesize
\begin{tabular}{@{}ll@{\hspace{2em}}ll@{}}
\toprule
symbol & meaning & symbol & meaning\\
\midrule
$\phi$ & phase field, $+1$ liquid, $-1$ vapour & $\mathcal G$, $\mathcal G_{\bm w}$ & solution operator, its approximation\\
$\theta(x,z)$ & local Young angle of the wall & $\bm w$ & trainable weights\\
$\epsilon$, $\sigma$ & interface width, tension & $a$ & input channels\\
$W$, $g$, $h$ & double well, wall and indicator functions & $\phi_0$, $u_0$ & analytic cap prior, $\operatorname{artanh}\phi_0$\\
$\mathcal E$, $E$ & continuous, discrete free energy & $\hat u$ & network output (correction)\\
$\mathcal V$, $V_0$ & liquid volume, its target & $P$, $Q$ & lifting and projection\\
$\bm X$, $\bm X^*$ & drop centroid, holding position & $\mathcal K_l$, $W_l$ & global kernel, pointwise map of block $l$\\
$R_0$ & radius of the free drop & $d_v$, $L$ & latent width, number of blocks\\
$E_0$ & free-drop energy ($2\upi R_0$, $4\upi R_0^2$) & $k_{\max}$ & retained Fourier modes\\
$\lambda_V$, $\lambda_x$ & multipliers (pressure, holding force) & $r$ & CP rank of the TFNO\\
$F_x$ & virtual-work force & $J$ & Haar levels of the WNO\\
$L$ (chord) & wetted chord on the step line & $\mu$ & penalty weight\\
$\theta_{\rm app}$ & apparent angle (circle fit) & $\delta E$ & $(E-E_{\rm best})/E_0$\\
$\mathcal P$ & pattern distribution & IoU & intersection-over-union of $\phi>0$\\
\bottomrule
\end{tabular}
\end{table}

\section{Architecture details}\label{sec:archdetails}
\begin{figure}[htbp]
\centering
\fitbox{
\begin{tikzpicture}[font=\small]
\def\W{6.4}\def\H{3.6}
\fill[cGrid!35, draw=cInkTwo] (0,0) rectangle (\W,\H);
\fill[cGrid!35, draw=cInkTwo] (0,\H/2) rectangle (\W/2,\H);
\fill[cGrid!35, draw=cInkTwo] (0,3*\H/4) rectangle (\W/4,\H);
\fill[cGreen!35, draw=cGreen!60!black] (0,7*\H/8) rectangle (\W/8,\H);
\fill[cGreen!20, draw=cGreen!60!black] (\W/8,7*\H/8) rectangle (\W/4,\H);
\fill[cGreen!20, draw=cGreen!60!black] (0,3*\H/4) rectangle (\W/8,7*\H/8);
\fill[cGreen!20, draw=cGreen!60!black] (\W/8,3*\H/4) rectangle (\W/4,7*\H/8);
\node[font=\scriptsize] at (\W/16,15*\H/16) {$a_3$};
\node[font=\scriptsize] at (3*\W/16,15*\H/16) {$d_3$};
\node[font=\scriptsize] at (\W/16,13*\H/16) {$d_3$};
\node[font=\scriptsize] at (3*\W/16,13*\H/16) {$d_3$};
\node[font=\footnotesize] at (3*\W/8,7*\H/8) {$d_2$};
\node[font=\footnotesize] at (\W/8,5*\H/8) {$d_2$};
\node[font=\footnotesize] at (3*\W/8,5*\H/8) {$d_2$};
\node[font=\small] at (3*\W/4,3*\H/4) {$d_1$};
\node[font=\small] at (\W/4,\H/4) {$d_1$};
\node[font=\small] at (3*\W/4,\H/4) {$d_1$};
\draw[decorate, decoration={brace, amplitude=4pt}] (0,\H+0.1) -- (\W,\H+0.1) node[midway, above=4pt, font=\footnotesize] {$N_x=128$};
\draw[decorate, decoration={brace, amplitude=4pt, mirror}] (-0.1,0) -- (-0.1,\H) node[midway, left=5pt, font=\footnotesize, align=center] {$N_y+8$\\$=72$};
\begin{scope}[xshift=6.9cm, yshift=3.2cm]
  \fill[cGreen!35, draw=cGreen!60!black] (0,0) rectangle ++(0.4,0.3); \node[lbl, anchor=west, align=left] at (0.5,0.15) {approximation $a_3$ ($16\times9$): weights $W_a$,\\ $d_v\times d_v$ at each location};
  \fill[cGreen!20, draw=cGreen!60!black] (0,-0.9) rectangle ++(0.4,0.3); \node[lbl, anchor=west, align=left] at (0.5,-0.75) {three detail bands $d_3$: weights $W_d$,\\ $3\times d_v\times d_v$ at each location};
  \fill[cGrid!35, draw=cInkTwo] (0,-1.8) rectangle ++(0.4,0.3); \node[lbl, anchor=west, align=left] at (0.5,-1.65) {finer details $d_1,d_2$: set to zero\\ before the inverse transform};
  \node[lbl, anchor=north west, align=left] at (0,-2.3) {one level: $a_{j+1}=\tfrac12(v_{00}+v_{01}+v_{10}+v_{11})$\\ and three differences $d_{j+1}$ per $2\times2$ block;\\
    orthonormal, exactly invertible; in 3D,\\ $2\times2\times2$ blocks and seven detail bands};
\end{scope}
\end{tikzpicture}}
\caption{Haar bands of the padded two-dimensional latent field after $J=3$ levels. The WNO kernel mixes the channels of
the coarsest approximation $a_3$ and of the three coarsest detail bands $d_3$ with weights that depend on the location
on the $16\times9$ coarse grid; the finer details are set to zero, so that the kernel is a smooth, local, multi-scale
operator and local detail is carried by the pointwise path $W_l$.}\label{fig:si_haar}
\end{figure}
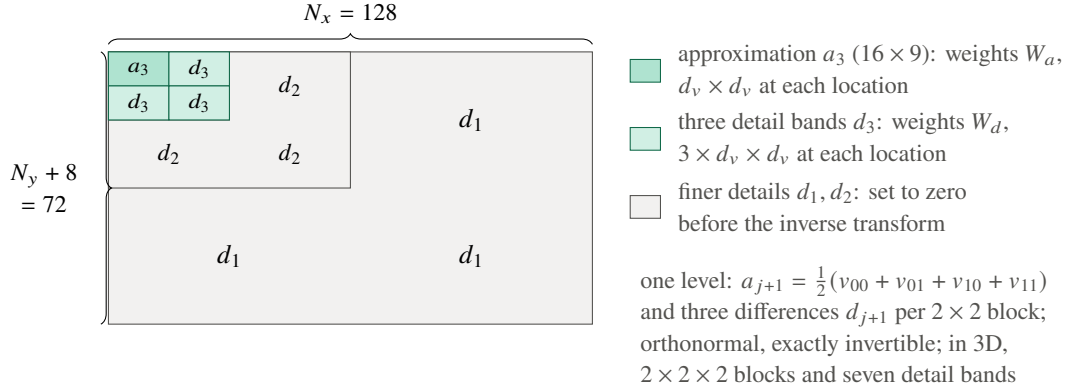

Table~\ref{tab:hyper} lists the grids and hyperparameters of the two- and three-dimensional configurations;
figure~\ref{fig:si_inputs} shows the input layout and the padding, and figure~\ref{fig:si_haar} the wavelet bands used by the
WNO kernel (complementing \mfig{fig:kernels}). Algorithm~\ref{alg:sampler} lists the pattern sampler of
\mfig{fig:patterns}.

\begin{table}[htbp]
\centering
\caption{Grids and hyperparameters of the operators. Parameter counts are real numbers (complex weights count twice).}\label{tab:hyper}
\small
\begin{tabular}{@{}lcc@{}}
\toprule
 & two dimensions & three dimensions\\
\midrule
grid $N_x\times N_y\,(\times N_z)$ & $128\times64$ & $64\times32\times64$\\
drop radius $R_0$, interface width $\epsilon$ & $16$, $1$ & $12$, $1$\\
input channels & 4 & 5\\
latent width $d_v$, blocks $L$ & $32$, $4$ & $20$, $4$\\
padding of $y$ inside the operator & 8 cells & 8 cells\\
FNO/TFNO retained modes $k_{\max}$ & $24\times16$ & $12\times8\times12$\\
TFNO CP rank $r$ & 32 & 48\\
WNO Haar levels $J$, coarsest grid & $3$, $16\times9$ & $3$, $8\times5\times8$\\
lifting $P$ & \multicolumn{2}{c}{$1{\times}1$ conv.\ to $d_v$, GELU, $1{\times}1$ conv.}\\
projection $Q$ & \multicolumn{2}{c}{$1{\times}1$ conv.\ to $4d_v$, GELU, $1{\times}1$ conv.\ to 1}\\
parameters FNO / TFNO / WNO & 6.30\,M / 42.8\,k / 2.37\,M & 14.75\,M / 39.7\,k / 4.10\,M\\
\bottomrule
\end{tabular}
\end{table}

\begin{figure}[htbp]
\centering
\fitbox{
\begin{tikzpicture}[font=\small]
\foreach \i/\c/\t in {0/cOrange/{$\cos\theta(x)$, constant in $y$},1/cGrid/{$y/N_y$},2/cGrid/{$(x-X^*)/R_0$},3/cGreen/{$\phi_0$, analytic cap}} {
  \begin{scope}[xshift=\i*0.35cm, yshift=-\i*0.35cm]
    \fill[\c!30, draw=cInkTwo, line width=0.6pt] (0,0) rectangle (4.0,2.0);
  \end{scope}
  \node[font=\scriptsize, anchor=north west, inner sep=1pt] at (\i*0.35+0.05,2.0-\i*0.35) {\t};
}
\node[ttl, anchor=south west] at (0,2.4) {(a) input channels (2D)};
\begin{scope}[xshift=7.4cm, yshift=-1.05cm]
  \fill[cBlue!15, draw=cInkTwo, line width=0.6pt] (0,0) rectangle (4.0,2.0);
  \fill[pattern=north east lines, pattern color=cInkTwo] (0,2.0) rectangle (4.0,2.5);
  \draw[cInkTwo, line width=0.6pt] (0,2.0) rectangle (4.0,2.5);
  \node[font=\footnotesize] at (2.0,1.0) {latent $v$, $d_v\times128\times64$};
  \node[font=\footnotesize, fill=white, inner sep=1pt] at (2.0,2.25) {zero padding, 8 cells};
  \fill[cPhob] (0,-0.18) rectangle (4.0,0);
  \node[lbl, anchor=north] at (2.0,-0.2) {wall below row $0$ (not padded)};
  \node[ttl, anchor=south west] at (0,3.45) {(b) padding inside the operator};
  \node[lbl, anchor=west, align=left] at (4.2,1.2) {$y$ is not periodic: pad at the top\\ before $\mathcal K_l$, crop after the\\ last block};
\end{scope}
\end{tikzpicture}}
\caption{(a) Input channels of the two-dimensional operator: the wall field extended uniformly in $y$, the normalised
height, the horizontal distance from the holding position and the analytic prior. (b) The latent field is zero-padded by
eight cells at the top before the global kernels, because the $y$ direction is not periodic, and cropped after the last
block; the wall lies below the first row.}\label{fig:si_inputs}
\end{figure}

\begin{algorithm}[htbp]
\caption{Pattern sampler $\mathcal P$ (\mfig{fig:patterns})}\label{alg:sampler}
\begin{algorithmic}[1]
\Require grid $N_x$ ($\times N_z$), drop radius $R_0$, random generator
\State family $\gets$ uniform choice among \{uniform, step, stripes $\parallel z$, patches\} (2D) or these with stripes $\parallel x$ and checkerboard (3D)
\State $\theta_1,\theta_2\sim U[35^\circ,150^\circ]$; $c_1=\cos\theta_1$, $c_2=\cos\theta_2$
\If{uniform} $f\gets0$
\ElsIf{step} $f\gets\tfrac12\big[1+\tanh\big((x-x_s)/0.75\big)\big]$
\ElsIf{stripes} draw period $P\in[5,32]$, fraction $\in[0.2,0.8]$, phase; $f\gets$ smoothed periodic indicator
\ElsIf{checkerboard} draw $P\in[8,32]$ and phases; $f\gets\tfrac12\big[1+\operatorname{sgn}\sin\tfrac{2\upi x}{P}\,\operatorname{sgn}\sin\tfrac{2\upi z}{P}\big]$
\Else\ \ $f\gets$ smoothed step of Gaussian-filtered white noise (filter scale $k\in[3,10]$) \Comment{patches}
\EndIf
\State $\cos\theta\gets c_1+(c_2-c_1)f$
\State $X^*,Z^*\sim U[-R_0/2,R_0/2]$; for a step, shift the pattern so that the step lies within $\pm1.2R_0$ of $X^*$
\State \Return $(\cos\theta,\bm X^*)$
\end{algorithmic}
\end{algorithm}

\section{Training runs}\label{sec:runs}
\begin{figure}[htbp]
\centering
\includegraphics[width=\textwidth]{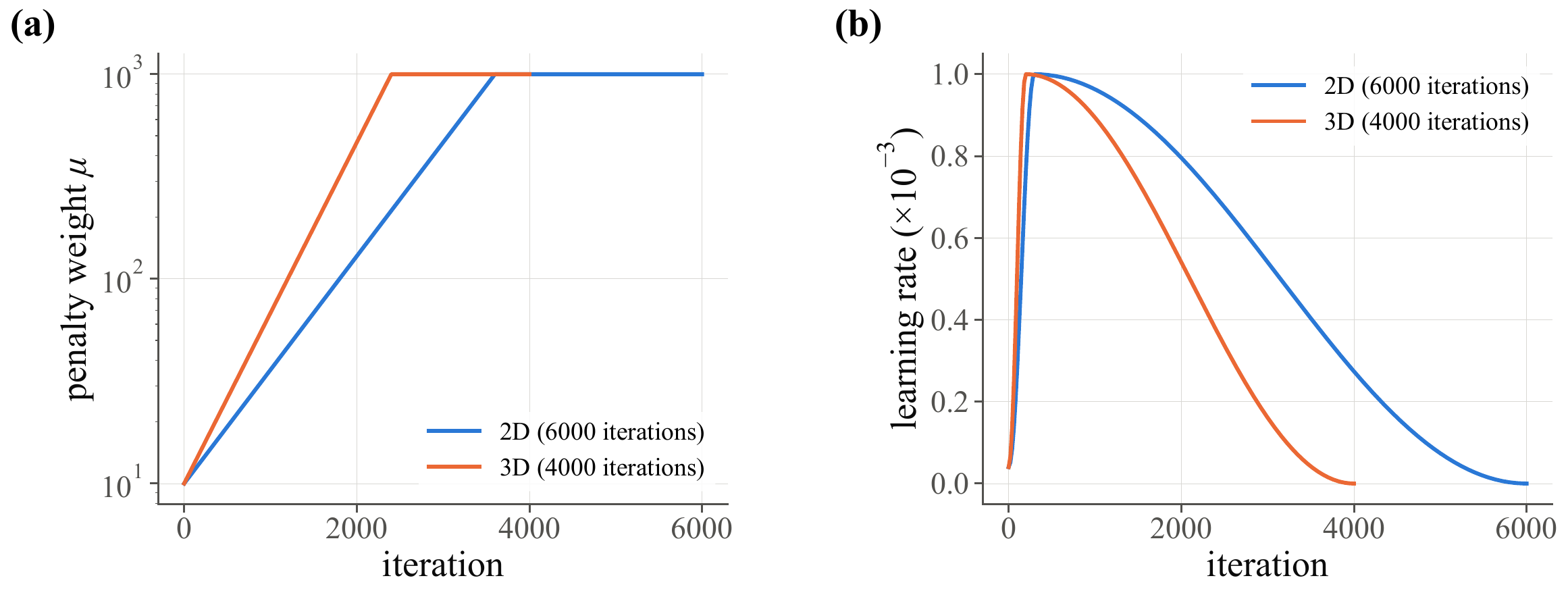}
\caption{Training schedules (FNO runs; identical for all operators). (a) Penalty weight $\mu$. (b) Learning rate of the
one-cycle schedule (5\% warm-up, cosine decay).}\label{fig:si_sched}
\end{figure}

\begin{figure}[htbp]
\centering
\includegraphics[width=\textwidth]{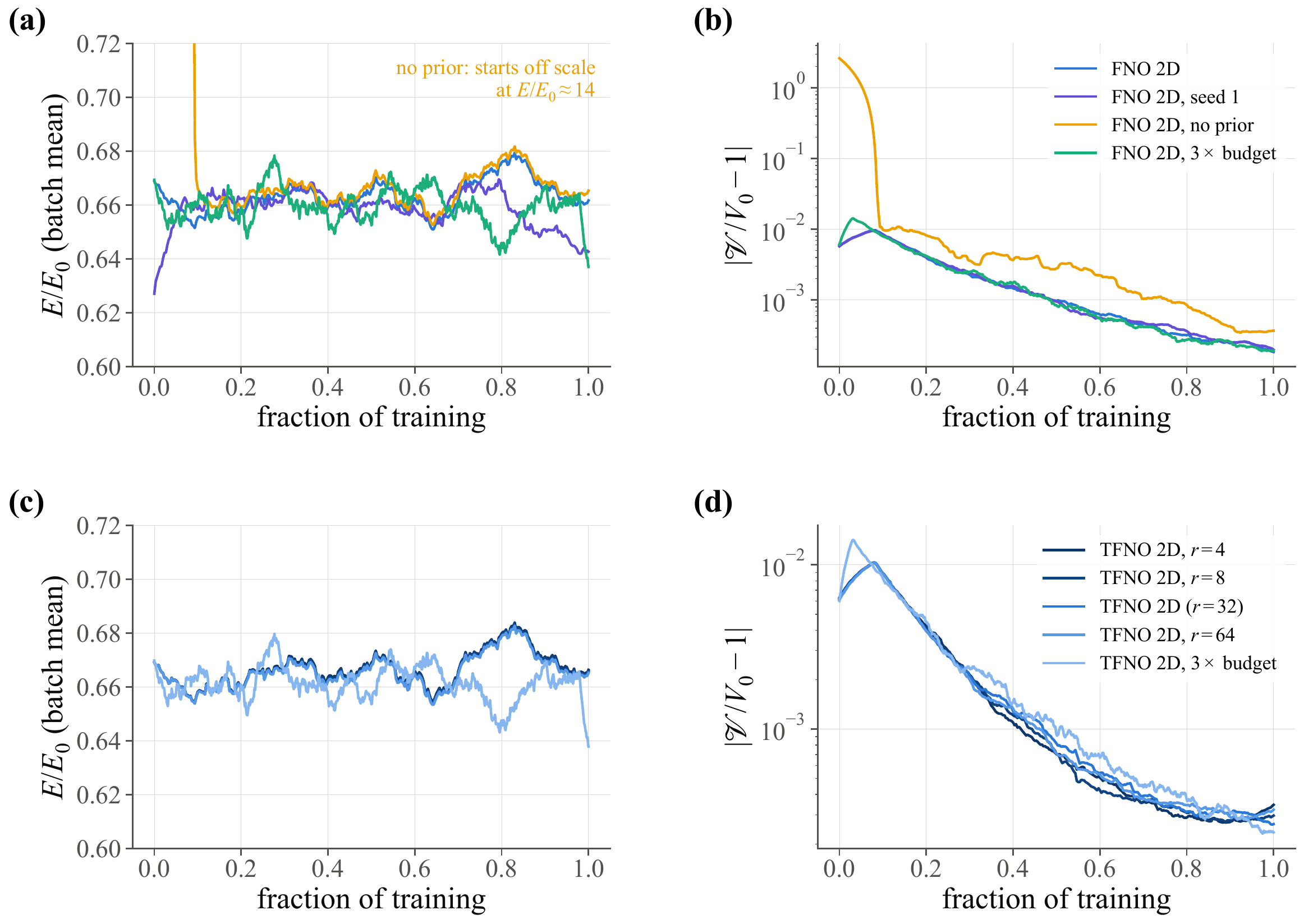}
\caption{Training histories of the two-dimensional ablation runs against the fraction of training completed (running
mean over 40 logged iterations). (a,b) FNO: main run, second seed, no prior, three times the budget. (c,d) TFNO: CP ranks
$4$--$64$ and three times the budget. (a,c) Batch-mean energy; (b,d) volume residual.}\label{fig:si_hist}
\end{figure}

Table~\ref{tab:si_runs} lists every training run of the study, with its cost. All runs use the same schedules
(figure~\ref{fig:si_sched}): a geometric continuation of the penalty weight from $10$ to $10^3$ over the first $60\%$ of
the iterations and a one-cycle learning rate. Figure~\ref{fig:si_hist} compares the histories of the ablation runs of
\msec{sec:ablation}. Without the prior the network starts from $\phi\approx0$ everywhere, with an energy fourteen times
that of the free drop and a volume error of order one, and recovers a drop within the first tenth of training; its volume
residual stays about twice as large as that of the runs with the prior until the end. The CP rank has no visible effect on
the histories, consistent with panel (a) of \mfig{fig:ablation}.

\begin{table}[htbp]
\centering
\caption{Configuration and cost of every training run. Instances: number of distinct random problems seen (iterations $\times$ batch). All runs use a peak learning rate of $10^{-3}$ and the penalty continuation $\mu:10\to10^3$. Params in thousands; training wall-clock time on the CPU threads listed.}\label{tab:si_runs}
\footnotesize
\begin{tabular}{@{}lrrrcccrr@{}}
\toprule
run & iterations & batch & instances & seed & prior & threads & params (k) & time (h)\\
\midrule
FNO 2D & 6000 & 8 & 48{,}000 & 0 & yes & 4 & 6{,}301.2 & 0.90\\
TFNO 2D ($r=32$) & 6000 & 8 & 48{,}000 & 0 & yes & 4 & 42.8 & 0.67\\
WNO 2D & 6000 & 8 & 48{,}000 & 0 & yes & 4 & 2{,}369.1 & 0.86\\
FNO 2D{,} seed 1 & 6000 & 8 & 48{,}000 & 1 & yes & 4 & 6{,}301.2 & 0.84\\
FNO 2D{,} no prior & 6000 & 8 & 48{,}000 & 0 & no & 4 & 6{,}301.2 & 1.02\\
FNO 2D{,} $3\times$ budget & 18000 & 8 & 144{,}000 & 0 & yes & 2 & 6{,}301.2 & 4.08\\
TFNO 2D{,} $3\times$ budget & 18000 & 8 & 144{,}000 & 0 & yes & 2 & 42.8 & 3.53\\
TFNO 2D{,} $r=4$ & 6000 & 8 & 48{,}000 & 0 & yes & 4 & 13.9 & 0.67\\
TFNO 2D{,} $r=8$ & 6000 & 8 & 48{,}000 & 0 & yes & 4 & 18.0 & 0.98\\
TFNO 2D{,} $r=64$ & 6000 & 8 & 48{,}000 & 0 & yes & 4 & 75.8 & 0.98\\
FNO 3D & 4000 & 2 & 8{,}000 & 0 & yes & 4 & 14{,}749.6 & 1.92\\
TFNO 3D & 4000 & 2 & 8{,}000 & 0 & yes & 4 & 39.7 & 1.20\\
WNO 3D & 4000 & 2 & 8{,}000 & 0 & yes & 4 & 4{,}100.0 & 2.26\\
\bottomrule
\end{tabular}
\end{table}

\section{Additional accuracy statistics}\label{sec:acc}
\begin{figure}[htbp]
\centering
\includegraphics[width=\textwidth]{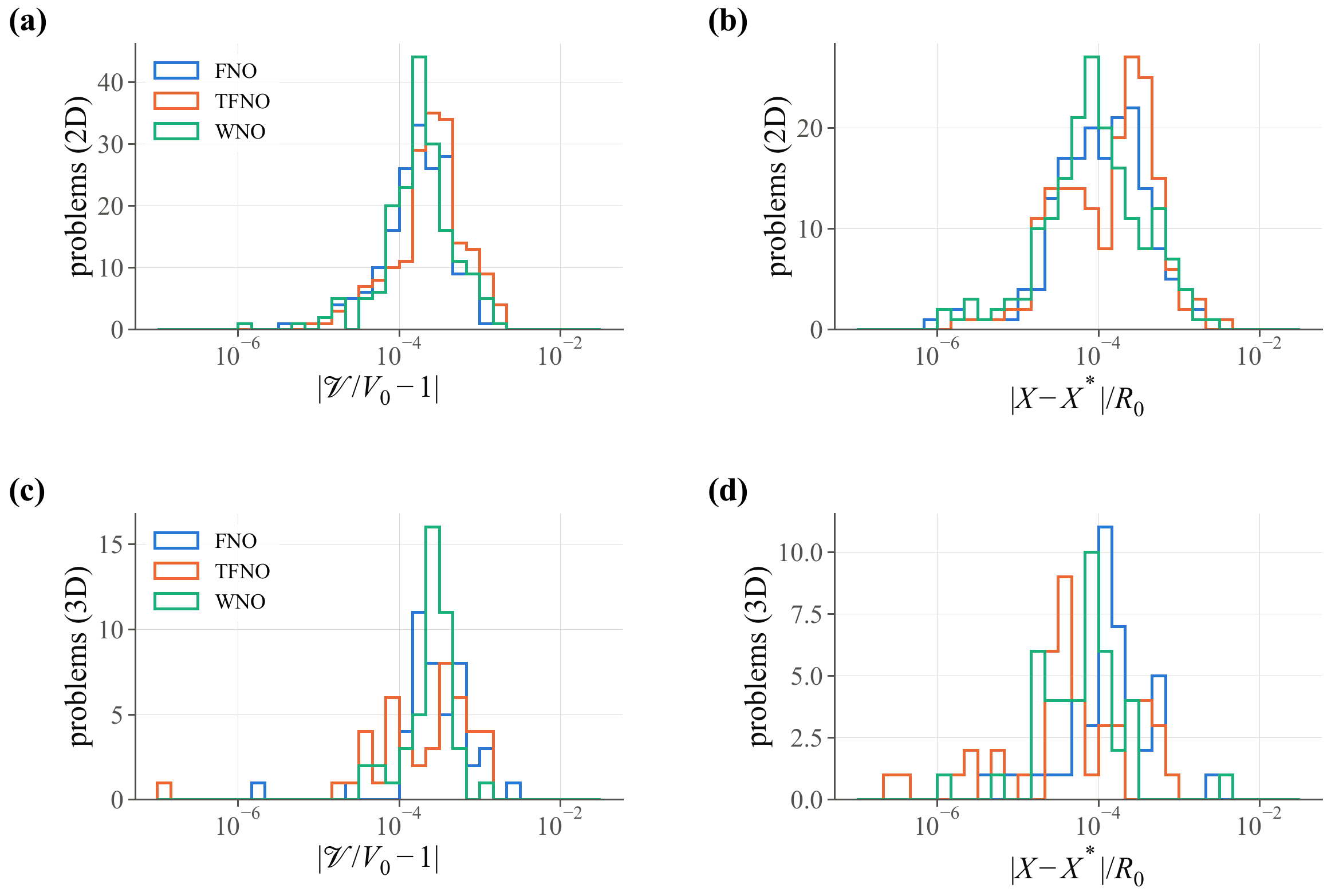}
\caption{Constraint residuals of the predictions on all held-out problems: (a,c) volume, (b,d) centroid; (a,b) two
dimensions, (c,d) three dimensions. Residuals below $10^{-7}$ are shown at $10^{-7}$.}\label{fig:si_constraints}
\end{figure}

\begin{figure}[htbp]
\centering
\includegraphics[width=\textwidth]{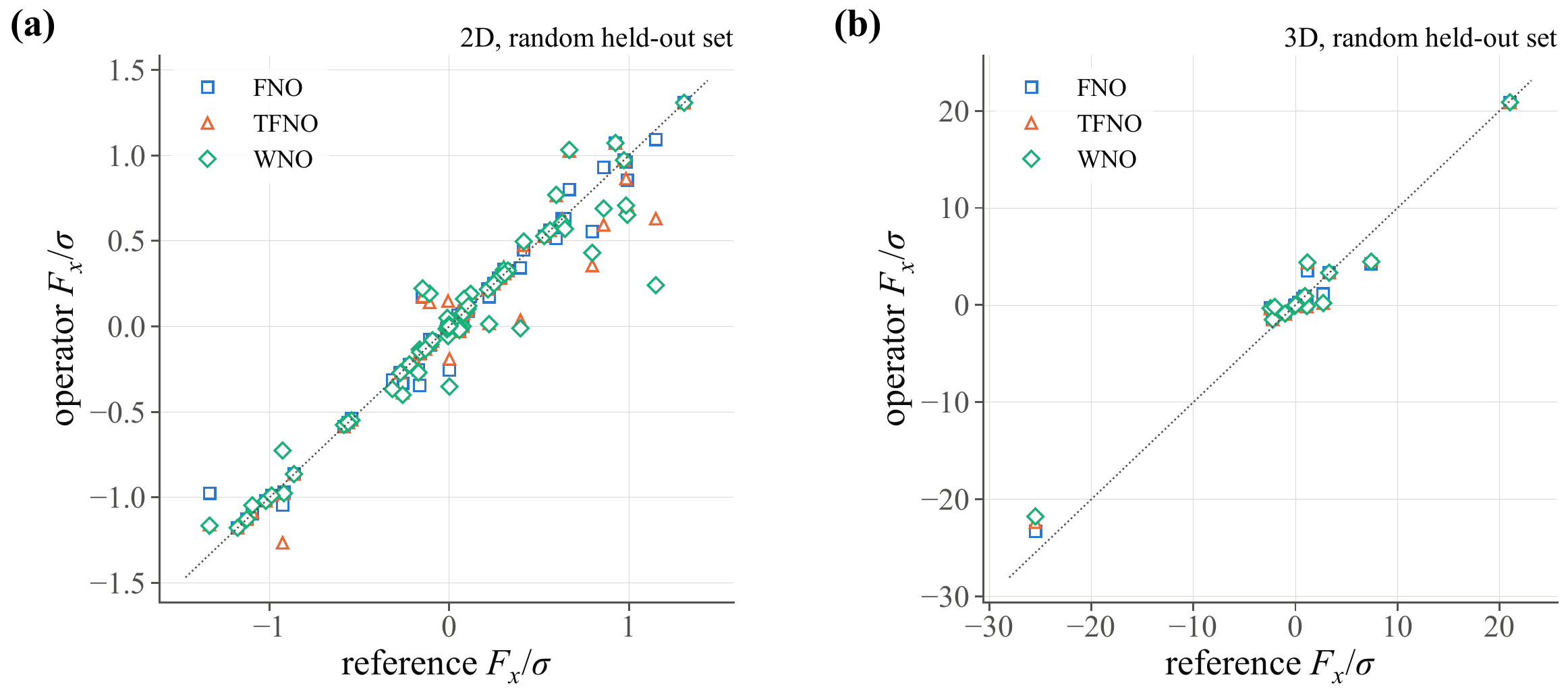}
\caption{Virtual-work force $F_x$, equation~(\xref{eq:hf}), of the operator predictions against that of the reference
minimiser on the random held-out problems: (a) two dimensions, (b) three dimensions. The dotted line is equality.}\label{fig:si_force}
\end{figure}

\begin{figure}[htbp]
\centering
\includegraphics[width=\textwidth]{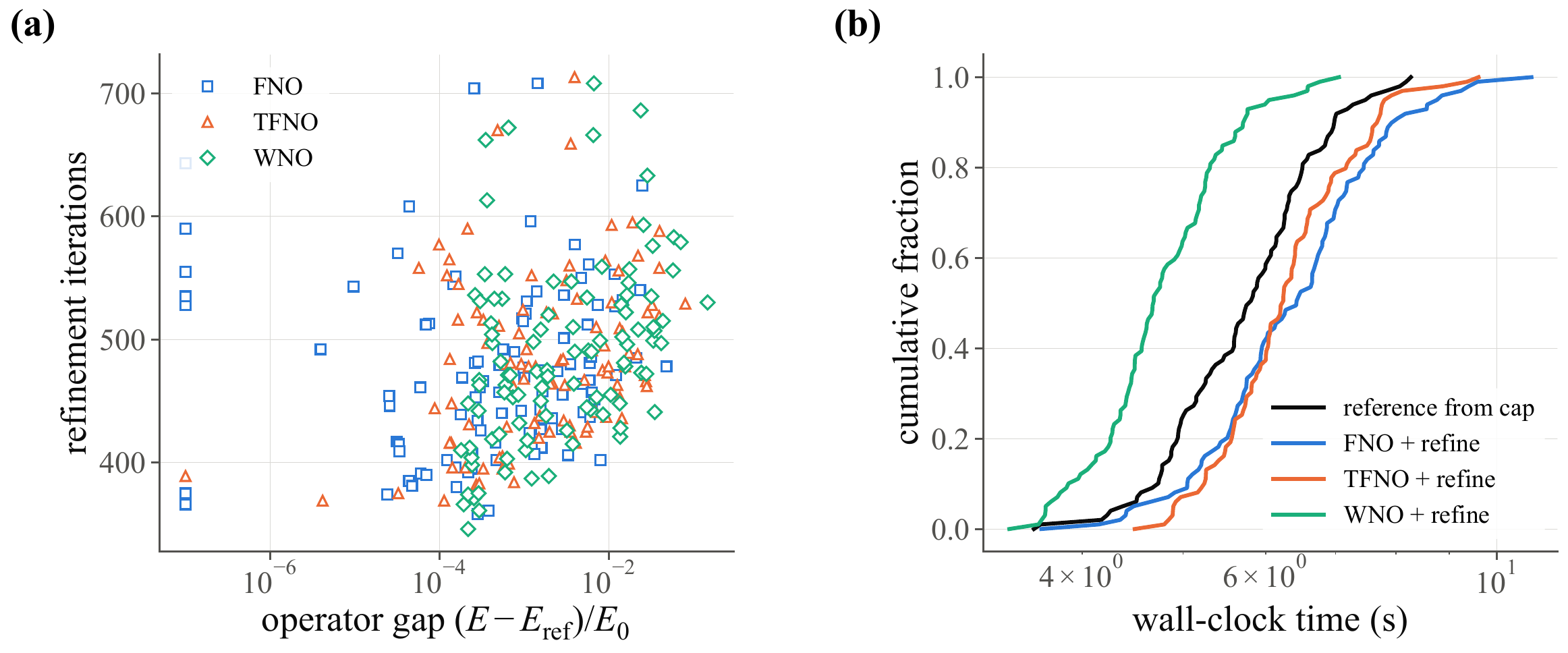}
\caption{Refinement of the two-dimensional predictions on the random held-out set. (a) Quasi-Newton iterations of the
refinement against the energy gap of the prediction. (b) Cumulative distribution of the wall-clock time of the reference
solve from the analytic cap and of the refinement of each operator's prediction.}\label{fig:si_refine}
\end{figure}

Table~\ref{tab:si_acc} extends {\color{linkblue}table~\xref{tab:accuracy}} of the main paper to every run and
adds the 90th percentile of the energy gap, the constraint residuals and the error of the virtual-work force.
Table~\ref{tab:si_fam} breaks the random held-out set down by family, table~\ref{tab:si_uniform} gives the apparent angles
on uniform walls behind panel (a) of \mfig{fig:verify}, and table~\ref{tab:si_pol} the cost of the refinement.
Figure~\ref{fig:si_constraints} shows that the penalty enforces the volume and the centroid to a few parts in $10^4$ for
every operator and both dimensions, so that the energy gaps of the main paper are not an artefact of violated
constraints. Figure~\ref{fig:si_force} compares the virtual-work force of each prediction with
that of the reference on all random problems, and figure~\ref{fig:si_refine} the cost of a refinement started from a
prediction with that of a solve started from the analytic cap.

\begin{table}[htbp]
\centering
\caption{Accuracy of every run on the random held-out set: energy gap (median and 90th percentile), relative field error, intersection-over-union of the liquid regions, volume and centroid residuals, and absolute error of the virtual-work force $F_x$ (in units of $\sigma$) with respect to the reference. Gaps of the main runs are taken to the best known minimum $E_{\rm best}$, those of the ablation runs to the reference minimum.}\label{tab:si_acc}
\footnotesize
\begin{tabular}{@{}lccccccc@{}}
\toprule
run & med.\ $\delta E$ & P90 $\delta E$ & $\|\delta\phi\|$ & IoU & $|\delta\mathcal{V}|/V_0$ & $|\delta X|/R_0$ & $|\delta F_x|$\\
\midrule
FNO 2D & $6.7\times10^{-4}$ & $6.8\times10^{-3}$ & 0.057 & 0.965 & $1.7\times10^{-4}$ & $1.0\times10^{-4}$ & 0.001\\
TFNO 2D ($r=32$) & $2.0\times10^{-3}$ & $2.2\times10^{-2}$ & 0.120 & 0.936 & $2.5\times10^{-4}$ & $1.2\times10^{-4}$ & 0.002\\
WNO 2D & $1.9\times10^{-3}$ & $3.3\times10^{-2}$ & 0.118 & 0.939 & $1.6\times10^{-4}$ & $9.4\times10^{-5}$ & 0.002\\
FNO 2D, seed 1 & $6.2\times10^{-4}$ & $8.1\times10^{-3}$ & 0.060 & 0.966 & $2.0\times10^{-4}$ & $9.6\times10^{-5}$ & 0.001\\
FNO 2D, no prior & $3.2\times10^{-3}$ & $1.3\times10^{-2}$ & 0.138 & 0.929 & $2.8\times10^{-4}$ & $2.4\times10^{-4}$ & 0.003\\
FNO 2D, $3\times$ budget & $1.8\times10^{-4}$ & $2.6\times10^{-3}$ & 0.026 & 0.986 & $2.1\times10^{-4}$ & $6.3\times10^{-5}$ & 0.000\\
TFNO 2D, $3\times$ budget & $9.0\times10^{-4}$ & $8.0\times10^{-3}$ & 0.063 & 0.966 & $1.9\times10^{-4}$ & $8.6\times10^{-5}$ & 0.001\\
TFNO 2D, $r=4$ & $2.3\times10^{-3}$ & $2.4\times10^{-2}$ & 0.126 & 0.939 & $2.6\times10^{-4}$ & $1.4\times10^{-4}$ & 0.002\\
TFNO 2D, $r=8$ & $2.0\times10^{-3}$ & $2.3\times10^{-2}$ & 0.123 & 0.941 & $1.9\times10^{-4}$ & $1.0\times10^{-4}$ & 0.002\\
TFNO 2D, $r=64$ & $1.7\times10^{-3}$ & $2.2\times10^{-2}$ & 0.118 & 0.941 & $2.3\times10^{-4}$ & $8.3\times10^{-5}$ & 0.001\\
\midrule
FNO 3D & $1.0\times10^{-3}$ & $2.2\times10^{-2}$ & 0.067 & 0.956 & $2.5\times10^{-4}$ & $1.1\times10^{-4}$ & 0.104\\
TFNO 3D & $2.0\times10^{-3}$ & $2.9\times10^{-2}$ & 0.080 & 0.944 & $3.7\times10^{-4}$ & $4.7\times10^{-5}$ & 0.075\\
WNO 3D & $1.8\times10^{-3}$ & $3.5\times10^{-2}$ & 0.069 & 0.954 & $2.5\times10^{-4}$ & $5.6\times10^{-5}$ & 0.113\\
\bottomrule
\end{tabular}
\end{table}

\begin{table}[htbp]
\centering
\caption{Random held-out set broken down by pattern family: number of problems, median energy gap $\delta E=(E-E_{\rm best})/E_0$ and median IoU of each operator.}\label{tab:si_fam}
\footnotesize
\begin{tabular}{@{}llrcccccc@{}}
\toprule
 & family & $n$ & \multicolumn{2}{c}{FNO} & \multicolumn{2}{c}{TFNO} & \multicolumn{2}{c}{WNO}\\
\cmidrule(lr){4-5}\cmidrule(lr){6-7}\cmidrule(lr){8-9}
 & & & $\delta E$ & IoU & $\delta E$ & IoU & $\delta E$ & IoU\\
\midrule
2D & uniform & 21 & $4.4\times10^{-5}$ & 0.988 & $1.6\times10^{-4}$ & 0.983 & $3.0\times10^{-4}$ & 0.985\\
2D & step & 26 & $1.1\times10^{-3}$ & 0.954 & $3.3\times10^{-3}$ & 0.915 & $5.5\times10^{-3}$ & 0.909\\
2D & stripes $\parallel z$ & 20 & $1.0\times10^{-3}$ & 0.970 & $2.7\times10^{-3}$ & 0.944 & $2.8\times10^{-3}$ & 0.949\\
2D & patches & 33 & $1.7\times10^{-3}$ & 0.942 & $5.3\times10^{-3}$ & 0.910 & $8.5\times10^{-3}$ & 0.910\\
\midrule
3D & uniform & 4 & $1.3\times10^{-4}$ & 0.986 & $7.4\times10^{-4}$ & 0.985 & $3.4\times10^{-4}$ & 0.985\\
3D & step & 3 & $2.1\times10^{-4}$ & 0.984 & $1.1\times10^{-3}$ & 0.975 & $5.3\times10^{-4}$ & 0.981\\
3D & stripes $\parallel z$ & 4 & $8.8\times10^{-3}$ & 0.865 & $1.3\times10^{-2}$ & 0.857 & $1.4\times10^{-2}$ & 0.860\\
3D & stripes $\parallel x$ & 1 & $2.2\times10^{-2}$ & 0.752 & $2.8\times10^{-2}$ & 0.741 & $3.5\times10^{-2}$ & 0.746\\
3D & patches & 4 & $1.1\times10^{-2}$ & 0.873 & $1.3\times10^{-2}$ & 0.871 & $1.4\times10^{-2}$ & 0.869\\
3D & checkerboard & 4 & $5.7\times10^{-3}$ & 0.903 & $8.1\times10^{-3}$ & 0.899 & $1.1\times10^{-2}$ & 0.898\\
\bottomrule
\end{tabular}
\end{table}

\begin{table}[htbp]
\centering
\caption{Uniform walls in two dimensions: apparent contact angle (degrees) measured by a circle fit to the interface of the reference minimiser and of each operator, and the energy gap $\delta E$ of each operator.}\label{tab:si_uniform}
\footnotesize
\begin{tabular}{@{}rrrrrccc@{}}
\toprule
$\theta_Y$ & ref. & FNO & TFNO & WNO & $\delta E$ FNO & $\delta E$ TFNO & $\delta E$ WNO\\
\midrule
40 & 37.7 & 40.4 & 42.8 & 40.7 & $6.1\times10^{-4}$ & $1.5\times10^{-3}$ & $1.1\times10^{-3}$\\
50 & 48.4 & 50.4 & 52.8 & 50.5 & $3.5\times10^{-4}$ & $9.7\times10^{-4}$ & $7.1\times10^{-4}$\\
60 & 59.4 & 60.2 & 62.2 & 60.4 & $7.0\times10^{-5}$ & $5.6\times10^{-4}$ & $3.1\times10^{-4}$\\
70 & 68.3 & 70.0 & 71.4 & 70.2 & $1.3\times10^{-4}$ & $3.3\times10^{-4}$ & $3.0\times10^{-4}$\\
80 & 79.8 & 79.9 & 80.7 & 80.0 & $2.5\times10^{-5}$ & $1.4\times10^{-4}$ & $1.9\times10^{-4}$\\
90 & 90.0 & 90.1 & 90.2 & 89.9 & $1.0\times10^{-6}$ & $3.7\times10^{-5}$ & $1.9\times10^{-4}$\\
100 & 101.4 & 100.1 & 99.8 & 99.8 & $9.9\times10^{-5}$ & $2.0\times10^{-4}$ & $3.4\times10^{-4}$\\
110 & 109.2 & 109.7 & 109.5 & 109.6 & $1.0\times10^{-6}$ & $5.5\times10^{-6}$ & $2.0\times10^{-4}$\\
120 & 118.4 & 119.8 & 119.4 & 119.5 & $4.9\times10^{-5}$ & $5.3\times10^{-5}$ & $2.3\times10^{-4}$\\
130 & 130.6 & 130.1 & 129.4 & 129.4 & $1.0\times10^{-6}$ & $1.8\times10^{-4}$ & $2.9\times10^{-4}$\\
140 & 141.5 & 140.3 & 139.3 & 139.3 & $1.0\times10^{-6}$ & $1.6\times10^{-4}$ & $4.6\times10^{-4}$\\
150 & 150.6 & 150.4 & 149.1 & 149.1 & $1.3\times10^{-4}$ & $9.4\times10^{-4}$ & $5.9\times10^{-4}$\\
\bottomrule
\end{tabular}
\end{table}

\begin{table}[htbp]
\centering
\caption{Refinement of the operator predictions by the reference minimiser: number of problems, median number of quasi-Newton iterations and wall-clock time of the refinement, median time of the reference solve from the analytic cap, and median relative field difference between the refined prediction and the reference.}\label{tab:si_pol}
\footnotesize
\begin{tabular}{@{}lllrrrrc@{}}
\toprule
 & operator & set & $n$ & iterations & $t_{\rm refine}$ (s) & $t_{\rm ref}$ (s) & $\|\phi_{\rm refined}-\phi_{\rm ref}\|$\\
\midrule
2D & FNO & uniform & 12 & 368 & 4.0 & 3.7 & $4.8\times10^{-8}$\\
2D & FNO & step & 51 & 478 & 6.8 & 5.8 & $6.0\times10^{-8}$\\
2D & FNO & stripes & 16 & 409 & 4.1 & 4.3 & $7.0\times10^{-8}$\\
2D & FNO & random & 100 & 466 & 6.4 & 5.8 & $6.7\times10^{-8}$\\
2D & TFNO & uniform & 12 & 348 & 5.2 & 3.7 & $2.7\times10^{-8}$\\
2D & TFNO & step & 51 & 490 & 6.8 & 5.8 & $5.0\times10^{-8}$\\
2D & TFNO & stripes & 16 & 414 & 5.6 & 4.3 & $4.2\times10^{-8}$\\
2D & TFNO & random & 100 & 478 & 6.2 & 5.8 & $6.3\times10^{-8}$\\
2D & WNO & uniform & 12 & 343 & 4.7 & 3.7 & $9.0\times10^{-8}$\\
2D & WNO & step & 51 & 508 & 6.8 & 5.8 & $4.3\times10^{-8}$\\
2D & WNO & stripes & 16 & 407 & 3.9 & 4.3 & $5.3\times10^{-8}$\\
2D & WNO & random & 100 & 480 & 4.7 & 5.8 & $5.3\times10^{-8}$\\
\midrule
3D & FNO & uniform & 4 & 428 & 87.1 & 49.4 & $1.0\times10^{-7}$\\
3D & FNO & step & 9 & 550 & 118.3 & 87.1 & $7.4\times10^{-8}$\\
3D & TFNO & uniform & 4 & 390 & 82.7 & 49.4 & $9.5\times10^{-8}$\\
3D & TFNO & step & 9 & 527 & 116.9 & 87.1 & $7.2\times10^{-8}$\\
3D & WNO & uniform & 4 & 414 & 78.8 & 49.4 & $5.4\times10^{-8}$\\
3D & WNO & step & 9 & 548 & 104.2 & 87.1 & $9.2\times10^{-8}$\\
\bottomrule
\end{tabular}
\end{table}

\section{Galleries of the TFNO and WNO predictions}\label{sec:gal}
\begin{figure}[htbp]
\centering
\includegraphics[width=0.92\textwidth]{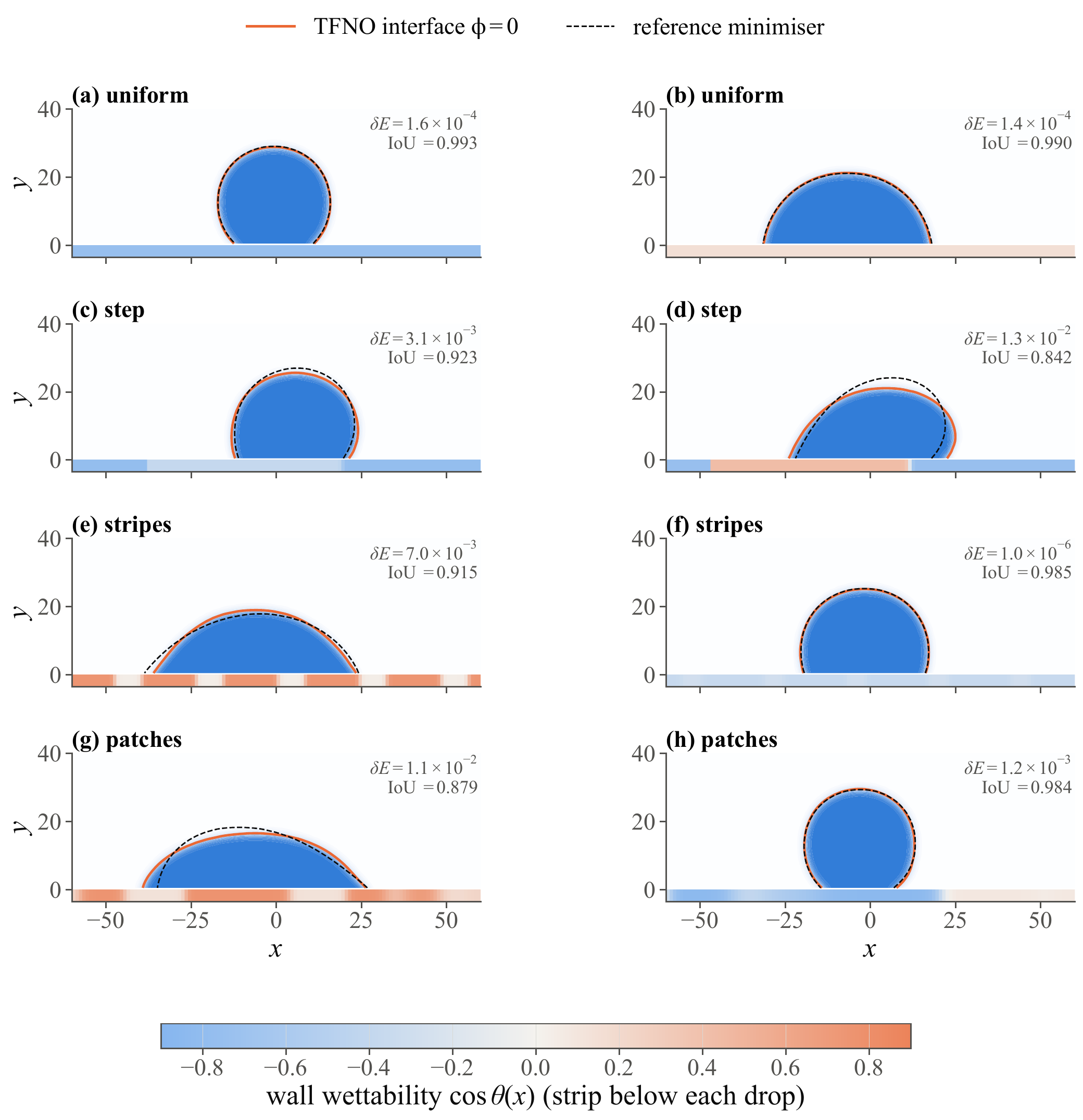}
\caption{As \mfig{fig:gallery2d}, for the TFNO.}\label{fig:si_g2t}
\end{figure}

\begin{figure}[htbp]
\centering
\includegraphics[width=0.92\textwidth]{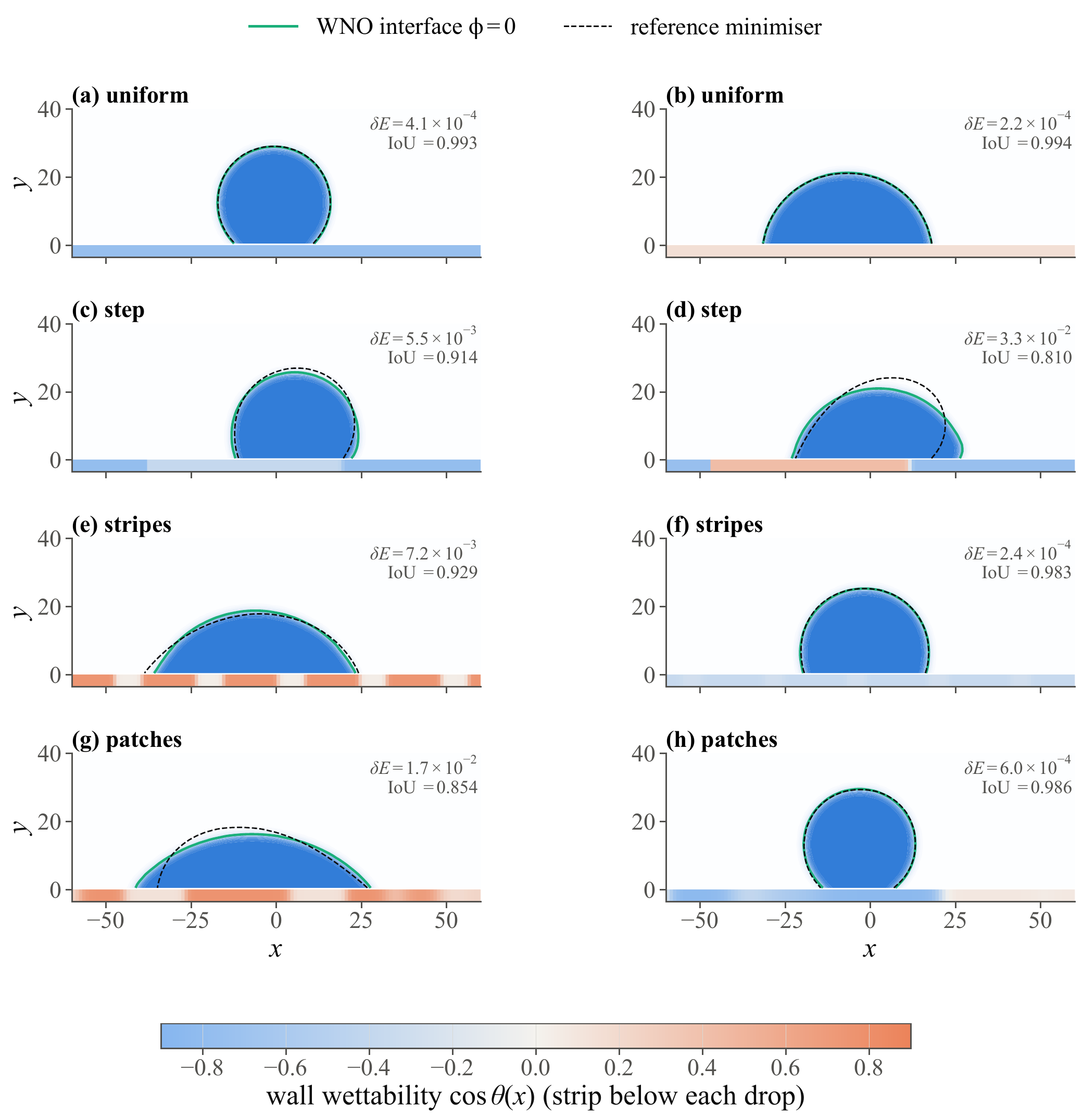}
\caption{As \mfig{fig:gallery2d}, for the WNO.}\label{fig:si_g2w}
\end{figure}

\begin{figure}[htbp]
\centering
\includegraphics[width=\textwidth]{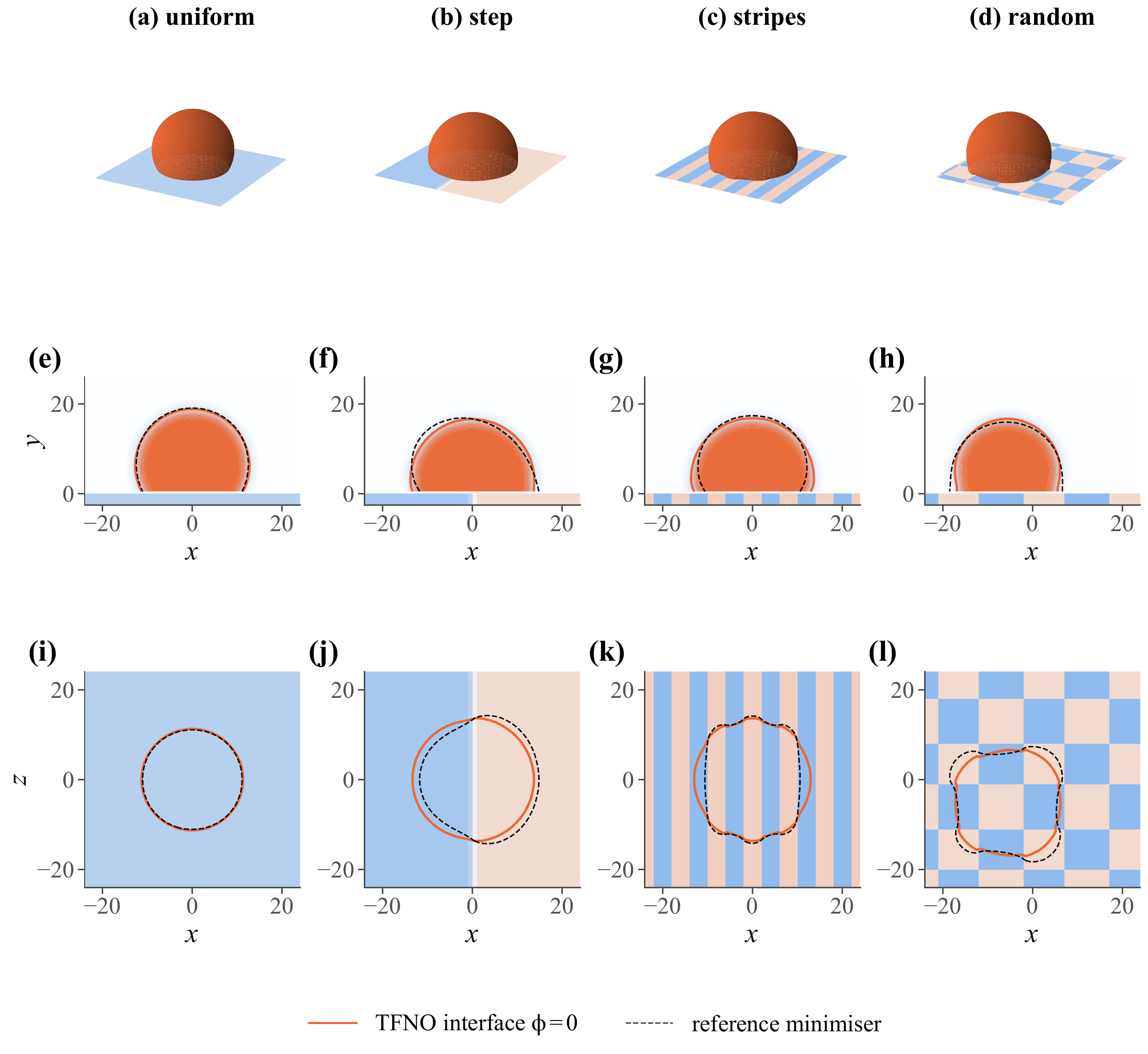}
\caption{As \mfig{fig:gallery3d}, for the TFNO.}\label{fig:si_g3t}
\end{figure}

\begin{figure}[htbp]
\centering
\includegraphics[width=\textwidth]{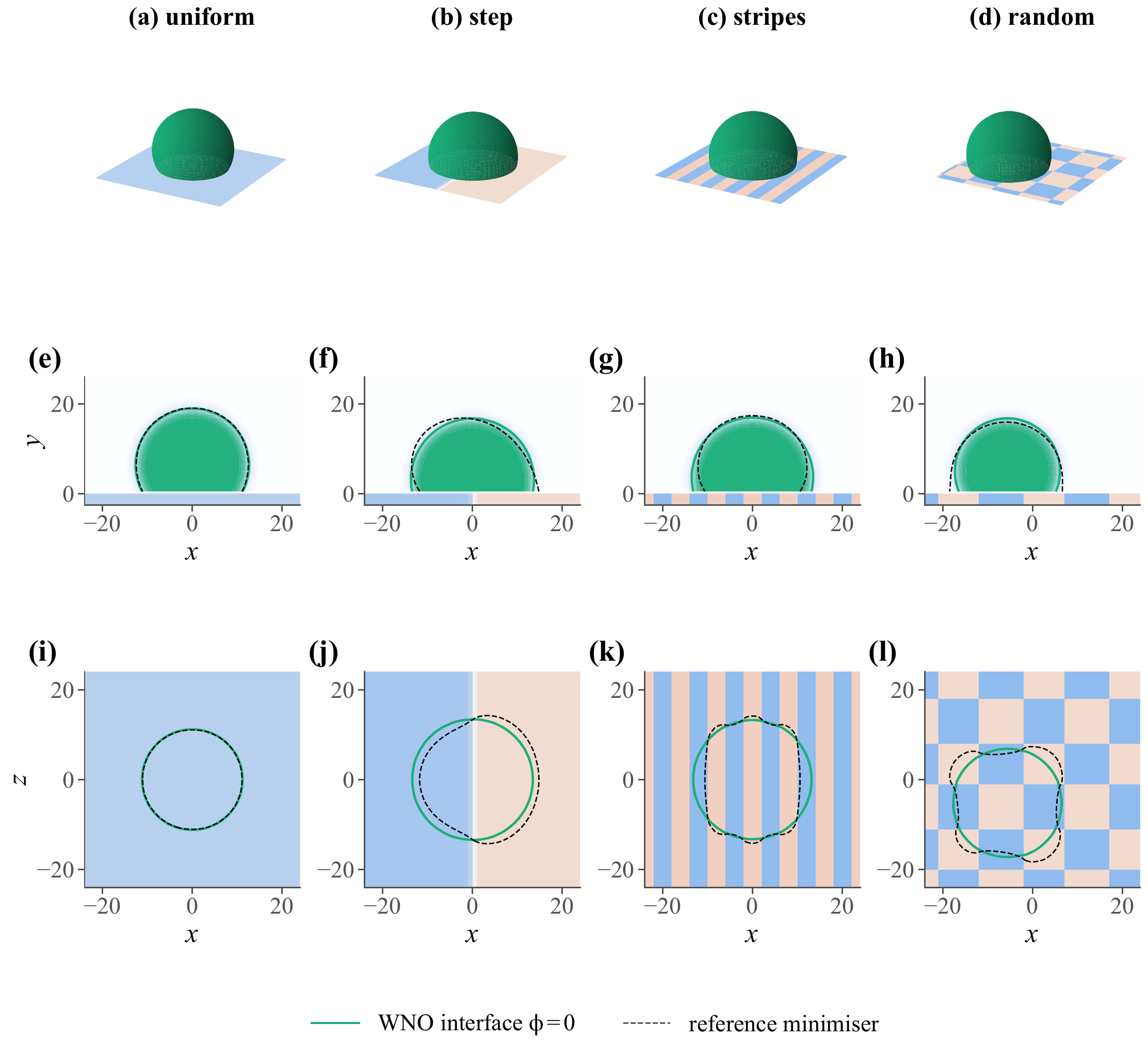}
\caption{As \mfig{fig:gallery3d}, for the WNO.}\label{fig:si_g3w}
\end{figure}

Figures~\ref{fig:si_g2t}--\ref{fig:si_g3w} repeat \mfig{fig:gallery2d} and \mfig{fig:gallery3d} for the other two
operators on the same held-out problems.

\section{Stiffness diagnostic}\label{sec:stiff}
The diagnostic of panel (a) of \mfig{fig:diagnostics} measures how stiff the error of a prediction is with respect to the energy:
the Rayleigh quotient of the error with the Hessian of the discrete energy at the reference minimum, compared with that
of interface-localised noise and of the translation mode of the interface (algorithm~\ref{alg:stiff}). The Hessian is never
formed: each quotient needs one Hessian-vector product by double automatic differentiation of the energy.

\begin{algorithm}[htbp]
\caption{Stiffness of the operator error (panel (a) of \mfig{fig:diagnostics})}\label{alg:stiff}
\begin{algorithmic}[1]
\Require reference minimum $\phi_{\rm ref}$, prediction $\phi_{\bm w}$, wall field
\Function{Ray}{$v$}
  \State $\bm h\gets\nabla_\phi\big(\nabla_\phi E[\phi_{\rm ref}]\cdot v\big)$ \Comment{Hessian-vector product by autograd}
  \State \Return $(v\cdot\bm h)/(v\cdot v)$
\EndFunction
\State $w\gets1-\phi_{\rm ref}^2$ \Comment{weight localised on the interface}
\State $s_{\rm op}\gets\Call{Ray}{\phi_{\bm w}-\phi_{\rm ref}}$
\State $s_{\rm white}\gets\Call{Ray}{w\,\xi}$, \ $\xi$ white Gaussian noise
\State $s_{\rm smooth}\gets\Call{Ray}{w\,\bar\xi}$, \ $\bar\xi$ noise averaged over $7\times7$ cells
\State $s_{\rm transl}\gets\Call{Ray}{\phi_{\rm ref}(x+1)-\phi_{\rm ref}(x-1)}$ \Comment{rigid shift of the interface}
\State \Return $s_{\rm op}$, $s_{\rm white}$, $s_{\rm smooth}$, $s_{\rm transl}$
\end{algorithmic}
\end{algorithm}